%% file: wyuan.tex
\documentclass[numberedappendix]{emulateapj}
\usepackage[]{apjfonts}
\usepackage{graphics}
\usepackage{lscape}
\def\av{$\alpha_{\nu}$}
\def\alpox{$\alpha_{\rm ox}$ }
\def\alpoxe{$\alpha_{\rm ox}$}
\def\alprx{$\alpha_{\rm rx}$ }
\def\alprxe{$\alpha_{\rm rx}$}
\def\alpro{$\alpha_{\rm ro}$ }
\def\alproe{$\alpha_{\rm ro}$}
\def\alpr{$\alpha_{\rm r}$ }
\def\alpre{$\alpha_{\rm r}$}
\def\alpx{$\alpha_{\rm x}$ }
\def\alpxe{$\alpha_{\rm x}$}
\def\alpwvl{$\alpha_{\lambda}$ }
\def\alpwvle{$\alpha_{\lambda}$}
\def\agm{$\alpha_{\rm 327M}^{\rm 1.4G}$ }

\def\amm{$\alpha_{\rm 151M}^{\rm 327M}$ }
\def\amme{$\alpha_{\rm 151M}^{\rm 327M}$}
\def\chisq{$\chi^2$ }

\def\gnh{$N_{\rm H}^{\rm Gal}$ }
\def\gnhe{$N_{\rm H}^{\rm Gal}$}
\def\hb{H$\beta$ }
\def\hbe{H$\beta$}
\def\ha{H$\alpha$ }
\def\hae{H$\alpha$}
\def\feii{\ion{Fe}{2} }
\def\feiie{\ion{Fe}{2}}
\def\oiii{[\ion{O}{3}] }
\def\oiiie{[\ion{O}{3}]}
\def\othb{[\ion{O}{3}]$\lambda5007/{\rm H}\beta$ }
\def\othbe{[\ion{O}{3}]$\lambda5007/{\rm H}\beta$}
\def\lhb{$L_{{\rm H}\beta}$ }
\def\lhbe{$L_{{\rm H}\beta}$}

\def\rfe{$R_{4570}$ }
\def\rfee{$R_{4570}$}
\def\mbh{$M_{\rm BH}$ }
\def\mbhe{$M_{\rm BH}$}
\def\msun{\ensuremath{M_{\odot}} }
\def\msune{\ensuremath{M_{\odot}}}
\def\redd{$R_{\rm Edd}$ }
\def\redde{$R_{\rm Edd}$}

\def\llf{$L_{\rm 151MHz}$ }
\def\llfe{$L_{\rm 151MHz}$}
\def\rl{$R_{\rm 1.4}$ }
\def\rle{$R_{\rm 1.4}$}
\def\ps{$P_{\rm chance}^{\rm sp}$}

\def\pch{$P_{\rm chance}$}

\def\ucre{${\rm cts\,s^{-1}}$}
\def\ergs{${\rm erg\,s^{-1}\,cm^{-2}}$ }

\def\ulum{${\rm erg\,s^{-1}}$ }
\def\ulume{${\rm erg\,s^{-1}}$}

\def\umlume{${\rm erg\,s^{-1}\,Hz^{-1}}$}
\def\nh{$N_{\rm H}$ }

\def\unhe{${\rm cm^{-2}}$}
\def\kmps{${\rm km\,s^{-1}}$ }
\def\kmpse{${\rm km\,s^{-1}}$}

\def\whze{$\rm W\,Hz^{-1}$}

\def\whzsre{$\rm W\,Hz^{-1}\,sr^{-1}$}
\slugcomment{accepted for publication in ApJ}
\shorttitle{Radio-loud Narrow Line Seyfert\,1 galaxies}
\shortauthors{W. Yuan, et al.}

\begin{document}

\title{A Population of Radio-Loud  Narrow Line Seyfert\,1 Galaxies
with Blazar-like Properties?}

\author{
W. Yuan\altaffilmark{1},
H. Y. Zhou\altaffilmark{2,3,4},
S. Komossa\altaffilmark{4},
X. B. Dong\altaffilmark{2,3},
T. G. Wang\altaffilmark{2,3},
H. L. Lu\altaffilmark{2,3},
J. M. Bai\altaffilmark{1}
}

\altaffiltext{1}{National Astronomical Observatories/Yunnan Observatory,
Chinese Academy of Sciences, Kunming, Yunnan, P.O. BOX 110, P.R.China}

\altaffiltext{2}{Center for Astrophysics, University of Science and Technology of China,
Hefei, Anhui, 230026, P.R.China}

\altaffiltext{3}{Joint Institute of Galaxies and Cosmology, SHAO and USTC}

\altaffiltext{4}{Max-Planck-Institut f\"{u}r extraterrestrische Physik,
Postfach 1312, 85741 Garching, Germany}

\email{wmy@ynao.ac.cn}

\begin{abstract}
Blazars with strong emission lines were found to be  associated
mostly with  broad-line type 1 Active Galactic Nuclei (AGN).
Hitherto, evidence for blazars identified with Narrow Line
Seyfert\,1 (NLS1) AGN was limited to very few individual cases. Here
we present a comprehensive study of a sample of 23 genuine radio-loud
NLS1 galaxies which have the radio-loudness parameters,
the ratio of radio (21\,cm) to optical (4400\AA) luminosity,
greater than 100.
The sample, drawn from the SDSS
and FIRST, is homogeneous and the largest of this kind.
A significant fraction of the sample objects show interesting radio to X-ray
properties that are unusual to most of the previously known
radio-loud NLS1 AGN, but are reminiscent of blazars.
These include flat radio spectra,
large amplitude flux and spectral variability, compact VLBI cores,
very high brightness temperatures ($10^{11-14}$\,K)
derived from variability, enhanced optical emission in excess of the
normal ionising continuum, flat X-ray spectra, and blazar-like SEDs.
We interpret them as evidence for the postulated blazar nature of these
very radio-loud NLS1 AGN, which might possess at least
moderately relativistic jets.
We suggest that those steep spectrum radio-loud NLS1 AGN in the sample
are of the same population but with their radio jets aligned
at large angles to the lines-of-sight.
Intrinsically, some of the objects have relatively low radio power
and would have been classified as radio-intermediate AGN.

The black hole masses, estimated from the broad Balmer line width
and luminosity, are within $10^{6-8}$\,\msune, and the inferred
Eddington ratios are around unity.
Unless the black hole masses are largely under-estimated,
our result stretches the low mass end of the
black holes of luminous, fast accreting radio-loud AGN to
a smaller mass regime (the order of  $10^6$\,\msune)
in the black hole mass--radio-loudness space
where other normal AGN are seldom found.
The results imply that radio-loud AGN may be powered
by black holes with moderate masses ($\sim10^{6-7}$\,\msune)
accreting at high rates
(Eddington ratios up to unity or higher).
The host galaxies of a few nearby objects appear to be disk-like
or merger;  and some of the objects show
imprints of young stellar populations in their SDSS spectra.
We find that some of the objects, despite having strong emission lines,
resemble high-energy peaked BL Lacs in their SED with the synchrotron
component peaked at around the UV;
such objects constitute an intriguingly high fraction of the sample.
The radio sources of the sample are ubiquitously compact.
They are smaller than  at most several tens of kilo-parsecs,
suggesting a possible link with compact steep-spectrum radio sources.
Given the peculiarities of blazar-like NLS1 galaxies,
questions arise as to whether they are plain downsizing extensions
of normal radio-loud AGN,
or whether they form a previously unrecognised population.
\end{abstract}

\keywords{galaxies: active --- galaxies: Seyfert  -- galaxies: jets
-- quasars: general -- X-rays: galaxies -- radio continuum: galaxies}

\section{Introduction}

As a minority of AGN, radio-loud (RL) AGN
differ from their radio-quiet (RQ) counterparts mainly in possessing
prominent radio jets and/or lobes that produce strong radio radiation.
In observations,
a widely used division between RL and RQ AGN
is the radio-loudness parameter $R\approx$10,
defined as the ratio between the radio 5\,GHz to
optical $B$-band luminosity \citep[][]{kel89,sto92},
though it is still a controversy
whether $R$ has a bimodal distribution below and above this value.
RL AGN are an important laboratory
to study the formation of relativistic jets,
which is not yet understood so far
\citep[e.g.][]{bla00,cel01,mei03}.
Nevertheless, it is widely thought
that radio-loudness (formation of radio jets)
is possibly depending on
the accretion rate/state \citep[e.g.][]{ho02,mac03,gre06,kor06},
the spin of the black hole
\citep[e.g.][]{bla00,mar03,sik07},
black hole mass \citep[\mbhe, e.g.][but see Woo \& Urry 2002
for a different view]{laor00,lacy01,dun03,mcl04,metca},
and host galaxy morphology \citep[e.g.][]{balm06,cap06,sik07},
or a combination of some of these factors \citep[e.g.][]{ball07}.

As an important sub-class of RL AGN, blazar is
a collective term for BL Lac objects and flat-spectrum radio quasars (FSRQs).
Blazars are characterised by flat radio spectra at above $\sim$1\,GHz,
fast variability, high and variable polarization,
superluminal motion, and high brightness temperatures
\citep[see e.g.][for reviews]{urry95,cel02}.
They are now believed to be RL AGN with the orientation of
relativistic jets close to the line-of-sight, and hence
their  non-thermal jet emission is highly Doppler boosted
\citep{bla78,urry95}.
Blazars have distinctive spectral energy distributions (SED)
that are characterised by two broad humps in the
$\log\nu-\log\nu f_{\nu}$ representation.
The low-energy hump is commonly interpreted as synchrotron emission,
while the high-energy one as inverse Compton emission scattering
off the same electron population which produces the synchrotron emission.
Depending on the peak frequency of the synchrotron hump,
BL Lac objects are divided into Low-energy-peaked BL Lacs (LBL)
with the peak around IR--optical wavelengths and
High-energy-peaked BL Lacs (HBL) with the peak around UV/soft X-ray energies
\citep{pado95}.
Classical FSRQs have generally the peak frequencies of the synchrotron hump
similar to or even lower than LBL.
Interestingly, it was suggested that the whole blazar family can be arranged in
a sequence from HBL to LBL and to FSRQ in the order of decreasing
synchrotron peak frequencies and increasing source power---the so called
blazar sequence\footnote{
The blazar sequence can be explained as, for instance,
the decrease of the energy of
electrons emitting at the SED peaks with increasing energy density
of the seed photons for Compton scattering \citep{ghi98}
}
\citep[][]{fos98}.
However,  exceptions were also reported in recent years
as the presence of powerful FSRQs with HBL-like SEDs
(though this remains a matter of controversy),
whose synchrotron hump is peaked around the UV/soft X-ray band
\citep[see e.g.][and references therein]{pado03}.
Following \citet{per98}, we term FSRQs with HBL-like SED
HFSRQs (High-energy-peaked FSRQs),
and those with LBL-like SED---the classical FSRQs---LFSRQs
(Low-energy-peaked FSRQs), as parallels to HBLs and LBLs.

Until recently, RL AGN (blazars) with strong broad
emission lines were found to ubiquitously have the Balmer line widths
greater than 2000\,\kmps in full width at half maximum (FWHM).
It was found that the line width is correlated with
the orientation of the radio axis, with narrower lines as well as
stronger optical \feii emission in objects
with smaller viewing angles to the radio axis \citep[e.g.][]{wills86,jack91}.
This was interpreted as arising from a disk-like emission line region.
A marked absence of radio-loud quasars (RLQs) was noted
below the 2000\,\kmps line-width cutoff \citep{wills86},
that  is the characteristic line-width range for
narrow line Seyfert\,1 galaxies---a sub-class of type\,1 AGN
in contrast to the classical Broad Line AGN (BLAGN).
The situation has changed since the detection of
radio emission from NLS1 galaxies.

Apart from narrow line-widths of the broad Balmer lines
\citep[originally defined as $<2000$\,\kmpse,][]{ost85,goo89},
NLS1 galaxies also show other extreme properties
compared to normal broad line AGN \citep[see][for a recent review]{kom08},
such as strong permitted optical/UV \feii emission lines
\citep{bg92,gru99,veron01},
steep soft X-ray spectra \citep{wang96,bol96,gru98},
and rapid X-ray variability \citep{lei99,kom00}.
Observational evidence suggests that NLS1 galaxies tend to have small
black hole masses and high Eddington ratios
(defined as the bolometric to the Eddington luminosity ratio),
\redd$\equiv L_{\rm bol}/L_{\rm Edd}\approx 1$
 \citep[e.g.][]{bor02,col04}.
In fact, they were found to locate at one extreme end,
opposite to BLAGN,
of eigenvector\,1 of the correlation matrix
which is believed to be driven primarily by \redd
\citep{bg92,sul00}.
NLS1 galaxies  were once thought to be radio-quiet.
This was simply a consequence of the fact that
NLS1 galaxies have a low radio-loud fraction \citep{zhou06,kom06b}
and the small number of NLS1 galaxies known till then.

Previous studies of the radio properties of NLS1 galaxies
 are all based on small samples.
These showed that the radio sources, if detected,
are compact (less than a few hundred parsec),
at modest radio power \citep{ulv95},
of steep radio spectra \citep{moran00},
mostly in the RQ regime \citep{ste03},
and of low RL fraction \citep{zhou02}.
Until recently, there were only several RL NLS1 galaxies
identified and studied individually
\citep[see][for a review and references therein]{kom06b}.
Studies using (small) samples of RL NLS1 galaxies were carried out by
\citet{kom06b} and \citet{wha06} for
non-radio selected and radio selected objects, respectively.
These studies confirmed that, as in `normal' NLS1 galaxies (their RQ counterparts),
RL NLS1 AGN are accreting at a high rate close to the Eddington limit.
More importantly, as pointed out by  \citet{kom06b},
with relatively low \mbh and high $R$, they occupy a previously sparsely
populated region in the \mbhe--$R$ diagram.
Most of the sources of that sample are compact steep spectrum sources.
It was confirmed that RL NLS1 galaxies are rare ($\sim7$\%),
and very radio-loud objects with $R>100$ are even more sparse
\citep[see also][]{zhou02,zhou06},
compared to about 10--15\% for normal BLAGN and quasars
\citep[e.g.][]{ive02}.
Several of these studies discussed the starburst versus AGN contribution
to the radio emission, and concluded that it is AGN dominated.

Of particular interest, a few outstanding RL NLS1 galaxies
came to be known to exhibit blazar properties.
RXJ\,16290+4007 \citep{sch00,gru04}
was known as a blazar and its X-ray emission is dominated by
synchrotron emission \citep{pado02},
or else by soft X-ray emission that is typical for NLS1 galaxies,
 perhaps from the accretion disk \citep{kom06b}.
The SED of PKS\,2004-447\footnote{However,
the NLS1 nature of PKS\,2004-447 is not certain;
see  \citet[][]{zhou03} for a short comment.} \citep{osh01}
was well modeled with the blazar type \citep{gal06}.
Using information on the radio flux variations,
very high radio brightness temperatures ($\gtrsim10^{13}$\,K)
were inferred for J0948+0022 \citep{zhou03}
and 0846+513\footnote{A formerly known blazar; see Appendix\,\ref{sect:indv}
for details.} \citep{zhou05}---both included in our sample,
arguing for relativistic beaming \citep[see also][]{wang06}.
They also show some other blazar-like behavior.
Recently, \citet{doi06,doi07} performed high-resolution
VLBI observations for several radio-loud NLS1 AGN
(three are included in our sample\footnote{J0948+0022,
J1633+4718, and J1644+2619.})
and found that they are unresolved with milli-arcsec resolutions,
setting direct lower limits on the
brightness temperatures in the range of $10^7-10^9$\,K.
They also found that inverted radio spectra are common in
the radio-loudest objects.
The authors suggested that Doppler beaming, presumably resulting from
highly relativistic jets, can explain naturally the observations.
In this respect, the most remarkable object of this kind is
perhaps 2MASX\,J0324+3410,
found in our recent work \citep{zhou07}.
It showed rapid variability in the radio, optical, and X-ray bands,
and was even claimed to be marginally detected in TeV $\gamma$-rays \citep{fal04}.
Its non-thermal SED clearly resembles that of HBLs.
Moreover, 2MASX\,J0324+3410 is hosted by a relatively small,
apparent disk galaxy with one-armed spiral or ring galaxy morphology.
These enigmatic objects present a challenge to
current models of both NLS1 galaxies and  blazars,
and render a unique opportunity to  study jet formation
in black hole systems accreting at high rates.
Apparently, they do not seem to belong to any types of AGN currently known.
An immediate question is whether they are some
exceptional individuals or there exists a population of such objects.
What are their general properties
in continuum and line emission, central engine and host galaxies?
What are their relations with other types of AGN?

Motivated by these issues, we compiled a  RL NLS1 galaxy sample
from the Sloan Digital Sky Survey (SDSS) database,
aimed at a systematic study with a large and homogeneous sample.
Since, as the first step,
we focus on genuine RL objects,
we consider in this paper objects with the radio-loudness parameter
$R>100$ (see \S\,\ref{sect:samplesel} for the definition of $R$)
for the following reasons.
Firstly,
the classification of AGN into RL and RQ is somewhat
ambiguous in the range $R=10-50$ due to the presence of (a small
number of) so called radio-intermediate AGN spanning in between the
bulks of the two classes, which may have a different origin from
genuine RL objects \citep[e.g.][]{fal96a}.
The same may also be true for NLS1 galaxies.
Secondly, in $R>100$ objects contamination of radio emission from
the host galaxy is negligible (see \S\,\ref{sect:disc_sed}).
Thirdly, their jet component, if present,
can be detected and studied relatively easily in
other wavebands than  the radio, such as optical and X-rays.
Last, but not the least,
they are the least studied objects {\em as a sample},
 due to their extreme rarity,
in contrast to less RL NLS1 galaxies as the bulk in the
samples of \citet{kom06b} and \citet{wha06}.
We defer to a later paper a comprehensive and complete treatment of
the SDSS--FIRST detected  NLS1 galaxies sample,
including radio-intermediate objects,
to address questions such as the RL--RQ
dichotomy for NLS1 galaxies.
Our results show that there exists a population
 of NLS1 galaxies that are blazar-like, despite their  rarity.

The compilation of the sample and the analyses of multi-waveband data
are described in \S\,\ref{sect:sample}.
The broad band continuum and
emission line properties are presented in
\S\,\ref{sect:cont} and \S\,\ref{sect:emline}, respectively.
In \S\,\ref{sect:hostgal} the properties of their host galaxies are presented.
We discuss other properties such as black hole mass and
accretion rate, and isotropic radio emission,
as well as the implications of the results in \S\,\ref{sect:discus}.
Throughout the paper, we assume a
cosmology with $H_{0}$= 70 km\, s$^{-1}$\,Mpc$^{-1}$,
$\Omega_{M}=0.3$, and $\Omega_{\Lambda}=0.7$.
Errors are quoted at the 68\% confidence level
unless mentioned otherwise.
We use the following conventions to denote power-law spectral indices:
\av\ in the frequency domain
[$S(\nu) \propto \nu^{\alpha_{\nu}}$, specifically \alpre$\equiv$\av\
in the radio band], and
\alpwvl in the wavelength domain
[$S(\lambda) \propto \lambda^{\alpha_{\lambda}}$];
and the X-ray photon index $\Gamma$, defined as
 $f_{\rm pho}(E) \propto E^{-\Gamma}$,
where $f_{\rm pho}(E)$ is X-ray photon number flux density.

\section{Sample compilation and multiwavelength data analysis}
\label{sect:sample}
\subsection{SDSS NLS1 galaxy sample}
\label{sect:nls1}

We have carried out a systematic search for NLS1 galaxies from the SDSS
spectroscopic samples and the first results have been published in
\citet{zhou06}. The procedures of data analysis
have been documented in that paper in detail
and only an outline is summarised here.
We carefully fit emission line spectra, AGN continua, and host galaxy
starlight in a self-consistent manner.
Firstly, galaxy starlight and AGN continuum,
as well as the optical \feii emission complex are modeled
(see Appendix\,\ref{sect:sdssspec} for a brief account)
and then subtracted.
The optical \feii multiplets are modeled with the \feii spectral data
given by \citet{veron04} for both the broad and narrow components.
Emission line spectra are fitted with the following models,
using a code similar to that described in \citet{dong05}.
The Balmer
emission lines are de-blended into a narrow and a broad component,
which are modeled by a Gaussian and a Lorentz profile,
respectively.
All narrow emission lines are fitted with a single Gaussian
except the \oiii$\lambda\lambda$4959/5007 doublet.
Each line of the \oiii doublet is fitted with two Gaussians,
one of which is used to account for a possible blue wing as seen in a few objects.
As shown in \citet{zhou06}, our analysis procedures yield
reliable and accurate measurements of emission line parameters.

The SDSS spectroscopic samples of both galaxies
and QSOs in the redshift range of $z<0.8$ were analyzed based on
the SDSS data release 5 \citep[DR5,][]{sdssdr5}.
Following \citet{zhou06}, we classify NLS1 galaxies as those having the ``broad"
component of H$\beta$ or H$\alpha$ which is detected at the
$10\,\sigma$ or higher confidence level and is narrower than
2200\,\kmps in FWHM. As discussed in \citet{zhou06}, objects
selected as such fulfill naturally the second of the
conventional criteria of NLS1 classification, i.e.\ \othb$<3$,
and the sample is reliable and uniform.
As a result, about 3300 NLS1 galaxies are found, superseding the
previously published 2011 drawn from the SDSS DR3
\citep[][hereafter Zhou06 sample]{zhou06}.

\subsection{Search for radio counterparts}
\label{sect:samplesel}

Having compiled a large NLS1 galaxies sample from the SDSS,
we search for their radio emission using data of the FIRST\footnote{
Faint Images of the Radio Sky at Twenty-centimeters}
survey \citep[][]{becker95}.
The survey has a spatial resolution of 5\arcsec and
typical 90\% source positional uncertainty ellipses
less than 1\arcsec.
The identification is performed by matching the
SDSS positions of NLS1 AGN against those of radio sources
following the procedure used by  \citet{lu07}.
This procedure was   designed to  search for
SDSS quasars detected in the  FIRST survey, and
is summarised here.
Since extended radio sources often have multiple (diffuse) components
in morphology and sometimes their positions are vague to define,
we treat extended sources differently from compact sources in search of
radio counterparts.
By compact FIRST source we mean that only one source is found
within 3\arcmin of the optical position and is  unresolved;
for them,  we use a 2\arcsec  matching radius.
We search for possible extended  radio sources associated with the NLS1
AGN in two steps.
First, candidates are selected if they match one of the following criteria.
(1) Only one resolved radio source is found within 3\arcmin
of the optical position, and the optical position is located
within the size of the radio source.
(2) Two radio sources are located nearly symmetrically around
the optical position within 3\arcmin.
(3) More than two radio sources are scattered around the
optical position within 3\arcmin.
Then we visually inspect $6\arcmin \times 6\arcmin$
cutouts of the FIRST images centered at an object to reject
false matches.
As discussed in  \citet{lu07},
this approach has proved to be effective for
finding extended radio sources associated with AGN;
 about 24\% of the radio quasars in the sample of \citet{lu07}
are resolved by the FIRST survey, among which half show
FR\,II type morphology \citep{fr74}.

This procedure yields a sample of SDSS NLS1 galaxies
detected in the FIRST survey.
Interestingly, we note that {\em all} the radio sources associated with
these NLS1 AGN are compact in morphology.
For each object we calculate the radio-loudness
parameter defined as
\rle$\equiv f_{\nu}({\rm 1.4\,GHz})/f_{\nu}(4400\,\mbox{\AA})$,
where the fluxes are in the rest frame of the
objects\footnote{It should be noted that our definition of
radio-loudness, \rle, is related to the previously commonly
used $R_{\rm 6cm} \equiv f_{\nu}(\rm 6\,cm)/f_{\nu}(4400\,\mbox{\AA})$
\citep{kel89} via \rle$=1.9 R_{\rm 6cm}$. Therefore, our radio-loudness
cutoff corresponds to $R_{\rm 6cm}\ga 50$.}.
The rest frame 4400\,\AA\ flux densities are
calculated from the SDSS g-magnitudes assuming a spectral slope
\av=$-0.5$.
An index of $-0.5$ is also assumed
for the K-correction for the radio fluxes.
We focus on very radio-loud objects only with \rle$>100$ in this paper,
and will present the whole sample elsewhere.

We note that
three NLS1 galaxies with apparent \rle$>100$ stand out for their extremely red
optical continua (\alpwvle$\ga 2.7$),
namely,  J095919.14+090659.4, J111354.66+124439.0, and J233903.82-091221.3.
Our original fit to the continuum spectrum with a fixed power law
slope \alpwvl= $-1.7$ yields substantial reddening of
$E(B-V)$=0.7--1.
Among them, J233903.82-091221.3 ($z$=0.66) has been found to show
convincing evidence for substantial dust extinction \citep{wang05}\footnote{
In addition to the extremely red optical continuum, its SDSS spectrum
shows signatures of reddening of broad emission lines,
detection of strong \ion{Ca}{2} and \ion{Mg}{2} absorption lines of
non-stellar origin \citep[see][for details]{wang05}.}.
Given the fact that the other two objects resemble closely J2339-0912
in their SDSS spectra, their optical light is most likely
subject to heavy extinction as well.
After correction for extinction,
their  radio-loudness parameters are largely reduced
 to \rle$\simeq 10-20$, below or merely close to the
RL/RQ dividing line.
We therefore exclude these three objects from our current sample.

\subsection{Very radio-loud NLS1 galaxy Sample}
\label{sample}

\begin{figure*}
\plotone{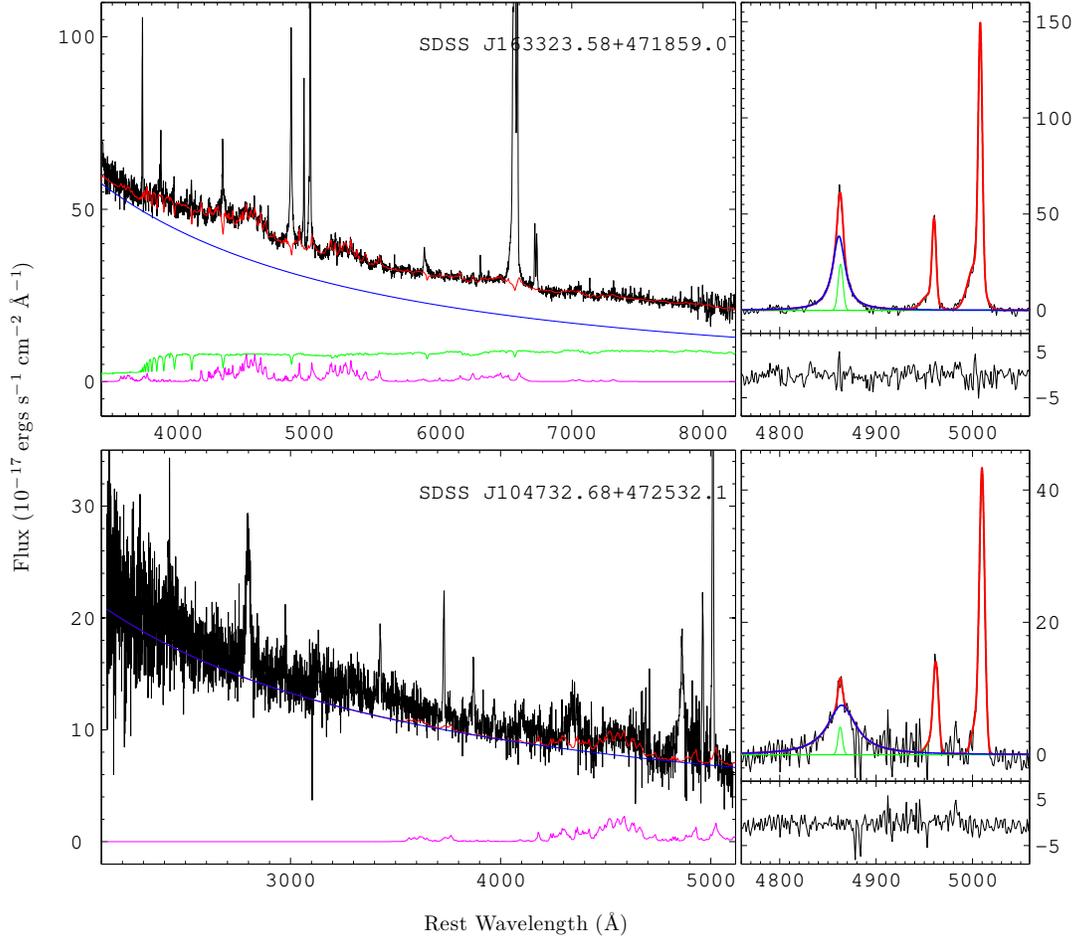}
\caption{ \label{fig:sdssspec}
Examples of the SDSS spectra
of two radio-loud NLS1 galaxies in our sample.
The left-hand side panels show the rest frame spectra and the fits
to the continuum with a nuclear power-law (blue color)
and optical \feii emission (magenta) model,
plus host galaxy starlight contribution (green) wherever non-negligible
(see \S\,\ref{sect:nls1} for a description of spectral analysis).
The panels on the right-hand side show a close-up of the
continuum-subtracted emission line spectra  and
the best fits  in the \hbe--\oiii region.
SDSS\,J1633+4718 (upper panel), a flat-spectrum radio source,
has the lowest radio-loudness and redshift,
and its spectrum is among the highest S/N;
it  has the largest fraction ($\sim$20\%)
of detectable host galaxy starlight contribution  among the sample.
SDSS\,J1047+4725 (lower panel),
a steep-spectrum radio source,
has the highest radio-loudness and redshift,
and its spectral S/N is among the lowest in the sample.
}
\end{figure*}

The above selection procedures result in 23 NLS1 galaxies having \rle$>100$,
which compose our sample.
By excluding less RL objects  with \rl$<100$
our sample can be regarded as very radio-loud to some extent.
We believe that the radio-loudness estimates of our sample
are not significantly affected by optical extinction,
as is discussed in \S\,\ref{sect:dis_opt}.
The objects are listed in Table\,\ref{tbl:sample}, along with some
relevant parameters of the continua and emission lines.
As demonstration, we show in Figure\,\ref{fig:sdssspec} examples of
the SDSS spectra and their best fit models for two representative objects;
they are chosen to
represent the ranges of some characterising parameters of the sample.
Measurable  starlight contribution is present
in only one object,  J1633+4718,  accounting for at most $\sim$20\% of
its observed spectral flux density at 5100\,\AA.
Most of our objects show strong \feii emission complexes
(see Section\,\ref{sect:emline}).
The \oiii lines are weak in general, all having \othb$<1.2$.
All of the objects fulfill the conventional
definition of NLS1 \citep{ost85,goo89}, and are therefore bona-fide NLS1 galaxies.

\begin{figure*}
\plottwo{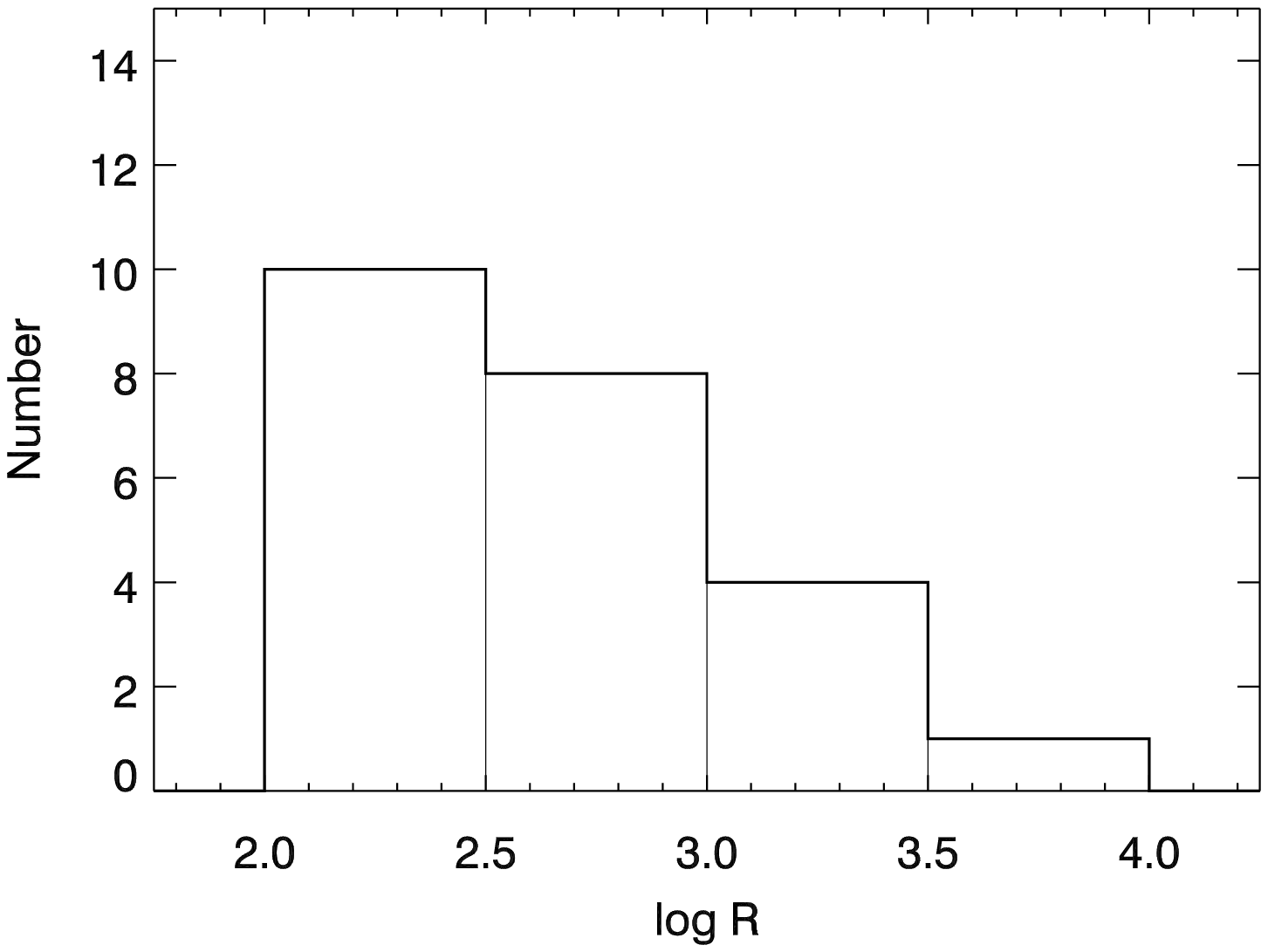}{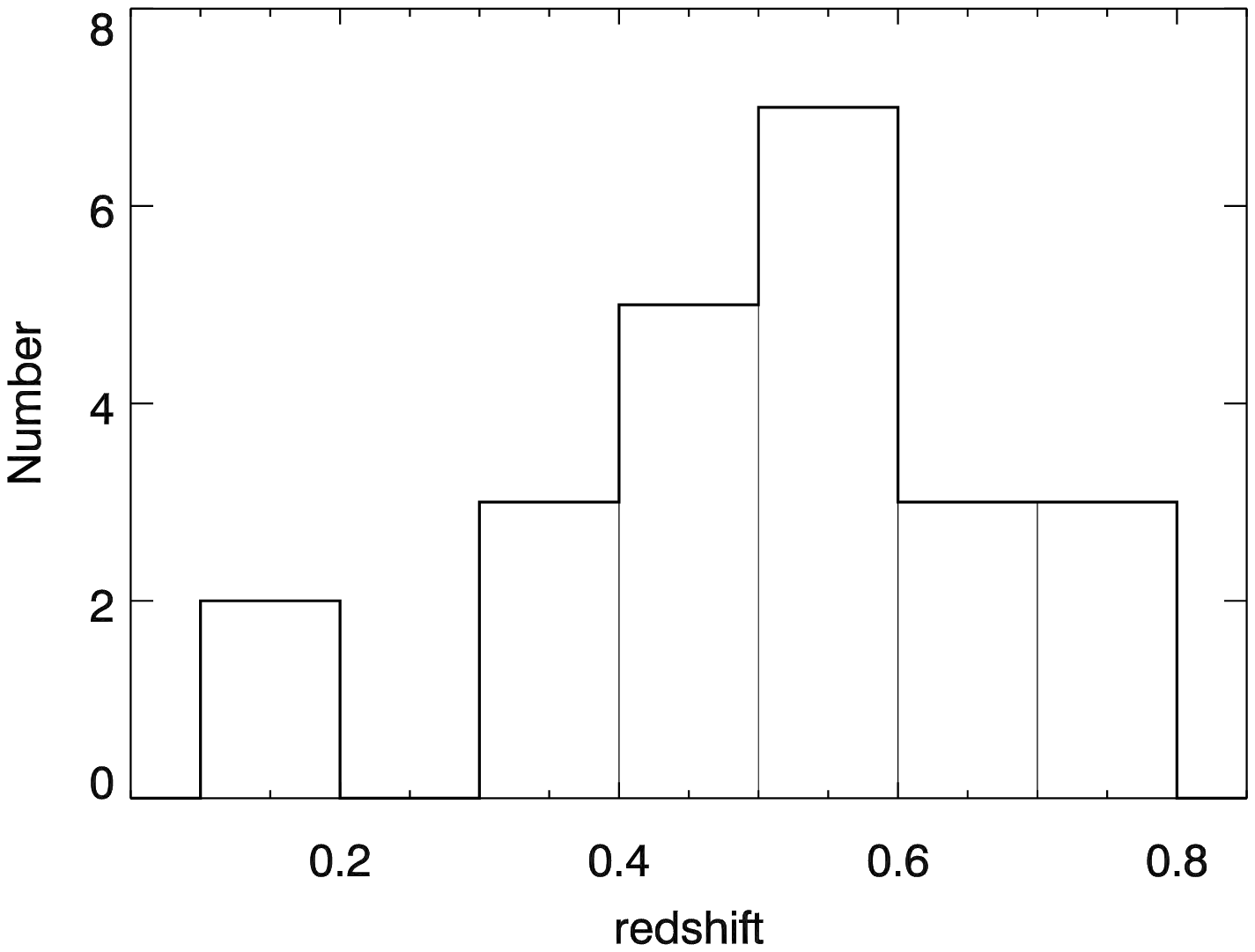}
\caption{ \label{fig:distrl}
Distribution of the radio loudness \rl (left panel)
and redshifts (right panel) of the radio-loud NLS1 galaxies sample.
}
\end{figure*}

The redshift and radio-loudness distributions are plotted in
Figure\,\ref{fig:distrl}.
Our sample has a relatively high redshift distribution
peaked at $z\sim0.5$ (the median).
We calculate
the $B$-band absolute magnitudes $M_{B}$ assuming
 \alpwvle=$-1.5$ for the K-correction,
which are listed in Table\,\ref{tbl:sample}.
The $M_{B}$ values range from $-20.8$ to $-25.6$, with a median of $-22.8$.
Among the sample, 10 objects are as bright as $M_{B}\leq -23$ and hence
can be classified as narrow-line type\,I quasars.
Three objects in the sample have been
previously identified with RL NLS1 galaxies and  studied in detail,
namely, J094857.32+002225.5 \citep{zhou03},
J084957.98+510829.0 \citep{zhou05}, and
J172206.03+565451.6 \citep{kom06a}.
In addition, several objects were known as general RL AGN
and have been studied previously,
whose properties are summarised individually in Appendix\,\ref{sect:indv}.

\subsection{X-ray and UV data analysis}
\label{sect:xray}

We have searched for X-ray emission from the sample objects using
ROSAT source catalogues of the RASS and pointed observations of
both the RXP \citep{voges99} and WGA \citep{wga95} catalogues.
The matching radii are 5 times the given positional errors
of ROSAT sources, which are typically 10\arcsec--20\arcsec.
For each match, we visually inspect their SDSS optical images
to avoid spurious matches in which the X-ray source is
actually associated with another nearby object within the matching radius.
Of the 23 objects, 12 were detected in the RASS
and 4 in pointed observations, among which 2 detected in both.
Thus a total of 14 objects---more than half of the sample---were
detected in X-rays with the ROSAT PSPC\footnote{Positional Sensitive Proportional Counter}.

All of the objects except J1633+4718 do not have
sufficient X-ray counts to allow proper spectral modeling.
Following \citet{sch96}, \citet{siethe}, and \citet{yuanthe}, we estimate
the X-ray photon index $\Gamma$
from the two hardness ratios\footnote{
Defined as $HR1=(C_2-C_1)/(C_{2}+C_{1})$,
$HR2=(C_4-C_3)/(C_{4}+C_{3})$,
where $C_1$, $C_2$,  $C_3$ and $C_4$ are the number of photon counts in the
0.1--0.4\,keV, 0.5--2.0\,keV, 0.5--0.9\,keV, and 0.9--2.0\,keV
bands, respectively.
}, wherever available,
assuming an absorbed power-law spectral model.
Both $\Gamma$ and absorption column density \nh can be estimated
as free parameters,
or the index alone assuming \nh to be the Galactic column density.
This method had been verified to be reliable and robust
under the assumption that intrinsic spectral shapes are
indeed a simple absorbed power-law \citep{bs94}.
The X-ray fluxes in the ROSAT PSPC  band (0.1--2.4\,keV)
are calculated from the count rates using the energy to counts
conversion factor (ECF) for a power law spectrum and Galactic absorption
(ROSAT AO-2 technical appendix, 1991).
The ECF for each X-ray source is explicitly calculated from
the ROSAT PSPC effective area\footnote{
We used those that are appropriate for the RASS
or pointed observations, accordingly.},
by using information on individual $\Gamma$ if obtainable,
or using the mean otherwise ($\Gamma=2.49$, see below).
We list some of the information about the X-ray observations
in  Table\,\ref{tbl:xray}.

For objects not detected in X-rays, we determine upper limits
on their X-ray fluxes from  count rate limits set by the RASS.
Since the X-ray backgrounds are very low in most of the RASS observations,
detection of a source is mostly determined by
the source photon counts that follow  Poisson statistics.
We set an upper limit of source counts as
12 for an object not detected in the RASS, as often adopted \citep[e.g.][]{yuan98}.
The corresponding count rate limit is then calculated by using the
effective exposure time at the source position as available
from the RASS exposure map.
Then the flux limit for each object is calculated
using the same method above and assuming $\Gamma$ as the mean.

Among the ROSAT detected objects, J1644+2619 was observed with Chandra
with a large number of photon counts accumulated.
Moreover, for J0849+5108 the hardness ratios
($HR1=0.97\pm0.11$,  $HR2=0.28\pm0.15$) yield
an abnormally flat spectrum $\Gamma=0.6^{+0.36}_{-0.52}$
\citep[see also][]{zhou05}.
For this object we also investigate its ROSAT data
in spite of the relatively low number of counts ($\sim60$).
We perform X-ray spectral and timing analysis for
J1633+4718, J1644+2619, and J0849+5108.
We briefly describe here the data reduction procedures and
present results in  \S\,\ref{sect:res-xray}.
We use version 12.2 of XSPEC  \citep{xspec} for spectral modeling.

{\bf J1633+4718} (RXJ\,16333+4719): The target was observed with the
ROSAT PSPC-b with an exposure time of 3732 seconds on July 24th,
1993 (Obs-ID: 701549; PI: N. Bade). We retrieved the ROSAT data from
the archive and use the XSELECT (version 2.3) tool of FTOOLS
\citep{ftools} for data analysis. The X-ray spectrum is extracted
from a circle of 150\,\arcsec radius, and a background spectrum is
extracted from an annulus with inner/outer radii of
195/300\,\arcsec.
There are $\sim976$ net source counts, and the count rate
is $0.26\pm0.01$\,\ucre. The spectrum in the 0.1--2.4\,keV band is
binned to have at least 30 counts in each bin.
J1633+4718 was also detected in the RASS with a weighted exposure
of  909\,s, yielding $\sim 185$ net source counts and a count rate
$0.20\pm0.02$\,\ucre.
We extract the RASS spectrum following the
procedure described in \citet{bell94}.

{\bf J1644+2619}: The object was targeted with the Chandra ACIS-S
(PI: S.\ Laurent-Muehleisen) with an exposure time
of 2946\,s on June 02, 2003.
The  data were retrieved from the Chandra data archive
and were reduced from the level-2 data set
following the standard procedure using CIAO (version 3.4).
There are 553$\pm44$ net source counts detected,
with a count rate of 0.188\,\ucre.
The 0.2--5\,keV spectrum (very few counts accumulated above 5\,keV)
is binned to have at least 20 counts in each energy bin.

{\bf J0849+5108}: The object was observed with the ROSAT PSPC-b in a pointed
observation as a target in 1993, April
with an exposure time of 4496\,s (Obs-ID: 700821).
There are only 63 net source counts detected
in the 0.1--2.4\,keV band.
We fit the unbinned spectrum
by minimizing the C-statistic \citep{cash79,nou89},
which is valid for Poisson statistics appropriate to the
low counts regime.

We also searched for UV data from the GALEX \citep{martin05} GR2/GR3 data
release\footnote{http://galex.stsci.edu/GR2/}.
Twelves objects in our sample have available photometric measurements
in the far and/or near UV imaging bands with an effective wavelength
of 1516\,\AA\ and 2267\,\AA, respectively.
Their UV images are mostly point source like.
The GALEX magnitudes (in the AB magnitude system)
are corrected for Galactic extinction
using $A_{\rm FUV}/E(B-V)=8.376$ and $A_{\rm NUV}/E(B-V)=8.741$,
following \citet{wyder05}.

\section{Broad band continuum radiation}
\label{sect:cont}
\subsection{GHz radio emission}
\label{sect:radio}

\subsubsection{Radio morphology and spectrum}

The FIRST radio images of our RL NLS1 sources are ubiquitously
unresolved at a resolution of $5.4$\arcsec, which is remarkable.
This sets upper limits on the projected size of
10--40\,kpc, depending on the redshift.
Among our sample there are several objects
observed with VLBI, including
J0948+0022 \citep{doi06}, J1633+4718 and J1644+2619 \citep{doi07},
and J1505+0326 \citep{dal98}, as reported in the literature;
they remain unresolved at resolutions ranging
from a few milli-arcsec to a few tens of milli-arcsec,
which correspond to several to several-tens parsec in physical scale.
These observations set constraints on the
brightness temperatures of
$T_{\rm B}>10^{8-9}$\,K \citep{doi06,doi07}.
Since all the radio sources in our sample are compact,
we use hereafter the average of the FIRST and
NVSS\footnote{The NRAO VLA Sky Survey} measurements
as their 1.4\,GHz flux densities in the following analysis;
this is to minimize the effects of fluctuations on the results
caused by large flux variations.
Similarly, at other frequencies,
in cases where multiple epoch measurements are
available at the same frequency, the average flux is used.

\begin{figure}
\includegraphics[width=0.9\hsize,angle=0]{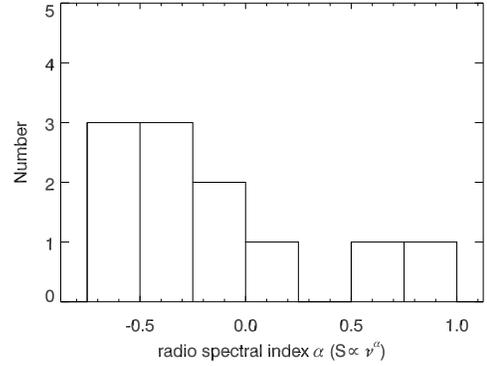}
\caption{Distribution of the radio spectral indices \alpr
around 5\,GHz [$S(\nu) \propto \nu^{\alpha_{\rm r}}$].
}
\label{fig:radslp}
\end{figure}

Multi-frequency and multi-epoch radio data were collected from
the NASA/IPAC Extragalactic Database (NED).
Of the 23 objects, 11 have flux measurements at 5\,GHz,
for which the 1.4\,GHz--5\,GHz spectral indices
\alpr are estimated, though the observations were not simultaneous.
The  5\,GHz fluxes and   \alpr are listed in Table\,\ref{tbl:sample},
and the histogram of \alpr is shown in Figure\,\ref{fig:radslp}\footnote{
For J0849+5108, the flux variations at 1.4\,GHz were substantial and
the spectral shape around 5\,GHz was flat or even inverted \citep{arp79};
we thus use \alpre=0 for this object in the analysis below.
}.
It can be seen that the radio spectra are systematically flat,
with a median of $-0.30$ and a mean of $-0.13$.
Adopting the conventional dividing line between
flat- and steep-spectrum sources, \alpre$=-0.5$,
8 out of the 11 objects with \alpr measurement turn out to be
flat-spectrum radio sources (\alpre$>-0.5$).
In particular, several objects show an inverted radio slope
(\alpr$>0$) near 5\,GHz and toward higher frequencies
based on simultaneous multi-frequency measurements, namely,
J0948+0022 \citep{reich00,zhou03,doi06},
J1633+4718 \citep{neum94},
J1644+2619 \citep{doi07},
and J1505+0326 \citep{tin05}.
Only 3 objects have \alpr steeper than, yet very close to, $-0.5$.

Among the remaining 12 objects without available slopes near 5\,GHz,
9 have lower frequency 327\,MHz--1.4\,GHz indices measurable
(see Section\,\ref{sect:lowfemi}), out of which
6 have  flat spectra (\agm$>-0.5$).
Thus, it is likely that most of these 12 objects
may have the same flat (or even flatter)
indices at higher frequencies as well,
unless there is a break near 1\,GHz,
as in Gigahertz-peaked spectrum sources (GPS).
This fact supports further the
flat spectral index distribution found above for our sample.
Though in part of the sample
the radio slopes are estimated from non-simultaneous data,
the current \alpr distribution should be correct in
the statistical sense; therefore, we expect that
the bulk of our sample objects should be flat-spectrum sources.
As is known, radio sources with flat spectrum,
which can be produced by relativistic jet models \citep[e.g.][]{bla79},
have a fairly good chance to be blazars \citep[e.g.][]{fug88}.

Now we can elaborate the radio-loudness parameter calculation
for our sample by using the averaged 1.4\,GHz flux densities and
the estimated \alpr for individual sources for the K-correction
(if not available,  \alpre=$-0.5$ is assumed as above).
We also make use of the measured optical slopes as presented
in \S\,\ref{sect:optcont}. The improved radio-loudness values
are listed in  Table\,\ref{tbl:sample}.
The differences between the previous estimates and the improved values
are small, typically within 10\% in $\log$\,\rle;
such differences are comparable with changes in \rl
caused by source flux variations typical of our objects (see below).

\subsubsection{Radio variability and brightness temperature}

A number of objects showed  significant flux variations
based on  multi-epoch observations taken at the same wavelengths
\footnote{
In cases where two observations have much different spatial
resolutions, we consider the variability to be genuine
only if the higher flux was measured at a higher resolution.
This is to avoid detection of spurious variations
caused by the possibility that a measured higher flux
may be contributed by contaminating extended/nearby emission
if the observation is made at a lower spatial resolution.
We take this approach even though this problem actually
has little effect to our objects, all of which are compact
on the scales concerned, e.g.\  several arc-seconds.
Thus, our results are conservative regarding the number of
variables, and variability amplitudes and timescales.}.
In Table\,\ref{tbl:radvar} we list those having
the significance of variations higher than $\sim3\,\sigma$
between two observations, where $\sigma$ is defined as
$\sigma=(S_1-S_2) /\sqrt{\sigma_{s1}^2+\sigma_{s2}^2}$.
We also give in the table the amplitudes of variations $\Delta\,S$,
the fractional amplitudes $\Delta S/\langle S \rangle$ where
$\langle S \rangle$ is the average of the two fluxes,
and the time spans  $\Delta t$ between the two observational epochs.
Their variability is consistent with the fact that
the variables have ubiquitously
flat or inverted radio spectra when available.
The fractional amplitudes given in  Table\,\ref{tbl:radvar}
have a median of 43\%.
In particular, several objects show large amplitude
 variations ($\sim$40--75\%) on relatively short timescales,
J0948+0022 \citep[reported by ][]{zhou03},
J0849+5108 \citep[reported by ][]{arp79},
and J1505+0326 within a few years, and
J1548+3511 by 44\% within 207 days
(140 days in the source rest frame).

Large amplitude variability on short timescales
has been commonly used to set lower limits on the
{\em apparent} brightness temperature ($T_{\rm B}$) of a radio source,
since the  size of the variable part of the source can be
constrained from the light-crossing time
\citep[e.g.][]{kro99,fanti83,gop84}.
Assuming that the variable portion of the radio flux is emitted from a
region with a size smaller than that corresponding to the light-crossing time,
which is the time span between the two observations,
the {\em apparent} brightness temperature is estimated to be
\footnote{We noted that there was an error
in our previous calculation of $T_{\rm B}$ in \citet[][]{wang06}
and \citet[][]{zhou03}, which was also pointed out by \citet{gho07}.
We have corrected it in this paper.
}
\begin{equation}
  T_{\rm B} \ga \frac{\Delta P_{\nu e}}{2 \pi^2 k \nu^2 (\Delta t)^2}
= \frac{2D^2_{L}\Delta S_{\nu}}{(1+z) \pi  k \nu^2 (\Delta t)^2}
\end{equation}
where $k$ is the Boltzmann constant, $D_{L}$ the luminosity distance,
 $\Delta S_{\nu}$ the  variable portion of the observed flux
density, $\Delta P_{\nu e}$ the corresponding radio power
at the emission frequency in the source rest frame,
$\nu$ the observing frequency and $\Delta t$ the time span
in the observer's frame.
The estimated $T_{\rm B}$ are listed in Table\,\ref{tbl:radvar}.

Of the  8 objects, 6 have estimated   $T_{\rm B}$
exceeding the equipartition brightness temperature
$\sim 3\times 10^{11}$\,K as proposed by \citet{read94},
which is expected to conform by most radio sources.
In particular,  4 of the objects, all of flat-spectrum,
have $T_{\rm B}$ exceeding the inverse Compton limit
$\simeq 10^{12}$\,K  \citep{kell69}.
The highest  $T_{\rm B}$ is found  to be $10^{14}$\,K in J1505+0326,
from two VLBI observations at 8.3\,GHz
with the variability significance of $3\sigma$.
The extremely high $T_{\rm B}$ values are commonly
explained as emission originating from relativistic jets
\citep{bla79,jones73}.
For instance, taking the inverse Compton limit as a conservative limit,
 the minimum Doppler factor can be estimated as
$\delta_{\rm min}=(T_{\rm B}/10^{12}\,K)^{1/3}$.
This results in $\delta_{\rm min}=$1.5--4.7
for the 4 objects (Table\,\ref{tbl:radvar}),
consistent with the range of the Doppler factors of relativistic jets
inferred for classical RL AGN \citep{ghi93,jiang98}.

\subsubsection{Summary of the GHz radio properties}

(1) All the  radio sources of our sample are compact,
being unresolved at the FIRST resolution of several arc-seconds.
(2) The majority of objects with available radio indices
are of flat-spectrum  (around 1.4--5\,GHz),
some even with inverted spectra.
(3) Most of the flat-spectrum sources show significant variability,
from which the brightness temperatures are inferred
to be as high as $T_{\rm B}>10^{11}$\,K, and even
exceeding the inverse Compton limit of $10^{12}$\,K in 4 objects.
These high $T_{\rm B}$ values may suggest the presence of
at least mildly relativistic beaming of the radio emission.

The properties of the radio emission at low frequencies
of our sample objects are summarised in Appendix\,\ref{sect:lowfemi},
which is important for understanding the intrinsic radio power
and radio-loudness of these objects,
as discussed in Section\,\ref{sect:intrpower}.

\subsection{Optical continuum emission}
\label{sect:optcont}

Our standard algorithm to select NLS1 galaxies as above does not
give information about the optical continuum shape.
However, for RL objects the optical continuum may deviate from
the  `canonical' shape in consideration of a
potential contribution of jet emission.
To properly take this effect into account,
we parameterise the optical continuum shape especially
for our RL sample.
We model the AGN continuum
in the 2500\AA--5500\AA\ rest frame range with a power-law or
a broken power-law, and repeat the above spectral fitting
 with the slope as a free parameter.
The fits are performed in the same way as described in \S\,\ref{sect:nls1}
assuming no intrinsic reddening.
The improved fits give a more realistic description of the AGN continuum;
while, on the other hand, we find that
the new fits have almost no effects on the parameterisation
of the emission line spectra.
The optical continuum fluxes
in the source rest frame are measured directly
from the power-law fits of the AGN continuum.

\begin{figure}
\includegraphics[width=\hsize]{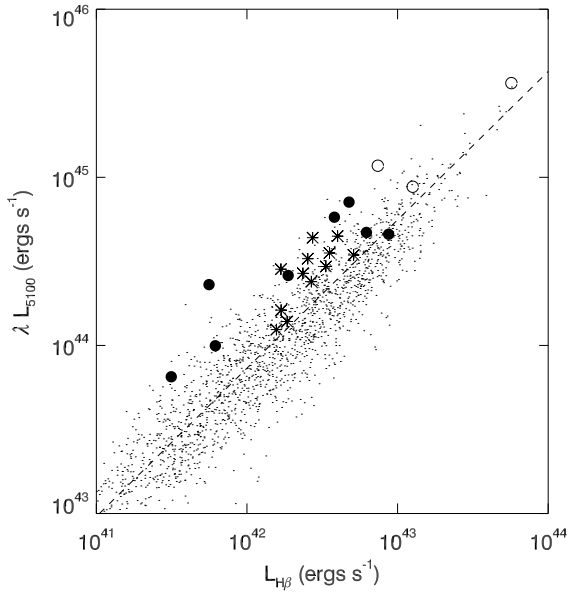}
\caption{Relationship between the nuclear monochromatic luminosity
at 5100\AA\ and the \hb luminosity for the radio-loud NLS1 AGN
of our sample (filled dots: flat-spectrum radio sources;
open circles: steep-spectrum radio sources;
asterisks: no available radio indices).
Over-plotted are NLS1 galaxies in the Zhou'06 sample (small dots),
predominantly radio-quiet,
as well as the best-fit relation (dashed).
It shows that the radio-loud NLS1 galaxies have systematically higher
observed luminosities in excess of that predicted
from the \hb luminosity of normal NLS1 galaxies.
}
\label{fig:lo-lb}
\end{figure}

It has been established  that
the optical luminosity $L_{\rm opt}$ of NLS1 AGN
is tightly correlated with the \hb line luminosity \lhb
based on large samples \citep{zhou06};
hence \lhb is a good indicator of $L_{\rm opt}$.
We test this scaling relation for RL NLS1 galaxies
as shown in Figure\,\ref{fig:lo-lb},
where the measured 5100\AA\ luminosities are plotted versus \lhbe.
As a comparison the $\sim 2\,000$
predominantly RQ NLS1 galaxies in the Zhou'06 sample
are over-plotted, along with the fitted
$\lambda L_{\lambda 5100}$-- \lhb relation
\citep[][their Eq.\,5]{zhou06}.
As can be seen, the RL NLS1 galaxies reveal the same trend of correlation,
but, of particular interest, lie systematically
above the known $L_{\rm opt}$--\lhb relation for normal NLS1 galaxies.
We estimate the ratio of the observed luminosity
to that expected from the \hb luminosity
($L_{\lambda 5100}$/$L_{\lambda 5100}^{H\beta}$)
using the $\lambda L_{\lambda 5100}$-- \lhb relation.
We find that the ratios
range from 0.93 to 5.2, with a mean of 1.9.
The ratios differ systematically between
flat- and steep-spectrum sources as groups,
with the mean being 2.4 for the former and 1.8 for the latter,
and 1.6 for those without radio indices.

\begin{figure}
\includegraphics[width=\hsize,height=0.9\hsize]{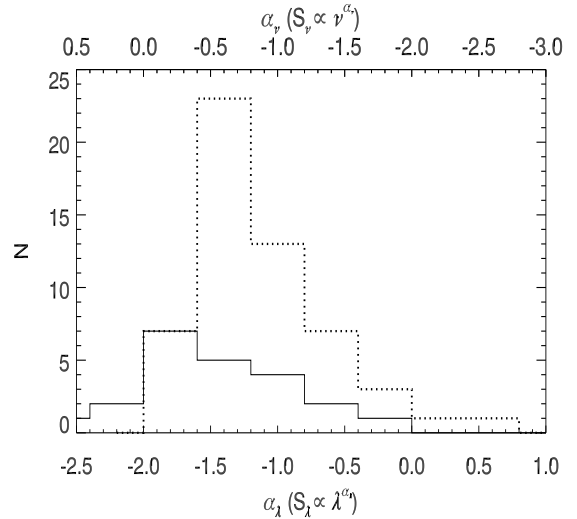}
\caption{
Histogram of the power-law slopes  of the optical continuum
(rest frame 2500--5000\AA) for our radio-loud NLS1 sample
(solid line; except for one object which is fitted
with a broken power-law).
As a comparison,
the dashed line represents the histogram of  \alpwvl for a
radio-quiet NLS1 sample selected from the SDSS
with redshift and luminosity distributions compatible to ours
\citep[from][]{ai08}.
}
\label{fig:optslp}
\end{figure}

The optical continua in the rest frame 2500\AA--5500\AA\ range
can be described with a power-law for most of the objects.
The fits are relatively poor for two objects, namely,
J1037+0036 and J1634+4809, which
show convex-shaped spectra in the optical--UV band with a
drop short-ward of \ion{Mg}{2}$\lambda$2800\AA\ in wavelength.
For one object, J1138+3653, the spectrum
cannot be fitted with a power-law due to a rapid drop in the blue part,
which results possibly from mild extinction and is discussed further
in \S\,\ref{sect:dis_opt}.
The best-fit slopes show a large range from modestly red
(\alpwvl=$-0.2$) to blue (\alpwvl=$-2.4$) spectra.
Their distribution is shown in Figure\,\ref{fig:optslp}
(excluding J1138+3653), which has a median of $-1.54$ and a mean of $-1.43$.
We compare this \alpwvl distribution with that of a RQ NLS1 sample,
which comprises 55 SDSS-selected objects with reliably measured optical slopes
\citep[dashed line in the figure, ][for objects with multi-epoch data
the averaged slopes are used]{ai08}.
It should be noted that that RQ sample has
the redshift and luminosity distributions indistinguishable
from those of our sample
[the two-sided Kolmogorov-Smirnov (K-S) test
probability of 0.48 and 0.52, respectively].
It can be seen that, though there is a broad range of overlap,
our very radio-loud NLS1 AGN appear to have systematically
bluer continua than the RQ objects.
The RQ sample have a median of $-1.24$ and a mean of $-1.15$.
The chance probability that the two distributions are the same is only
\pch$\la0.01$, as given by the K-S test.
A comparison with the smaller NLS1 galaxies sample of \citet{cons03},
which have UV--optical slope measurements with
the Hubble Space Telescope (HST),
yields a similar conclusion;
the closest value of their sample to ours is the continuum slope of
the median composite spectra, \av$\sim -0.79$ (\alpwvle$\sim -1.21$).

\subsection{X-ray radiation}
\label{sect:res-xray}

Assuming an absorbed power-law spectral model and
using the aforementioned hardness ratio method,
the effective photon indices are estimated for
most of the objects with valid ROSAT/PSPC hardness ratios
extracted from the ROSAT source catalogues.
The estimated absorption \nh values are found to be
well consistent with the Galactic values within errors,
indicating that there is no significant X-ray absorption
in these objects.
Hence, we adopt the photon indices $\Gamma$ (Table\,\ref{tbl:xray}) obtained by
assuming \nh=\gnh for better constraining the parameters.
Furthermore, for 5 of the objects no physically meaningful result
can be obtained by this method,
 because either the spectrum is complex and
highly deviating from the assumed absorbed power-law,
or the measured hardness ratios are displaced outside
the valid zone by Poisson fluctuations.
For objects with multiple X-ray data,
we use the weighted mean for $\Gamma$ and fluxes in the following analyses.
Results from detailed spectral analysis for the 3 objects ,
J1633+4718, J1644+2619, and J0849+5108, are summarised below.

\begin{figure*}
\centering
\epsscale{0.8}
\begin{minipage}[]{0.48\hsize}
  \includegraphics[width=0.8\hsize,angle=-90]{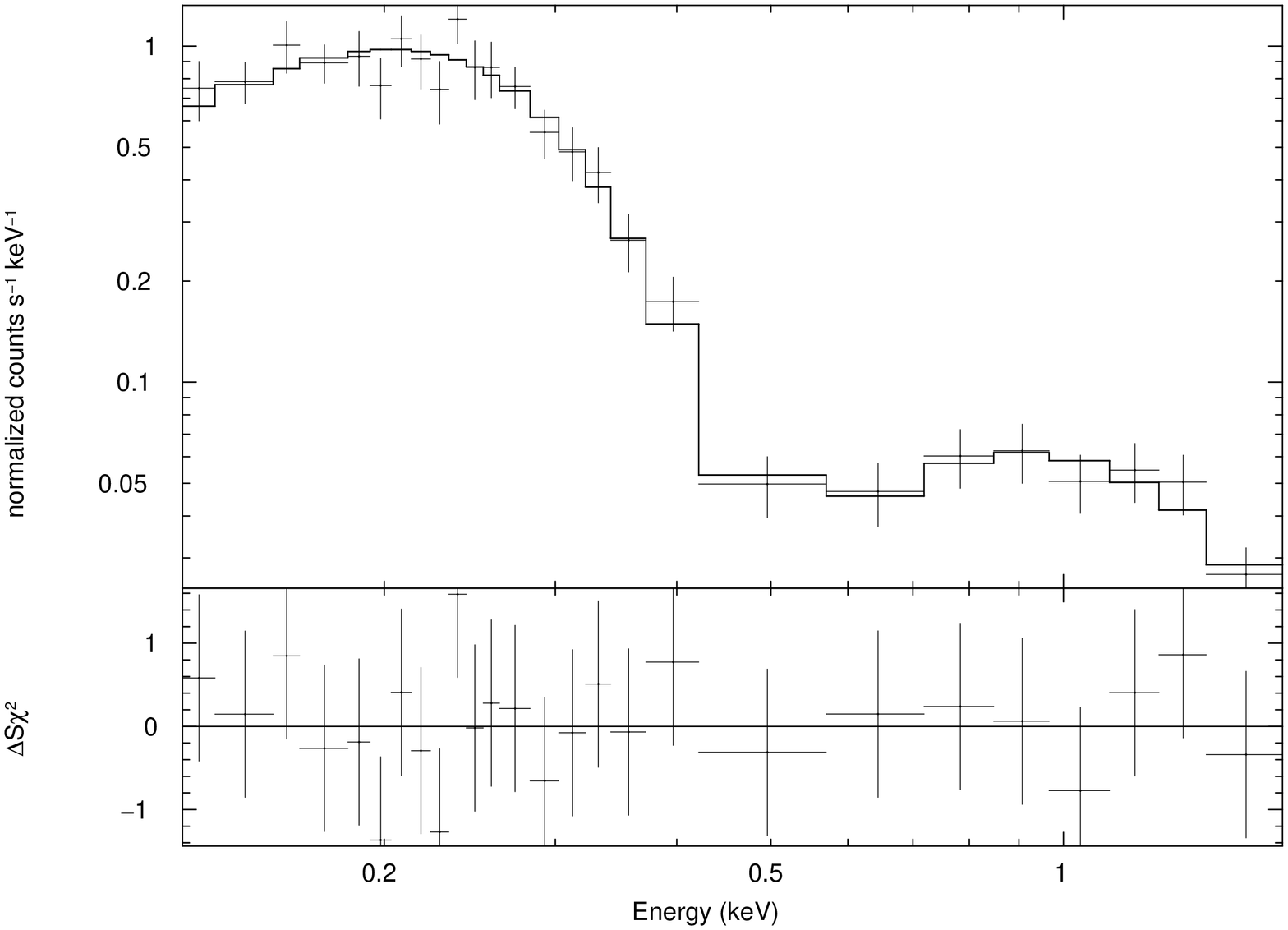}
\end{minipage}
\hfill
\begin{minipage}[]{0.48\hsize}
  \includegraphics[width=0.8\hsize,angle=-90]{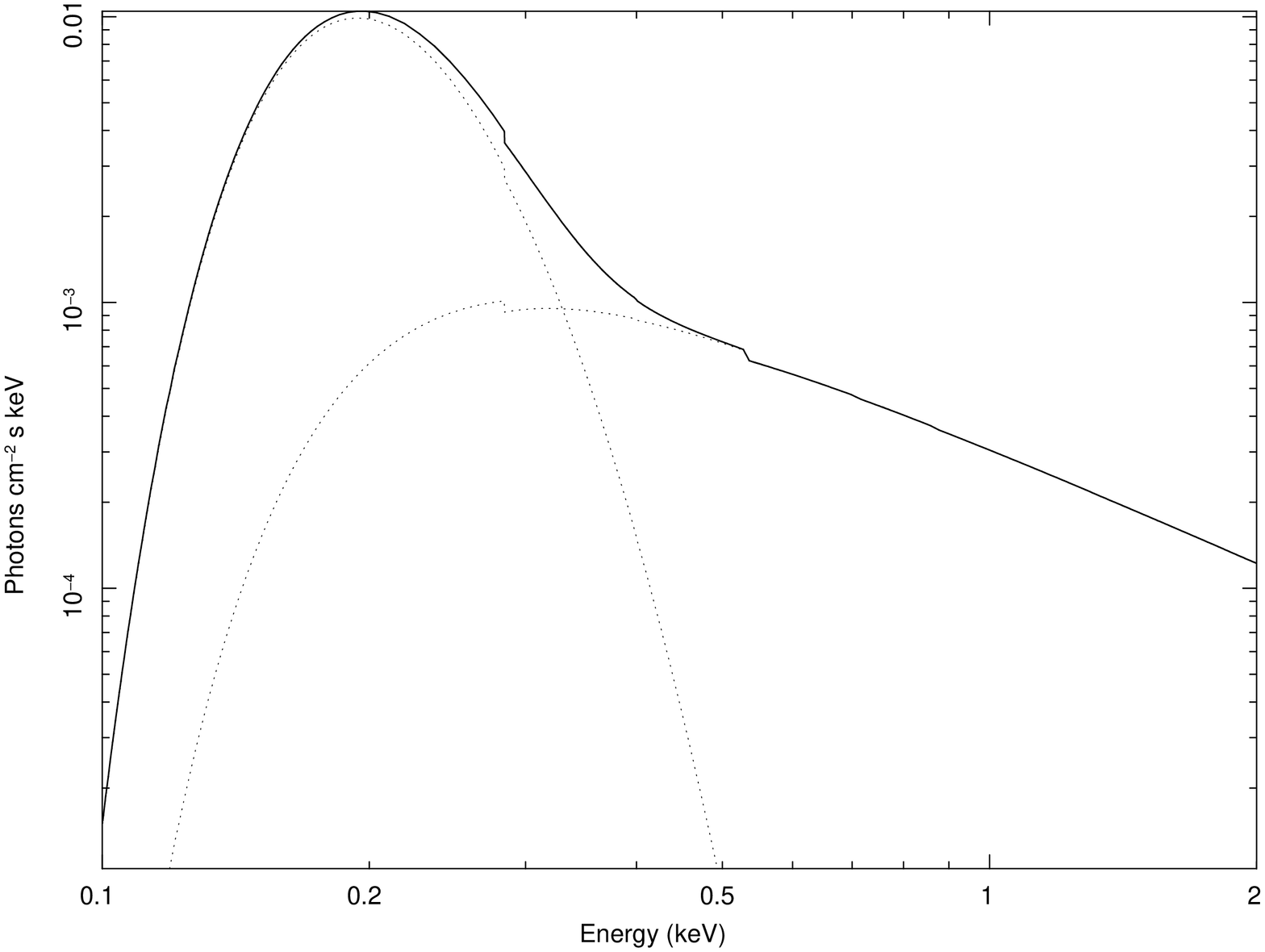}
\end{minipage}
\caption{X-ray spectral fit for J1633+4718.
Left panel: ROSAT PSPC spectrum,  the folded best-fit
model and the residuals;
right panel: best fit model of a power-law
plus a black-body (with Galactic absorption).
}
\label{fig:xsp_1633}
\end{figure*}

{\bf J1633+4718} (RX\,J16333+4719): The spectrum deviates
apparently from a simple absorbed power-law model, with prominent
soft X-ray emission dominating $E<$0.5\,keV and a hard tail at
$E>$0.5\,keV. The best fit is achieved using a model of a power-law
plus a redshifted blackbody component, and the absorption \nh is
found to be in excellent agreement with the Galactic
\gnh=$1.79\times 10^{20}$\,\unhe.
Hence \nh is fixed to \gnhe. The
data and the best fit model are shown in Figure\,\ref{fig:xsp_1633}.
The rest frame blackbody temperature is found to be
$32.5^{+8.0}_{-6.0}$\,eV and the photon index of the underlying
power law $\Gamma=1.37\pm0.49$ (90\% confidence for one parameter of
interest). The latter is less well constrained due to the limited
bandpass of the PSPC in the `hard' X-rays.
No temporal variability is found within the observation interval.
The spectrum measured in the RASS has almost identical spectral shape
as that of the pointed observation.
The best-fit parameters are in excellent agreement with those obtained
from the pointed observation, with $\Gamma=1.47^{+0.77}_{-0.92}$ and
the blackbody temperature $30^{+12}_{-10}$\,eV (90\% confidence level),
though the errors are large given the small source counts.
No variation in either the spectral shape or the flux is found between
the RASS and the pointed observation.

\begin{figure}
\includegraphics[width=0.8\hsize,angle=-90]{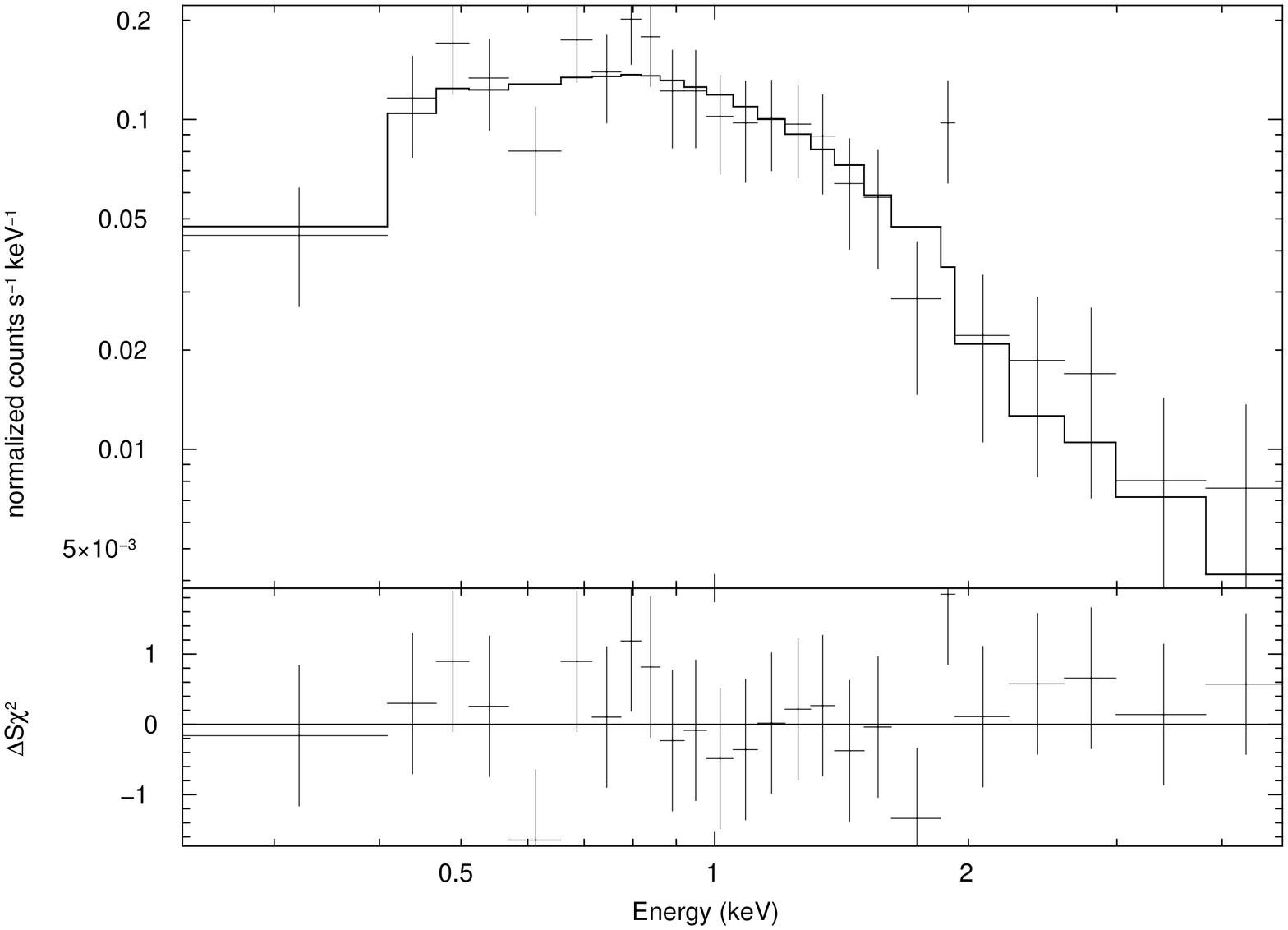}
\caption{Chandra ACIS X-ray spectrum of J1644+2619 and the best-fit model
as a steep power-law ($\Gamma=2.19\pm0.27$, 90\% errors)
with Galactic absorption.
}
\label{fig:xsp_1644}
\end{figure}

{\bf J1644+2619}:
A simple power-law model is found to be an excellent fit to the
spectrum (see Figure\,\ref{fig:xsp_1644}),
resulting in a reduced \chisq=13 (for 23 d.o.f).
The fitted absorption \nh is almost identical to the
Galactic value \gnh=$5.12\times10^{20}$\,\unhe.
The best-fit photon index is steep, $\Gamma=2.19\pm0.27$
(90\% confidence for one parameter of interest).
Adding an extra soft X-ray component (e.g. blackbody)
does not improve the fit, though with two more free parameters;
the power-law photon index becomes flatter,
$\Gamma\sim1.8^{+0.6}_{-0.3}$, but is subject to a large uncertainty range
which encompasses the single power-law model.
We therefore suggest that the simple power-law model with a relatively
steep index is the most likely spectral shape, though models
composed of a flatter power-law with a weak soft component
cannot be ruled out from the current data.
Compared to the observation in the RASS,  the X-ray source
was about 2.5 times fainter at the Chandra observation epoch,
though the photon index seems not to vary.

{\bf J0849+5108}:
The spectrum can be fitted with a power-law with absorption \nh
close to the Galactic value \gnh=$3.0\times10^{20}$\,\unhe.
Fixing column density at \gnh yields a photon index
$\Gamma=1.77^{+0.44}_{-0.60}$ (90\% confidence for one parameter of interest).
We note that the extremely hard hardness ratios given in the
source catalogue are incorrect, which may resulted from
inappropriate background subtraction around this object in
the presence of nearby, seemingly extended X-ray emission.

\begin{figure}
\includegraphics[width=\hsize]{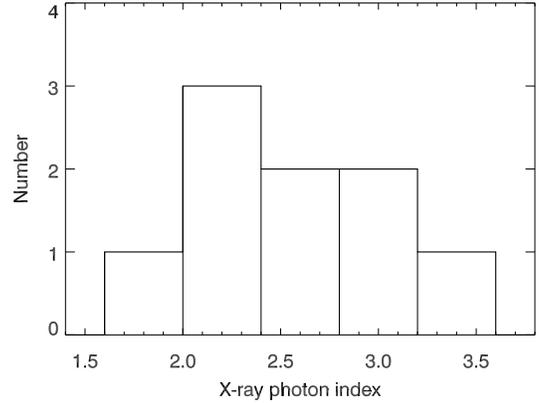}
\caption{
Distribution of the {\em effective} photon indices of representing
power-law of the X-ray spectra of our radio-loud NLS1 galaxies,
where available.
All the indices are measured with the ROSAT PSPC in 0.1--2.4\,keV,
except for J1644+2619 with the Chandra ACIS in 0.2-5\,keV.
For J1633+4718 we adopt $\Gamma=2.8$ obtained from the hardness ratios
to represent the overall spectral shape.
}
\label{fig:gam}
\end{figure}

Figure\,\ref{fig:gam} shows the distribution
of the {\em effective} X-ray photon indices $\Gamma$, which span
a wide range from 1.7 to 3.3.
Since the uncertainties of most of the derived $\Gamma$ are relatively large,
we try to disentangle the intrinsic distribution of  $\Gamma$ from
the measurement uncertainties,
using the Maximum-Likelihood method introduced by \citet{mac88}.
Assuming that the intrinsic distribution of  $\Gamma$ is Gaussian,
we find that the distribution is intrinsically broad, with
a standard deviation of $0.39^{+0.41}_{-0.19}$
and a mean of $2.60^{+0.37}_{-0.43}$
(at the 90\% confidence level for two interesting parameters).
We conclude that the soft X-ray continuum shape, as
described by the power-law photon index $\Gamma$,
exhibits a considerable variety among our RL NLS1 galaxies,
with both  flat and steep spectra.

It should be noted that the thus estimated $\Gamma$ from hardness
ratios are only an indicator of the overall spectral shape
in the 0.1--2.4\,keV range.
A steep $\Gamma$ value does not rule out
the presence of a flat `hard' X-ray component if both
a steep soft X-ray and a flat `hard' X-ray component co-exist.
As an example, for J1633+4718, a slope of $\Gamma\simeq 2.8$ is derived
from its hardness ratios if the spectrum is approximated as a power-law.
This is particularly true for NLS1 galaxies, since a soft X-ray excess
component is commonly seen in them.
Better quality X-ray spectra in a wider energy band than that of
the ROSAT/PSPC are needed to determine the true underlying power-law
in these objects.

\subsection{Broad band SED}

To quantify the broad band SED,
we calculate the commonly used effective spectral indices between the
 5\,GHz radio,  2500\,\AA\ optical/UV, and  2\,keV X-ray band, namely,
\alproe, \alpoxe, and \alprxe; they are defined as
$\alpha_{\rm 12}=-\log (S_{\nu1}/S_{\nu2})/\log (\nu_1/\nu_2)$,
where $S_{\nu1}$ and $S_{\nu2}$ are the fluxes at the frequencies
$\nu_1$ and $\nu_2$, respectively, in the object rest frame\footnote{
There are slight differences in the frequencies adopted in the
\alpro and \alpox definitions among the literature.
Whenever we cite \alpro and \alpox which have different definitions,
we first convert their values to those
according to the definition used in this paper.}.
The rest frame 2500\,\AA\ fluxes are computed from the $u$-band PSF magnitudes,
which  have the effective wavelength (3542\,\AA) very close
to the redshifted 2500\,\AA\ light (within a range of 500\,\AA);
for the K-correction, the measured optical slopes for individual objects are adopted
(for one object without the slope available, \alpwvl=$-1.5$ is assumed).
For the 3 objects for which X-ray spectral fits were enabled,
the X-ray flux densities at 2\,keV ($f_{\rm 2keV}$) are measured directly from
the best-fit spectral model;
while for the others $f_{\rm 2keV}$ are computed from the 0.1--2.4\,keV integrated fluxes.
In doing so, as well as the K-correction,
the derived $\Gamma$ values are used for individual sources whenever available,
or the mean $\langle \Gamma \rangle=2.6$ is used otherwise.
For those without X-ray detection, lower limits on \alpox and \alprx
are calculated using the upper limits on the X-ray fluxes
constrained by the RASS.
The obtained \alproe, \alprxe, and
\alpox are listed in Table\,\ref{tbl:arox}.

\begin{figure}
\includegraphics[width=\hsize,height=0.9\hsize]{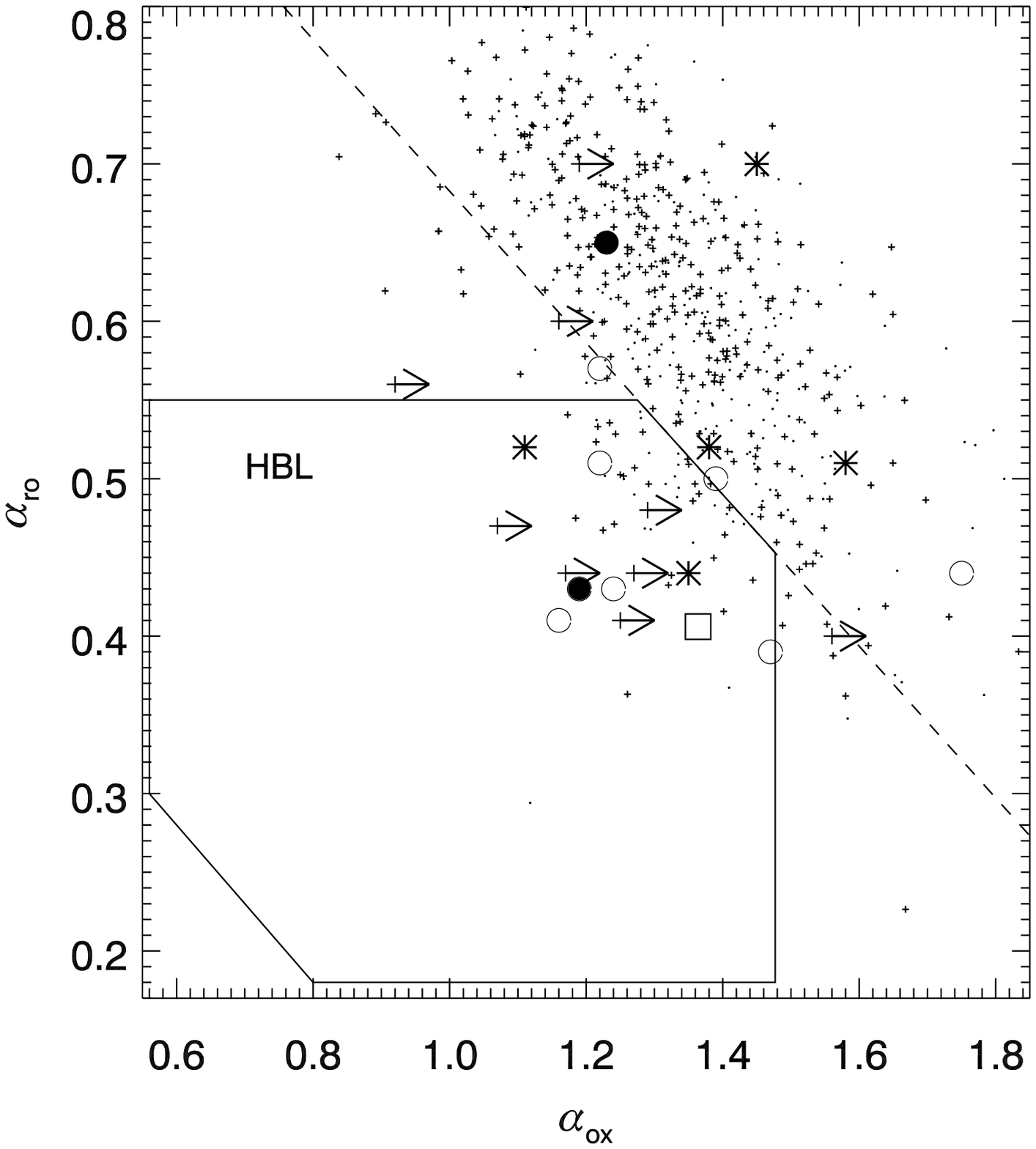}
\caption{
\alpro versus \alpox for the radio-loud NLS1 sample;
filled dots: objects with flat X-ray spectra ($\Gamma<2$);
open circles: objects with steep X-ray spectra ($\Gamma>2$);
asterisks: objects without available X-ray slopes.
Arrows represent lower limits of \alpox for X-ray non-detections.
For comparison, we over-plot ROSAT detected
radio-loud quasars   from \citet{bri97},
consisting of both FSRQs (small crosses) and SSRQs (small dots).
The blazar-NLS1 hybrid object 2MASX\,J0324+3410 discovered by \citet{zhou07}
is also indicated  for comparison (square).
The dashed line represents a constant \alprx=0.78, which is commonly
used to divide BL Lac objects into LBLs and HBLs
(see text for details).
A more conservative, schematic locus for HBL is also indicated
following  \citet[][solid line]{pado03}.
}
\label{fig:alprox}
\end{figure}

In  Figure\,\ref{fig:alprox} we plot our RL NLS1 galaxies
on the commonly used   \alproe--\alpox blazar diagnostic diagram,
separating them into three groups:
flat (underlying) spectra ($\Gamma<2.0$, filled dots),
steep spectra ($\Gamma>2.0$, open circles),
and those without $\Gamma$ estimates (asterisks).
For comparisons, we over-plot
RLQs from the ROSAT detected quasar sample of \citet{bri97}.
That sample is composed of both FSRQ and steep-spectrum radio quasars (SSRQ),
whose loci largely overlap with each other.
It can be seen that some of the objects do follow the trend
defined by FSRQ/SSRQs,
indicating their SEDs are similar to those of FSRQ/SSRQs.
Specifically, all the 3 objects in our sample known to have
a steep radio spectrum  are  well consistent with normal SSRQs.
It is known that FSRQs (and SSRQs) occupy almost the same region on the
\alproe--\alpox diagram as LBLs, owing to their similar SEDs
\citep{pado97a,bri96,sam96,fos98}.

However, a considerable fraction of the objects clearly offsets.
We expect that these outliers must have different SEDs from
those of classical FSRQ and LBLs.
By invoking the classification scheme of BL Lac objects,
we divide the \alproe--\alpox diagram into the loci of LBLs and HBLs.
A simple yet commonly used criterion separating  HBLs and LBLs is \alprxe;
a nominal value is set to be around $\alpha_{\rm rx}'\sim 0.75-0.78$
(defined between 5\,GHz and 1\,keV) suggested by various authors,
with HBLs having smaller (flatter) \alprx values.
We adopt  0.78 as the dividing line  \citep[as in, e.g.][]{pado03} ,
which corresponds to \alprxe=0.787 for the \alprx definition adopted in this paper
and is represented by the dashed line in  Figure\,\ref{fig:alprox}.
Following  \citet[][]{pado03},
a more elaborated locus for HBL is also indicated schematically,
which is populated by X-ray selected BL Lac objects \citep[e.g.][]{bri96}.
Of particular interest, most of the objects offsetting from
the classical FSRQ and LBL locus
actually fall within the HBL locus\footnote{
In a strict sense, the HBL locus in Figure\,\ref{fig:alprox}
represent the SEDs of the
`pure blazar' component only or dominated by the non-thermal continuum.
Our NLS1 objects are, however, strong-lined and
a significant contribution is expected from the thermal ionizing continuum
to the observed optical flux
($\sim50$\% on average, see \S\,\ref{sect:optcont}).
For rigorous comparisons with HBLs, we
should consider the 'corrected SED' corresponding to
the pure non-thermal continuum only.
However, such corrections are very small
($\Delta$\alpox$\sim-0.12$ and $\Delta$\alpro$\sim0.05$ on average)
for the bulk of the objects, and the results are
 essentially not affected.
}.
This indicates that the broad band SEDs of these NLS1 AGN
are similar to those of HBLs.
Following our terminology introduced in \S\,1,
we refer to them as NLS1 HFSRQs, and those with SEDs typical
of classical FSRQs (LBL-like) as NLS1 LFSRQs.
These NLS1 HFSRQs have ubiquitously flat
radio spectra whenever available, consistent with a
potential blazar nature.
Those without X-ray detection (arrows) have their
\alpro and \alpox limits consistent with the  \alpro and \alpox values
for either LFSRQs/SSRQs or HFSRQs.
Also, a few objects lie very close to the border line,
on both the HBL and the LBL sides, suggesting a smooth transition
between the two types.
As a comparison, we also mark in the plot the position of
2MASX\,J0324+3410, an extreme NLS1--blazar composite object,
which has a HBL-type SED with the synchrotron peak
in the UV/soft X-rays \citep{zhou07}.
Clearly, 2MASX\,J0324+3410 fulfills the HBL classification
in terms of its radio--optical--X-ray SED, as expected.
We note there are a few objects lying in the vicinity of
2MASX\,J0324+3410 in the  \alproe--\alpox diagram,
suggesting their similarities and thus the previously known
NLS1--blazar hybrid 2MASX\,J0324+3410 is not unique.

\begin{figure}
\epsscale{0.8}
\includegraphics[width=\hsize,height=0.9\hsize,angle=0]{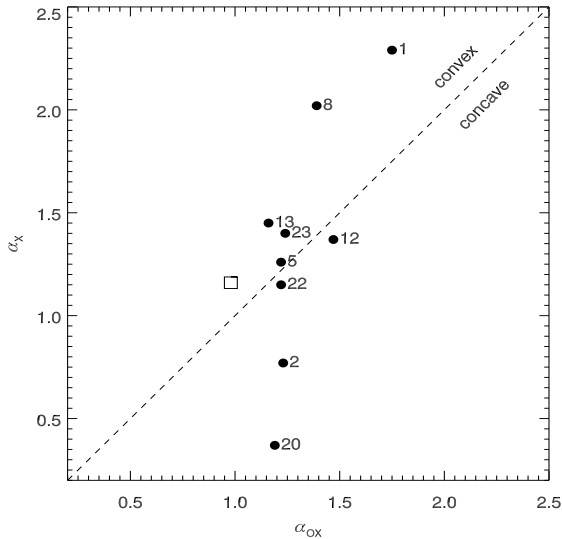}
\caption{X-ray spectral energy index $\alpha_{\rm X}$ ($\Gamma - 1$)
versus  \alpoxe.
The overall optical--to--X-ray spectral shapes are divided
into  convex shapes (\alpox$<\alpha_{\rm X}$) and
concave shapes (\alpox$>\alpha_{\rm X}$) by the dashed line
(\alpox$=\alpha_{\rm X}$).
The numbers mark the object IDs as given in Table\,\ref{tbl:sample}.
2MASX\,J0324+3410  is also plotted (square).
}
\label{fig:aox-ax}
\end{figure}

FSRQs/LBLs and HBLs are known to differ in their X-ray spectral shapes
as FSRQs/LBLs have relatively flat spectra while HBLs have steep spectra
\citep[e.g.][]{wor90,bri96,pado96,wol98};
specifically,  $\Gamma\sim2.0$ were found for  FSRQs/LBLs and
$\Gamma\sim 2.5$ or steeper for HBLs in the ROSAT band.
As such, the overall spectral shapes from the optical to the X-ray band
are concave in FSRQs/LBLs
(\alpox$>\alpha_{\rm x}$,
where $\alpha_{\rm x}\equiv \Gamma-1$ is the energy index),
while they are convex  (\alpox$<\alpha_{\rm x}$) in HBLs
\citep[e.g.][]{sam96,pado97a,per05}.
Parallelly, the X-ray emission of HFSRQs is hypothesized to
be dominated by synchrotron radiation with a steep X-ray spectrum,
as claimed to be seen in a few individuals \citep{pado02}.
For the NLS1 HFSRQs in our sample
with measured X-ray slopes, we find that
all but one (J1633+4718) have steep X-ray spectra ($\Gamma>2.0$).
We also examine the optical--to--X-ray spectral shapes
for the objects with available \alpx by plotting \alpx
versus \alpox  in Figure\,\ref{fig:aox-ax}.
As can be seen, the two objects (J1633+4718 and J0849+5108)
with flat   X-ray spectra
\alpxe$<1.0$  show prominently concave
optical-to-X-ray spectral shapes;
whereas most of the steep X-ray spectrum objects
are consistent with convex spectra,
including 2MASX\,J0324+3410 with $\Gamma\simeq 2.2$ \citep{zhou07}.
We conclude that
some of our radio-loud NLS1 AGN have SEDs similar to those of LBLs/RLQs,
while the remaining similar to HBLs/HFSRQs.
The X-ray spectral shapes are also broadly
consistent with these two types of SED.

\begin{figure*}
\plotone{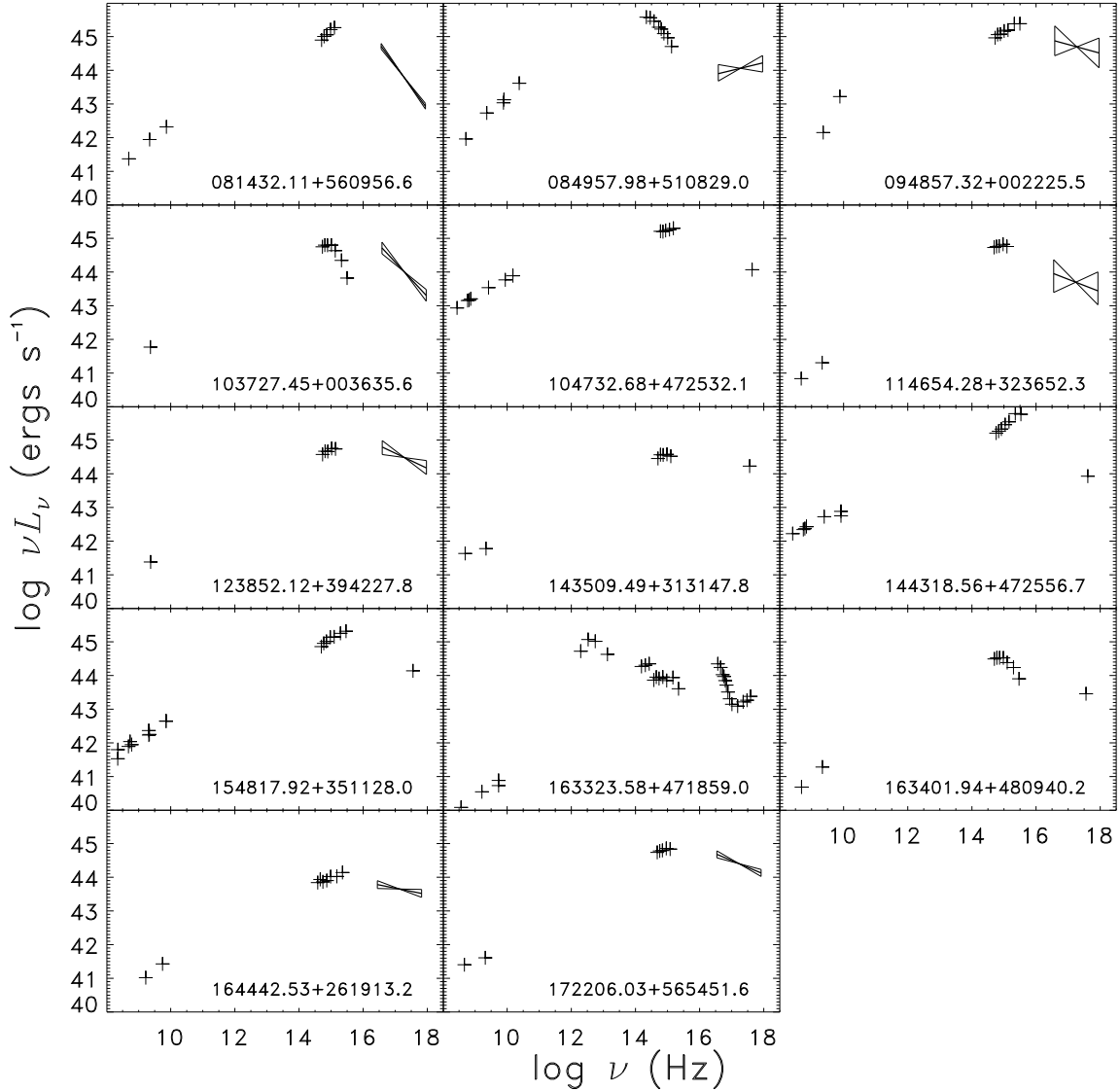}
\caption{
Broad-band spectral energy distribution
in the $\log \nu L_{\nu}-\log\nu$ representation
for radio-loud NLS1 AGN with available X-ray data. The luminosities
are at the emitting frequencies in the rest frame of the objects.
The radio and infrared data are collected from the NED
and supplementarily from recent radio surveys such as FIRST.
The optical data are  SDSS PSF magnitudes in the five bands
($u, g, r, i, z$).
The UV data are GALEX near- and far-UV magnitudes.
The optical and UV magnitudes have been corrected for Galactic extinction.
The host galaxy contamination should be negligible,
as shown in Section\,\ref{sample}.
For objects whose X-ray spectrum can be represented by a power-law,
the estimated slope and uncertainties are indicated as a bow-tie.
For J1633+4718, the X-ray spectrum is complex and the unfolded
spectrum is presented.
}
\label{fig:sed}
\end{figure*}

Finally, in  Figure\,\ref{fig:sed} we plot the broad band SED
in the $\log\nu-\log\nu L_{\nu}$ representation
for the 14 objects with X-ray detection.
Though there is only one broad radiation hump detected
(due to a lack of hard X-ray and $\gamma$-ray data),
they do reveal some similarities with the low-energy (synchrotron)
hump characteristic of blazars.
In particular, some of the HBL-like candidates are found to
have the radiation hump peaked around the UV, supporting the
results from the above \alproe--\alpox distribution;
these include
J1146+3236, J1238+3942, J1644+2619, and J1722+5654, etc.
They also show SEDs resembling that of  2MASX\,J0324+3410.

\section{Emission lines and AGN eigenvectors}
\label{sect:emline}

The  Eigenvector-1 space involves primarily a set of well known
correlations  among the following parameters:
the line width $FWHM$(\hbe), \rfee, \othbe, and soft X-ray spectral index \citep{bg92}.
\rfe is the relative strength of the \feii multiplets
in terms of the \feiie-to-\hb flux ratio,
$R_{4570} \equiv$ \feiie/H$\beta$,
where \feii is the flux
of the \feii multiples in the range of 4434--4684\AA, and
\hb is the total \hb flux  \citep{veron01}.
\othb is the strength of \oiii relative to that of \hbe.
Here we examine these correlations for our RL NLS1 galaxies,
and compare them with those for the large NLS1 sample of Zhou'06,
which are predominantly radio-quiet
(plotted as small dots in the figures in this section).

\begin{figure}
\includegraphics[width=\hsize,height=0.9\hsize,angle=0]{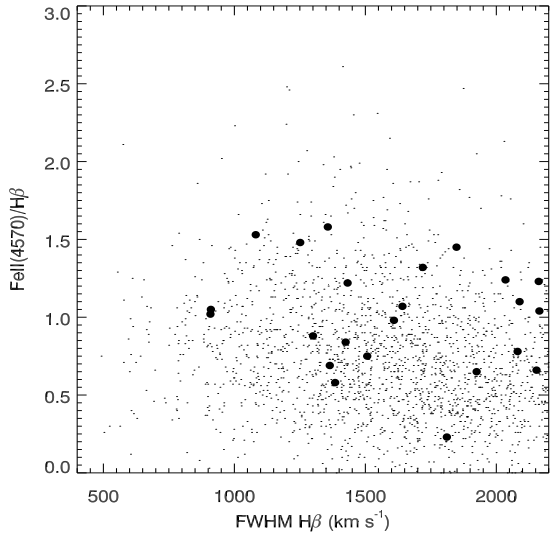}
\caption{Ratio of the \feii multiplets to the total \hb line flux,
 \rfe  versus \hb line-width in $FWHM$
for the radio-loud NLS1 galaxies (fill circles),
as well as normal NLS1 galaxies in the \citet{zhou06} sample (small dots).
}
\label{fig:feii-fwhm}
\end{figure}

\begin{figure}
\includegraphics[width=\hsize,height=0.9\hsize,angle=0]{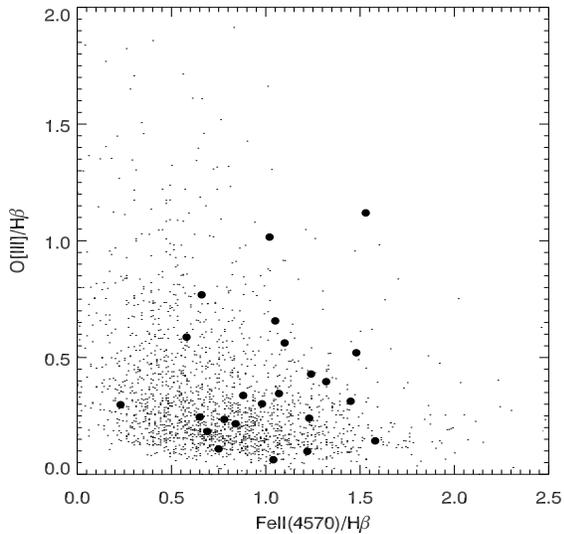}
\caption{
Ratio of the \feii multiplets to the total \hb flux,  \rfe  versus
the ratio of the \oiiie-to-\hb line flux.
Plot symbols are the same as in Figure\,\ref{fig:feii-fwhm}.
}
\label{fig:fe-o3}
\end{figure}

Figure\,\ref{fig:feii-fwhm} shows the correlation between
the \hb line-width and the relative strength of \feiie;
again, a significant anti-correlation is present for the RQ NLS1 sample
(with a large scatter though),
as shown in \citet{zhou06},
whereas no significant correlation is found for the RL objects.
A similar result is found for the correlation between  \rfe and
\othbe, as shown in Figure\,\ref{fig:fe-o3}.
While an anti-correlation between
the \feii and \oiii strength  is  evident for `normal' NLS1 galaxies,
there appears no such correlation for the RL objects.
 However, it should be noted that,
given the weakness of most of these correlations found for the RQ sample,
our RL NLS1 sample is too small to draw statistically
significant inferences.
Also, we note that
the high \rfe values for the RL objects span a
relatively narrow range of \rfe in the diagram, which may partly
account for the lack of the \rfee--\othb correlation.

Another well known Eigenvector-1 correlation is the
anti-correlation between soft X-ray photon index $\Gamma$
and \hb line-width \citep[e.g.][]{wang96,bol96}, that was found
to extend down to $FWHM\sim1000$\,\kmps \citep[][their Figure\,17]{zhou06}.
We examine this property for the 14 RL NLS1 galaxies with
measured  $\Gamma$, and find that they do not follow the
$\Gamma$--$FWHM$ trend defined by their RQ counterparts.
However, the sample size is small, and the scatter in the known
relation is huge.
Alternatively, the finding could be
attributed to a different
X-ray radiation mechanism dominating the ROSAT PSPC band
 in RL and RQ NLS1 galaxies (see below).

We also search for potential correlations between radio-loudness
\rl and  \rfe as well as \othb within our RL NLS1 sample;
 no significant correlations are found, however.

\begin{figure}
\includegraphics[width=\hsize,height=0.9\hsize,angle=0]{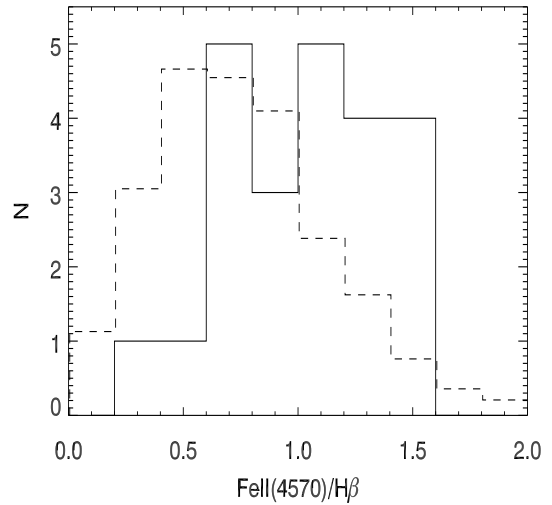}
\caption{
Distribution of \feiie/\hb for the radio-loud NLS1 galaxies (solid line).
For comparison, also plotted is the distribution
for the normal NLS1 galaxies in the \citet{zhou06} sample,
normalized to the size of the radio-loud sample (dashed line).
}
\label{fig:feii}
\end{figure}

In the process of optical spectral analysis,
we have noticed that the \feii multiplets are generally strong
in our RL NLS1 galaxies.
We plot in Figure\,\ref{fig:feii} the distribution of \rfee,
and,  as a comparison,
 also the \rfe distribution for the NLS1 galaxies in the Zhou'06 sample
(normalized to the  number of RL objects).
The two-sided K-S test testing if the two distributions
are drawn from the same population
yields a probability level of 0.2\% only;
it is evident that the RL NLS1 galaxies have  systematically higher
\rfe  values than their RQ counterparts.
The average \rfe is 1.03, compared to 0.82
for the `normal' NLS1 galaxies
with reliable \feii measurements in the Zhou'06 sample.
To examine whether the large \rfe values are caused by systematically
weak (total) \hb emission, we compare the \hb flux distribution
of the RL and RQ samples;
we find that the two \hb  flux distributions are indistinguishable
between the two samples (the K-S test).
Moreover, the RL NLS1 sample has $FWHM$(\hbe) and \othb distributions
indistinguishable from those of the Zhou'06 sample (the K-S test).
This result, together with the absence of the anti-correlations of
\rfee--FWHM(\hbe) and  \rfee--\othb
indicate that the high \rfe values for the RL sample
are real, rather than spuriously induced from dependence on other
parameters.
We thus conclude that the optical \feii emission in our
very RL NLS1 galaxies is on average stronger than that in RQ NLS1 galaxies.
In fact, this effect
has been noted in the sample of \citet[][their Figure\,3]{kom06b},
but the statistical significance is not as high
due to a factor of 2 smaller of the sample size.

We have identified four \oiii blue outliers in our sample,
with the \oiii lines significantly blueshifted
by $V$([\ion{O}{3}])$> 200$\,\kmps relative to [\ion{O}{2}]
\footnote{$V$([\ion{O}{3}]) are 220, 400, 480 and 520\,\kmps
for J1634+4809, J1505+0326,
J1305+5116, and J1443+4725, respectively}.
The phenomenon of blue outliers has been observed in a few radio
galaxies \citep[e.g.][]{tad01}, and
is especially frequent among NLS1 galaxies
\citep[][and references therein]{kom07}.
We find a fraction of blue outliers with $V$([\ion{O}{3}])$> 200$\,\kmps
 of 17\% among our sample,
about a factor of 2.4 higher than that in the RQ NLS1 sample of \citet{kom07}.
These authors also discovered that, for RQ blue outliers, the velocity
offsets are correlated with ionization potential and
line width. Such correlations persist for RL blue outliers.
In all four we detect blueshifted [\ion{Ne}{5}] lines with velocity
shifts up to $\sim$2300\,\kmps (J1443+4725).
The high frequency of blue outliers in RL NLS1 galaxies may indicate that
jet-cloud interactions contribute to, or dominate, the
\oiiie$\lambda5007$ blueshifts in the RL sources.

Most of the objects  show symmetric profiles of the \oiiie$\lambda5007$ line.
In three objects, particularly J1633+4718,
the \oiii doublets show a blue wing that can be modeled by an extra
Gaussian component (see Figure\,\ref{fig:sdssspec}).
This indicates the presence of possible outflows
associated with the narrow line region.
In addition,  J103123.73+423439.3 shows double-peaked profiles
in both \oiiie$\lambda5007$ and \hbe.

\section{Host galaxies}
\label{sect:hostgal}
The host galaxies of  radio-loud NLS1 AGN are poorly understood due to
very sparse observations.
The only object, as far as we are aware of, observed with the
HST is 2MASX\,J0324+3410.
Its HST image reveals a ring or one-armed galaxy morphology,
indicating a possible galaxy interacting/merging history  \citep{zhou07}.
As for our current sample,
given the relatively high redshifts, most of their host galaxies
cannot be examined with the SDSS images.
The exceptions are J1633+4718 and J1644+2619,
which have redshifts of 0.116 and 0.144, respectively.
Interestingly, the host galaxy of J1633+4718 has a size of more than 20\,kpc
along the major axis and is in an interacting pair/merger system,
in which the second galaxy hosts a reddened starburst nucleus \citep{bade95}.
Both of the member galaxies appear to be gas rich with disturbed morphology,
and thus of late morphological type.
The host galaxy of J1644+2619, though being compact ($\sim15$\,kpc),
also seems to be disk-like.
We note that the black hole masses in both galaxies are small,
2.0$\times10^{6}$ and 8.4$\times10^{6}$\,\msun for
 J1633+4718 and J1644+2619, respectively (see \S\,\ref{sect:mbh}).

The SDSS spectrum of J1633+4718 includes a significant contribution
of the host galaxy light within the fiber of 3\arcsec\ diameter,
which appears to be dominated by a young stellar population,
as can be seen in Figure\,\ref{fig:sdssspec}
(the contamination from the starburst nucleus of 4\arcsec\
away should be negligible).
Besides, marginal imprints of high-order Balmer absorption lines,
though being week,
are likely present in the SDSS spectra of
other three objects at higher redshifts
(J1146+3236, J1505+0326, and J1634+4809; $z = 0.4-0.5$),
indicating possible young stellar populations
in their host galaxies.

We thus suggest that some of the radio-loud NLS1 AGN
(at least those with a small black hole mass; see \S\,\ref{sect:mbh})
may reside in gas-rich, disk galaxies,
and some reside in galaxies with possible young stellar populations.
This result  contrasts clearly with
classical RLQs and BL Lac objects,
that preferentially reside  in giant elliptical
or massive bulge-dominated galaxies,
or strongly passively evolving galaxies
\citep[e.g.][]{bah97,dun03,flo04,odo05}.
While it is, to some extent,
in line with the finding of \citet{ho01} and \citet{ho02}
for the Seyfert galaxies in their samples,
which are much less luminous and more sub-Eddington
than our radio-loud NLS1 AGN studied here, however.
As is shown in \S\,\ref{sect:mbh},
most of the objects in our sample have relatively small black hole masses
(\mbhe$\sim10^{6-8}$\,\msune).
Hence, considering the \mbhe--spheroid mass relation
known for nearby galaxies \citep{mag98},
our finding about the host galaxies of radio-loud NLS1 AGN
is not unexpected.
Deeper observations are needed to reveal the host galaxy properties
for the remaining radio-loud NLS1 AGN with larger \mbhe,
which are not an easy task given their relatively high redshifts, however.

\section{Interpretation and discussion}
\label{sect:discus}

\subsection{The sample}
\label{sect:dis_sample}
Our objects fulfill the conventional definition of NLS1 galaxies,
and show typical NLS1 characteristics such as strong optical \feii emission.
They also share similar distributions in black hole mass and
Eddington ratio with `normal' NLS1 galaxies (see \S\,\ref{sect:mbh}).
Therefore, they should be bona-fide NLS1 galaxies.
The 5\,GHz radio luminosities $L_{\rm 5GHz}$ range from
$10^{24.17}$ to $10^{26.96}$\,W\,Hz$^{-1}$ with a median of
$10^{25.23}$.
Thus, most of our objects would remain qualified as radio-loud
even if the alternative criterion of RL AGN is applied,
which is based solely on radio luminosity,
$L_{\rm 5GHz} \ga 10^{25}$\,W\,Hz$^{-1}$
as proposed by some authors \citep[e.g.][]{mil90,kel94}.
Of our sample, 19 were spectroscopically targeted
in the SDSS as quasar candidates based on their optical colors,
and 3 were targeted serendipitously;
only one object was selected solely as a ROSAT/FIRST counterpart.
Hence our very RL NLS1 sample can be regarded as
an essentially optically-selected one,
similar to other SDSS radio-quiet NLS1 samples.

The sample spans a luminosity range from Seyfert galaxies
to  quasars, with roughly half of the sample in each category.
Compared to the `normal' NLS1 galaxies in the Zhou'06 sample,
which have an average \hb luminosity
$\langle$\lhbe$\rangle =3.0 \times 10^{42}$\,\ulume,
our RL objects have relatively higher luminosities
($\langle$\lhbe$\rangle=6.0\times 10^{42}$\,\ulume,
see also Figure\,\ref{fig:lo-lb}).
This difference is likely a consequence of the trend that
RL objects are more likely to be found in AGN with
more massive black holes (and thus higher luminosities),
as suggested recently \citep[e.g.][]{best05}.
We find that such very RL objects represent only 0.7\%
of NLS1 galaxies drawn from the SDSS,
though this fraction may be somewhat uncertain
given all kinds of selection effects introduced in the surveys.

Compared to the other two previously
published RL NLS1 samples of \citet{kom06b} and
\citet{wha06}, which comprise 11 and 16 objects respectively,
our current sample differs mainly in that it has a
much higher radio-loudness cutoff.
Consequently, the sample is biased for selecting flat-spectrum sources
dominated by beamed radio emission.
In addition, our sample has a statistically higher redshift
distribution than those two samples.
In comparison, only three and four objects
from those two samples satisfy \rle$>100$,  respectively,
among which, two from the former and one from the latter sample
are included in our current sample.

\subsection{Broad band continuum emission}
\subsubsection{Radio emission: relativistic beaming and jets}
\label{sect:disc_sed}

The radio luminosities of our sample objects
are well above those expected from
the most radio-luminous starburst galaxies
$\sim 10^{22.3-23.4}$\,W\,Hz$^{-1}$ \citep{smith98}.
In addition, no imprints of vigorous starbursts are found
in the optical SDSS spectra of sources of our sample;
these spectra sample the central region of
the host galaxies within a radius of 3--11\,kpc.
None of the objects  are found to be associated with
IRAS sources except for J1633+4718, for which
the infrared emission is expected (at least predominantly)
from the starburst nucleus 4\arcsec\ away, rather than from
the active nucleus, however.
These results suggest that for essentially {\em all} of the objects
the radio emission is of AGN origin, rather than of starburst origin.

Large amplitude flux variations up to $\sim 70\%$ have been observed
in some of our objects with flat radio spectra.
To make sure that this is not spurious due to
contamination of nearby sources,
we visually examined the FIRST and NVSS images of the variable objects.
We find that for 6 of the 8 objects no other sources are detected
within 3\arcmin;
for the remaining 2, faint sources are present
but are all well separated from them.
Therefore, the observed variations are not caused
by flux contamination from nearby sources.
In addition,
we argue in this paper (see Appendix\,\ref{sect:riss} for details) that
those observed variations are largely intrinsic, rather than
extrinsic that is caused by refractive interstellar scintillation (RISS).

We have shown that most of the sources known to be flat-spectrum
in our sample have very high brightness temperatures $10^{11-13}$\,K;
in 4 of the objects $T_{\rm B}$ even exceeds the
inverse Compton limit of $10^{12}$\,K.
Such high  brightness temperatures indicate that
 the radio emission is highly Doppler beamed
and  hence must originate from (at least mildly) relativistic jets,
as in blazars.
Among the 4 with $T_{\rm B} > 10^{12}$\,K,
J0849+5108 is a known OVV/HPQ blazar
\cite[e.g.][and see Appendix\,\ref{sect:indv}]{arp79,sitko84} and
the high brightness temperature is well expected;
in turn, this gives us confidence in the results for the
other objects.
J0948+0022 has been previously suggested to be so by \citet{zhou03}.
For the remaining two, J1505+0326 and J1548+3511,
the evidence for relativistic beaming is presented for the first time.
For J1505+0326, this result is consistent with
its VLBI images made at 2.3\,GHz,
which are dominated by a very compact component,
marginally resolved at a few tens milli-arcsec resolution \citep{dal98}.
Besides, two objects,  J1633+4718, and J1644+2619 have also been
suggested to be blazar-like based on VLBI observations \citep{doi07}.
Thus, in a total of 6 objects of our sample
there is evidence for  relativistic beaming.
Assuming that the brightness temperatures of $10^{11-12}$\,K
are probably also an indication of beaming,
the number of such objects will increase to 9, i.e.\ all
the 8 flat-spectrum sources and one with unknown radio index.

Our suggestion for the presence of (at least mildly) relativistic jets
is also supported by polarization detected in a few
individual objects, e.g. J0849+5108 (see Appendix\,\ref{sect:indv})
and 2MASX\,J0324+3410 \citep{neum94}.
Furthermore,  this is consistent with the observed large amplitude
variability in the optical in a few objects monitored, e.g.\ J0849+5108,
J0948+0022 \citep{zhou03}, since normal NLS1 galaxies are found to be
much less variable in the optical band \citep{ai08}.

These results are most naturally explained in terms of
beamed emission; and thus
we postulate that relativistic jets are likely present in most,
if not all, of the flat-spectrum, very radio-loud NLS1 AGN.
However, it should be noted that part of the observed properties above
may also be explained by some other non-blazar, compact sources.
For instance, flat and inverted spectra are also typical of
Gigahertz-Peaked Spectrum (GPS) sources, though they
generally lack of large amplitude flux variations \citep[e.g.][]{odea98}.
Moreover, flat spectra are often seen also in the radio sources
of some `normal' Seyfert galaxies,
though mostly with much less power compared to the objects
studied here \citep[e.g.\ ][]{ho01a}.
Incorporating the properties of these flat-spectrum NLS1 AGN
in other wave-bands,
we regard the blazar-like scenario as the most likely origin
of their multi-waveband (non-thermal) emission in this paper,
though other possibilities cannot be ruled out completely at this stage.
We note that  the majority of the sources of \citet{kom06b} sample
(\rle$>10$) are steep-spectrum sources,
in which strong beaming is generally not expected.
The difference is largely attributed to the
much higher radio-loudness cutoff of our sample, as mentioned above.

\subsubsection{Optical continuum: dust extinction
and jet contribution?}
\label{sect:dis_opt}

The majority of our sample objects (16/23)
have optical continuum slopes  bluer than \alpwvle=$-1$.
If the continuum is intrinsically blue in all objects
(assuming \alpwvle=$-1.7$) but is reddened by extinction,
a slope of \alpwvl=$-1$ corresponds to an amount of
extinction E(B-V)$\simeq0.2$, i.e. a correction factor of only $\sim2$
in the $B$-band flux.
So, even if this was the case,  most of our objects
would still have \rle$\ga$100, above the radio-loudness cutoff
of our sample.
There are 5 objects at low redshifts ($z<0.38$)
having both the \ha and \hb lines measurable,
including J1138+3653, the only one showing a significant
deviation from a power-law optical continuum.
J1138+3653 has a broad line Balmer decrement of \hae/\hbe=4.43,
indicative of  mild extinction.
Adopting the intrinsic  \hae/\hb ratio of 3.37
[with a dispersion of 0.06\,dex in $\log$(\hae/\hbe)]
for blue radio-loud quasars (free from dust extinction)
found in our recent work \citep{dong08}
and the Galactic extinction curve, we find an internal
extinction correction $A_{\rm B}\simeq 0.97$ [E(B-V)=0.236],
and an extinction corrected \rl a factor of 2.4 lower,
leading to an intrinsic radio-loudness \rle$\simeq 86$.
For the remaining 4 objects,
the  broad line  \hae/\hb ratios range from 2.8--3.8 with
a mean of 3.4, that are well consistent with the intrinsic
\hae/\hb distribution and hence with little or no extinction;
therefore their radio-loudness \rl measurements as above
are almost un-affected.
Particularly, J1031+4234 shows a red optical spectrum with \alpwvl=0.55,
indicating that in at least some of the objects the
red optical spectrum is intrinsic, rather than resulting from
dust extinction.
Another example is the blazar J0849+5108 with the reddest slope,
\alpwvle=$-0.20$,
that is close to the previous measurement at its high light state.
Its slope steepens with decreasing luminosity and changes rapidly
on timescales on several months, and hence the red slope is preferentially
explained as being intrinsic\footnote{Alternatively, the red slope
is  explained as contamination from a foreground red galaxy on
the line of sight \citep{stickel89}, though we consider this unlikely.
}
\citep{arp79}, similar to red BL Lac objects \citep{ste76}.

Observations have revealed that in a large fraction of FSRQs
the optical light can be contributed significantly by the
non-thermal emission from relativistic jets that is highly beamed
\citep[e.g.][]{whi01}.
We have discovered that the observed optical fluxes are
systematically in  excess of those predicted from
the broad \hb luminosity,  presumably from an accretion disk,
assuming that radio-loud NLS1 AGN have the same conditions
in the broad line region
as normal NLS1 AGN (see \S\,\ref{sect:optcont}).
In addition, the optical slopes of our radio-loud objects
are  systematically bluer than that of radio-quiet NLS1 AGN.
These facts point to the presence of a second component
superposed on the normal AGN-like (thermal) continuum.
In light of the similar radio properties of our RL NLS1 galaxies to FSRQs,
we attribute the excess optical emission to arising from the non-thermal
radiation from jets, as in FSRQs, i.e.\ the extension  to higher energies of
the radio-to-infrared synchrotron radiation.
This model explains naturally the observed difference
in the observed-to-predicted luminosity ratios
between flat- and steep-spectrum NLS1 AGN.
In flat-spectrum sources,
 the relativistic beaming effect is strong---as
suggested above---and hence the jet component over-shines the  thermal optical component;
while in steep-spectrum sources, the former becomes comparable to
or weaker than the latter.

The observed optical continua have a variety of slopes,
ranging from moderately red to blue.
They are reminiscent of the
optical emission of BL Lac objects, which have both red and blue
continua \citep[e.g.][]{ste76},
corresponding to the low- and high-peak frequency of
the synchrotron hump of LBLs and HBLs, respectively.
In fact, this comparison is particularly relevant in light of our finding that
the RL NLS1 AGN have both LBL- and HBL-like SEDs:
the blue optical continuum may be explained as contribution from
the synchrotron emission peaked
at high energies (UV or higher);
while the red slope may be contributed by that peaked in the infrared.
In turn, this is consistent with our above inference
that the red slopes in our objects are mostly intrinsic,
rather than resulting from extinction.

\subsubsection{X-ray radiation and broad band SED}

So far, we have shown that relativistically beamed jet emission,
as found in blazars,
provides a good explanation of the observed radio and optical
properties in flat-spectrum RL NLS1 AGN.
This scenario is also consistent with the fact that
some of the objects show SEDs similar to FSRQs/SSRQs,
while the others similar to HFSRQs
 in the \alproe, \alpoxe, and \alprx parameter space
(Figure\,\ref{fig:alprox}).
In fact, a few of these objects
were simply classified as FSRQs or SSRQs in previous studies.
It is thus natural to speculate that the
similar radiation mechanisms are operating in these objects as in
normal FSRQs/SSRQs and HFSRQs
for producing the broad-band non-thermal emission.

From the soft X-ray spectral shapes alone it is not
possible to distinguish the origin of the X-ray emission, i.e.\
that associated with accretion disk as in normal RQ NLS1 galaxies
or non-thermal jet emission as in blazars,
since RQ NLS1 galaxies also show a similar $\Gamma$ distribution
with a large scatter \citep[e.g.][]{bol96,gru99,lei00,zhou06}.
The only exception is perhaps the flat hard X-ray component
($\Gamma=1.37\pm0.49$) in J1633+4718, whose best estimate value
is hardly found in RQ NLS1 galaxies but is typical
of FSRQs.

One way is to search for possible enhancement in the X-ray emission
of RL NLS1 AGN compared to RQ objects in a statistical sense,
which, if is the case, may be
considered as an indirect evidence for the presence of a
non-thermal jet X-ray component.
Here, we simply use the X-ray detection rate in the RASS
as an indicator of possible
enhancement in the X-ray emission of our objects over normal NLS1 galaxies
by comparing the RL and RQ samples.
This is plausible since our RL NLS1 sample is essentially an optically
selected one (see discussion in \S\,\ref{sect:dis_sample}),
as the SDSS RQ NLS1 samples.
The RASS detection rate for RQ NLS1 galaxies is $\approx 31\%$
from the SDSS DR3 NLS1 sample \citep{zhou06};
while it is 50($\pm11$)\% (11/22) for the present RL sample
(excluding one solely RASS-selected target).
If we consider the objects (both detections and upper limits)
falling within the HBL locus only
(Figure\,\ref{fig:alprox}; excluding 2MASX\,J0324+3410),
the detection rate is even higher 58($\pm14$)\% (7/12).
Therefore, the higher RASS detection rate in our RL sample
compared to the RQ NLS1 sample is suggestive of
the presence of a significant excess X-ray component,
presumably from the relativistic jets.

In Appendix\,\ref{app:xray}, we discuss in detail
the possible mechanisms of the X-ray radiation of our RL NLS1 AGN,
which are summarised here.
In all of the objects, the X-rays produced in the accretion process
as in normal RQ NLS1 galaxies are likely not negligible and
make a considerable contribution.
In addition, we suggest that the non-thermal X-ray emission from
relativistic jets may also play a similarly important role.
Such a component is most likely to originate from inverse Compton
radiation and has a flat X-ray spectrum
(in J1633+4718 and possibly in J0849+5108),  as seen in FSRQs,
for objects having the LFSRQ-like SEDs (NLS1 LFSRQs);
whereas it is possibly the high-energy tail of synchrotron radiation with
a steep spectrum for objects with the HFSRQ-like SEDs (NLS1 HFSRQs).

\subsubsection{Dichotomy of LFSRQ- and HFSRQ-like NLS1 AGN}

In Figure\,\ref{fig:alprox}, there are about 7--12 objects,
or 1/3--1/2 of the sample,  falling within the box-shaped locus
and thus having HBL/HFSRQ-like SEDs
(the uncertainty comes from those with lower \alpox limits).
Considering that radio/optically-selected (`broad'-line)
FSRQs have predominately LBL/LFSRQ-like SED,
such a high fraction of HFSRQ-like objects is surprising
for our sample, which is essentially optically/radio-selected.
The figure also shows that there are no objects resembling extreme
HBL/HFSRQs (the lower-left corner of the HBL locus),
which have the synchrotron component peaked in the X-ray regime.
In fact, nearly all the candidate HFSRQ-type SEDs have
the synchrotron humps peaked likely around the UV band.

Strong emission lined--FSRQs with HBL-type SEDs
have recently been reported
as a result of the identification of radio--X-ray (ROSAT) selected
sources \citep[][]{pado97b,per98,pado03}.
One major distinction with our objects
is that those objects have much broader lines with $FWHM>2000$\,\kmpse,
typical of classical FSRQs.
The fraction of HBLs/HFSRQs is sample-selection dependent;
the authors proposed a fraction
(HBLs + HFSRQs) of about 10--15\% among
the whole  blazar population  and
about $30$\% in the RASS--Green Bank (RGB) sample  \citep{pado03}.
The RGB sample is comparable in the way of selection
to our RASS-detected sub-sample here, which
 comprises 12 objects;
among them, 8 objects are possibly of the HFSRQ-type
(Figure\,\ref{fig:alprox}), making the fraction as high as $\sim2/3$.
It seems that the HFSRQ fraction of our radio-loud NLS1 sample is higher than
that of HFSRQs in \citet{pado03}.

\begin{figure}
\includegraphics[width=\hsize,height=0.9\hsize,angle=0]{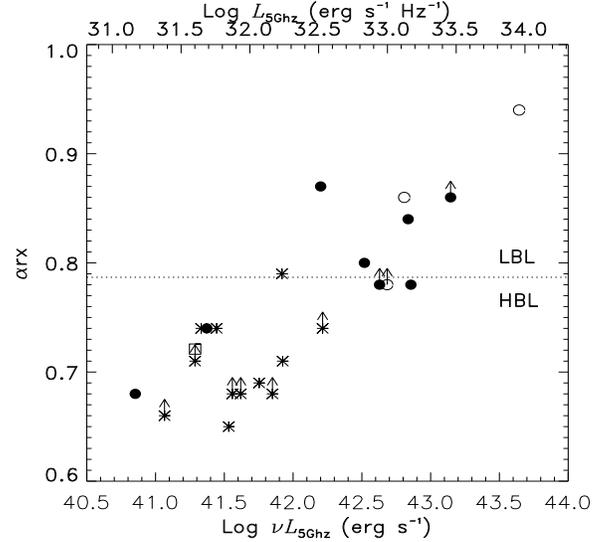}
\caption{Relationship of \alprx and 5\,GHz radio luminosity.
Symbols: arrows indicate \alprx upper limits for objects
without X-ray detection;
filled dots: flat spectrum radio sources;
open circles: steep spectrum radio sources;
asterisks: sources without available radio spectral indices.
2MASX\,J0324+3410  is also plotted (square).
The dotted line corresponds to the simple division between LBLs and HBLs
using \alprxe.
}
\label{fig:arx-lr}
\end{figure}

Within the unified blazar sequence,
the energy of the synchrotron peak was found to be
well anti-correlated with 5\,GHz radio luminosity\footnote{
$L_{\rm 5GHz}$ is thought to be a good tracer of the bolometric luminosity.},
and with \alprxe,
which was suggested to be a tracer of the peak energy \citep{fos98}.
We plot in Figure\,\ref{fig:arx-lr}  \alprx versus
 5\,GHz  radio luminosity for the sample, with those objects without
X-ray detection indicated by arrows.
It clearly reveals a good correlation;
the Spearman correlation test incorporating lower limits of \alprx
using ASURV \citep{iso86,iso90}
yields a significant correlation at the probability level \ps=0.4\%.
It can be seen that the transition between LFSRQs and HFSRQs
(using \alprxe$\approx 0.78$) happens at around $\nu L\nu\sim10^{42}$\,\ulume,
consistent with what was previously found between  HBLs and LBLs
\citep[][see their Figure\,7]{fos98}.
If \alprx does trace the synchrotron peak energy,
this result shows that
our radio-loud NLS1 AGN do follow  the blazar sequence well.
In fact, the objects at the low luminosity end are among the
least radio luminous blazars known so far,
falling close to or even below
the low-end of the FSRQ luminosity function according to
unification schemes
\citep[$L_{\rm 5GHz}\sim10^{31.5}$\,\umlume,][]{urry95}.
However, there seems to be no or at most a weak
correlation between \alprx and \hb luminosity (\ps=0.1).
Such a blazar sequence also explains naturally the high fraction
of HFSRQs, as the majority of our sample
have lower radio powers compared to classical FSRQs.
In fact, the objects of our sample have intrinsic powers mostly in the
range for FR\,I type radio sources (see \S\,\ref{sect:intrpower}).

\subsection{Intrinsic radio power and radio-loudness}
\label{sect:intrpower}

Optically thin radio emission at low frequencies is thought to be isotropic,
and thus can be used as a measure of the intrinsic power of the radio sources
free from Doppler boosting.
The low frequency power of radio galaxies can be separated into
two distinct classes at the fiducial value
$L_{\rm 178MHz}\simeq 2\times 10^{25}$\,W\,Hz$^{-1}$\,sr$^{-1}$:
the FR\,Is with the radio emission decreasing
outwards away from the nucleus, and the  FR\,IIs which have
lobes with prominent hot spots and bright edges
\citep{fr74}.
It has been suggested that RLQs are found in
FR\,II sources, and the distinction of the FR\,I and  FR\,II types
is likely determined by the accretion process
of the central engine \citep[e.g.][]{baum95},
as well as by the density of the environment \citep{gop02}.
Despite a lack of adequate observations on their radio morphology,
it would be informative to compare
the isotropic radio power of our RL NLS1 galaxies with the FR types.

We estimate the rest frame 151\,MHz luminosities, \llfe, for our
objects, based on their radio data at low frequencies, which are
summarised in Appendix\,\ref{sect:lowfemi}.
For 4 objects with 151\,MHz fluxes, \llf are calculated directly
by using the 151--327\,MHz indices for K-correction.
Among the rest, 11 objects have fluxes at 327\,MHz,
for which \llf are estimated  by extrapolating the spectra down to
151\,MHz assuming the
average \amme=$-0.34$  found from the above 4.
Using the steepest ($-0.62$) and the flattest ($+0.09$) \amm found above
would increase and decrease \llf by a factor of up to 1.4 and 1.8,
respectively; and thus the estimated \llf are relatively insensitive to
the choice of  \amme.
For the 15 objects with either detected or estimated 151\,MHz fluxes,
\llf range from
$4.5\times 10^{24}$ to $3.5\times 10^{27}$\,\whze, with a median of
$7.1\times 10^{25}$\,\whze.
Among these, three have luminosities in the FR\,II range
(J1047+4725, J1305+5116, and J1443+4725, all
detected in the 151\,MHz survey);
while the remaining majority have similar luminosities
as FR\,Is or are close to the transition line.
With estimated  \llf=$9\times10^{24}$\,\whze,
2MASX\,J0324+3410 also has a luminosity typical of a FR\,I galaxy.
However, it is not clear whether the clear separation in both
radio power and morphology for radio galaxies
is applicable to RL NLS1 galaxies,
considering their apparently peculiar, compact morphology.

It has been suggested that, in classical RL AGN,
isotropic radio emission at low frequencies can be used as a good
indicator of the power of the jets \citep[e.g.][]{raw91}.
If this is applicable to RL NLS1 galaxies,
the above results imply that
our sample has much lower jet power compared to classical RLQs,
which have radio power and morphology predominantly
in the  FR\,II category \citep[e.g.][]{raw91,baum95,liu06}.
We will address this issue in a detailed study in the future.

The radio-loudness, by its definition, is largely an
apparent rather than intrinsic parameter
for the sources in which the beaming effect is important
above 1\,GHz in the radio but less so in the optical band.
As a consequence, some radio-loud AGN would have been classified as
radio-intermediate or even radio-quiet, which
might also possess weak and mildly relativistic jets in at least some
\citep[e.g.][]{fal96b,blu03,bru05,wang06},
were their jets not directed close to our line-of-sight
\citep{fal96a,jar02}.
A more physically meaningful parameter is
the `intrinsic' radio-loudness free from the beaming effect.
We calculate the intrinsic radio-loudness using the un-beamed radio
component, that can be extrapolated from a low-frequency flux
and dominates fluxes at several GHz after the jet emission
is de-boosted.
The intrinsic  radio-loudness at 5\,GHz,
$R_{\rm 5GHz}^{\rm intr}$, is estimated  by using 327\,MHz fluxes
for those having this flux measurement.
We find $R_{\rm 5GHz}^{\rm intr}$ in the range 34--2600
(with a median of 110)
assuming a spectral slope of $-0.7$ for the un-beamed component,
or 17--1380 (with a median of 56) assuming a slope of $-1$.
It should be noted that such estimates are subject to large uncertainties,
partly due to the above finding that
the optical light is also partially beamed in some of the objects.
Nevertheless, they seem to indicate that
a sizable fraction of the sample, including 2MASX\,J0324+3410,
are likely radio-intermediate
\citep[e.g. $<$250 was suggested by][for flat-spectrum sources]{fal96a}.
Among the sample, there are almost no
genuine intrinsically RQ objects with $R\ll 10$, however.

\subsection{Black hole masses of radio-loud NLS1 galaxies}
\label{sect:mbh}
\subsubsection{Estimation of black hole mass and Eddington ratio}

It has become possible recently to estimate the black hole masses \mbh of AGN
from single epoch spectroscopic observational data
\citep[e.g.][]{kas00,wan99},
though the uncertainties of both systematics and scatter remain large
\citep[][]{vp06,col06}.
Assuming that the motion of the line-emitting clouds are virialised,
the black hole mass can be determined by
$M_{\rm BH} = f R_{\rm BLR} \Delta v^2 / G$,
where $f$ is a scale factor depending on the kinematics and geometry
of the BLR,
$\Delta v$ the emission line width, and
 $R_{\rm BLR}$ the BLR radius.
We assume that the commonly used $R_{\rm BLR}$--luminosity relation,
derived  via reverberation mapping for  nearby, normal AGN,
is applicable to NLS1 AGN.
We use the relation for
the \hb luminosity given by  \citet[][their Eq.\,1 and Table\,3]{kas05};
we intend not to use the continuum 5100\AA\ luminosity for
possible contamination from jet emission (see Section\,\ref{sect:optcont}).
A common practice, among a few others\footnote{A few other
scaling relations for \mbh estimation were recently presented
in the literature\citep[e.g.][]{onk04,gh05,gh07,vp06,col06},
that slightly differ from one another.
},
is to take $FWHM$ as $\Delta v$ and  $f=(\sqrt{3}/2)^2=0.75$,
assuming an isotropic distribution with random orbital inclinations
of the broad line clouds \citep{net90}.
We follow the practice in this work, the same as in  \citet{kom06b}.

The estimated black hole masses of our sample
range from $1\times 10^{6}$ to $1.5\times 10^{8}$\,\msun,
peaked at near the median $1.3\times 10^{7}$\,\msun.
The black hole mass distribution is well consistent with
that of the \citet{kom06b} sample.
However, by studying the systematic effects in \mbh estimation,
\citet{col06} argued that the use of $FWHM$ rather than line-dispersion
($\sigma_{\rm line}$) would introduce systematic bias in
the scaling factor $f$.
We also give alternative \mbh estimation by correcting this effect
using the formalism proposed by \citet[][their eq.7]{col06}.
For the objects in our sample (FWHM$<2200$\,\kmpse) specifically,
this correction sets a value of the scale factor $f\simeq$1.5,
about twice the value in the former estimation.
This results in \mbh estimates a factor of 2 larger
than the values obtained above,  ranging from $2\times 10^{6}$
to $3\times 10^{8}$\,\msun with
a median $2.6\times 10^{7}$\,\msun.
The black hole masses thus estimated are
listed in Table\,\ref{tbl:arox}, which are used
in the following analysis.

Having estimated the masses of the central black holes,
now we can derive the Eddington ratios, \redde.
For bolometric luminosity correction,
we assume  $L_{\rm bol}$=9$\lambda L_{\lambda5100}$,
as suggested by, e.g.\ \citet{war04}, \citet{ves04}, and \citet{kas00}.
The $\lambda L_{\lambda5100}$ luminosity is estimated from the
broad \hb line luminosity using Eq.(5) of \citep[][]{zhou06}
to eliminate the effect of the extra jet contribution.
It should be noted that such an estimate is subject to
large systematic uncertainties inherent in estimation of
\mbhe, as well as of bolometric luminosity.
Using the \mbh values obtained with the scale factor $f=0.75$
(see above), we find the Eddington ratios ranging from
\redd=0.52 to 2.6 and peaked  around a mean of 1.0,
similar to those found by \citet{kom06b} with a mean of 1.2.
However, when the correction of \citet{col06} is applied,
the \mbh values become about twice larger and, consequently,
the obtained \redd values become about twice smaller, i.e.\
ranging from 0.27 to 1.28 with an average of 0.52.

Now we compare the distributions of \mbh and \redd between
our RL sample and the normal NLS1 sample of Zhou'06.
Our RL sample has significantly higher
\mbh and \redd distributions compared to the RQ one.
This agrees with the general trend that,
as aforementioned in \S\,\ref{sect:dis_sample},
RL objects are more likely to be found in AGN with
massive black holes \citep[e.g.][]{best05}.
Given their narrow ranges of the line-width,
the black hole masses of NLS1 samples,
as determined from the \hbe/\ha line-width and luminosity,
is strongly correlated with luminosity, and in turn, with redshift for
flux limited samples.
This may explain the higher redshift distribution of our RL sample
compared to that of the normal NLS1 sample of Zhou'06 (see Figure\,\ref{fig:distrl}).

It should be pointed out that the virial \mbh estimation
is subject to assumption of an isotropic geometry of the BLR
with random orbital inclination of the clouds.
If the  BLR is flat, a small line width would be a result of
small inclination, and then \mbh would be largely underestimated
\citep[see, e.g.][]{wills86,dec08}.
Indeed there are some suggestions in the literature for a flattened
geometry of the BLR in some RL AGN \citep[e.g.][]{ves00,sul03,jar06};
but see \citet{pun07}.
This caveat concerns particularly those of our objects,
that have compact radio morphology and those that are blazar-like,
indicating a small viewing angle.
Given that the \mbh distribution of our sample is
indistinguishable from that of the RQ NLS1 sample
(within the same redshift range),
this comes back to the old question
whether NLS1 galaxies, as a population, have flattened BLRs and
are viewed preferentially face-on, rather than having small \mbh
\citep[e.g.][]{ost85,puc92,bian04}; but see e.g.\ \citet[][]{smi04}.

If RL NLS1 galaxies do possess relativistic jets
 and hence anisotropic radio emission, as we have shown here,
they can be used as a probe to this question since
their inclination can be inferred from radio observations
in a statistic manner.
However, with the relatively few data available currently,
we do not find any statistically significant trends of
the dependence of line-width (\mbhe) on probable inclination indicators,
such as radio-loudness or radio slope.
Therefore, though there is no compelling evidence supporting the
planar BLR nature of RL NLS1 from
the present data, we cannot rule out this possibility, either.
A much larger RL NLS1 sample covering a wide range of radio-loudness
is desirable to shed light to this important question in the future.

\subsubsection{Radio-loudness versus black hole mass}

As one of the fundamental relations in understanding AGN jets,
the dependence of radio-loudness $R$ on \mbh has been studied
extensively in recent years, with the advent of the easy
estimation of \mbh in AGN.
A $R$--\mbh correlation  was proposed by \citet{laor00} that
all RLQs in the PG quasar sample have \mbh$>3\times 10^8$\,\msun,
while the radio-quiet quasars have a much lower \mbh distribution.
This trend was confirmed in some studies \citep[e.g.][]{lacy01,mcl04,metca},
but was found to be weak or absent in others \citep{osh02,woo02}.
Furthermore,  by including Seyferts and low luminosity AGN,
\citet{ho02} found that $R$ is mostly correlated with \redd rather than
\mbhe.
In a recent comprehensive treatment of this question, \citet{sik07}
demonstrated that $R$ depends on both \redd and \mbhe, with two distinct
inverse $R$--\redd dependence sequences for AGN hosted by elliptical
(systematically higher $R$) and disk galaxies
(systematically lower $R$), respectively.
Regardless whether there is indeed a trend of dependence or not,
most aforementioned studies show that,
on the \mbhe--$R$ plane, there is a scarcity of RL AGN
with low \mbhe.
Even for those low-\mbh RL AGN suggested in the literature
\citep[e.g.][]{osh02,woo02}, there remains a controversy as to
the estimation of \mbh and $R$
\citep[see, e.g.][]{jar02,laor03}.

Obviously, as first pointed out by \citet{wang01} and \citet{zhou03},
radio-loud NLS1 AGN do {\em not} follow the above trend found for normal RL AGN
(provided the virial \mbh as estimated above are largely correct).
Moreover, \citet{kom06b} demonstrated explicitly that
RL NLS1 galaxies are located in a region of the \mbhe--$R$ diagram that
is sparsely populated by other types of AGN
($R\simeq 10^{1-3}$ and \mbh$< 10^8$\,\msune).

\begin{figure}
\plotone{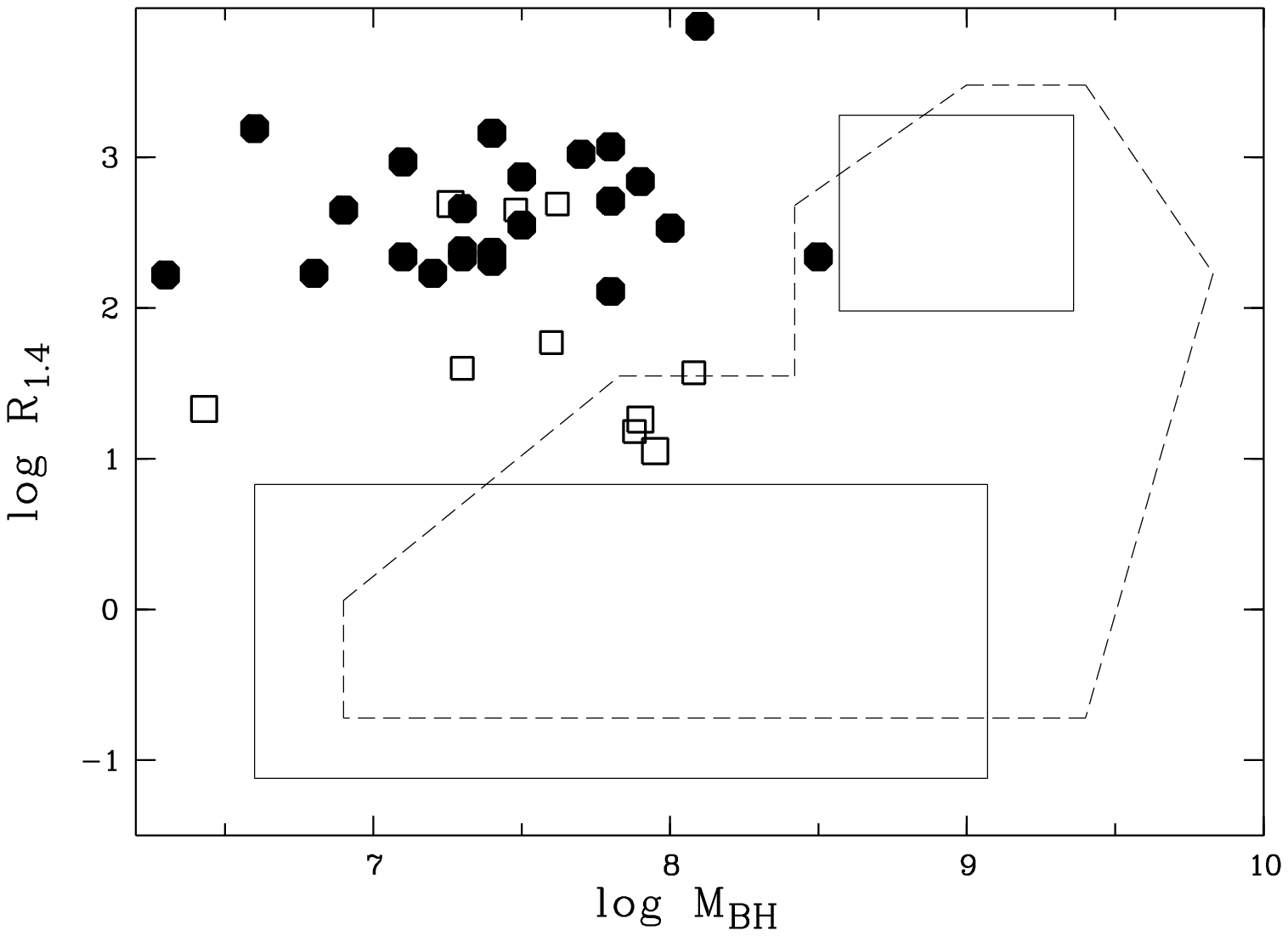}
\caption{
Dependence of radio-loudness on black hole mass for the
radio-loud NLS1 galaxies of our sample (filled dots) and those
from the sample of \citet[][open squares]{kom06b}.
Over-plotted are schematic areas populated by the bulk
of the radio sources of the samples of
\citet[][solid line]{laor00} and \citet[][dashed line]{lacy01}.
We note that a small number of (normal) RL AGN
with small \mbh and large  $R$ has also been claimed by \citet{woo02},
which lies mostly at higher  $R\simeq 10^{3-4}$ above ours in the diagram;
however, there remains a controversy as to
the estimation of those \mbh
\citep[see, e.g.][]{jar02,laor03}.
}
\label{fig:mhb-rdln}
\end{figure}

Similarly, we plot
\rl versus \mbh for our very RL NLS1 galaxies  in Figure\,\ref{fig:mhb-rdln}.
Also plotted are the loci of the objects in the samples of
\citet{laor00} and \citet{lacy01} for comparison.
Two pieces of results can be
inferred immediately from Figure\,\ref{fig:mhb-rdln}.
Firstly, within our sample, there is no correlation found between \rl and
\mbhe,  nor between \rl and \redd (\ps=0.44 and 0.27, respectively);
 however, we note that these results are rather uncertain given
the small sample size and the limited ranges in both \rl and \redde.
Secondly, our new result confirms and strengthens
that of  \citet{kom06b}, and improves upon it by extending
the RL NLS1 locus to even lower \mbh and higher \rle.
In particular, the bulk of our RL NLS1 galaxies  lies within a locus of
\mbh$\simeq 10^{6-7.5}$\,\msun and $R\simeq 10^{1-3}$,
which appears to be preferentially avoided by other types of RL AGN
[compared to, e.g. \citet{sik07}, see their Figure\,4],
except for low power Seyfert galaxies and LINERs;
however, those exceptions have exclusively low Eddington ratios
\redd$<10^{-2}$ and are drastically different
from NLS1 with high \redd \citep[see also][]{ho02}.
Compared to Figure\,4 in \citet{sik07},
our RL NLS1 galaxies appear to have
the largest radio-loudness up to \rle$\sim 10^{3-4}$
among all types of AGN within this \mbh range.
This means that the RL NLS1 galaxies show strong excess radio emission
with respect to their \mbhe,
when the suggested radio power--\mbh relation
for normal AGN  is compared\footnote{
We note that several (normal) RL AGN with  small \mbh similar to ours
and with even larger $R$ ($\simeq 10^{3-4}$)
have also been claimed by \citet{woo02},
though there remains a controversy as to the estimation of \mbh
\citep[see, e.g.][]{laor03}.
Our sample appears to differ from theirs in the sense that
it fills up the gap between  $R\simeq 10$ and $10^{3}$ in the
small \mbh regime in their \mbhe--$R$ diagram.
}
\citep[e.g.][]{lacy01}.
Finally, we note that the above two results remain valid
even if  the Doppler de-boosted radio-loudness values
 (see \S\,\ref{sect:intrpower}) are used instead,
which are in the range of $R\simeq 10^{1-3}$.

Hence, the existence of RL NLS1 galaxies as a {\em population}
with small \mbhe, if estimated correctly, fills up the previously sparsely
populated region in the \mbh--$R$ parameter space for normal RL AGN.
This result indicates that there indeed exist AGN with high
radio-loudness ($R\sim 10^{2-3}$) but intermediate \mbh ($10^{6-8}$\,\msune).
Though there is the systematic uncertainty in estimating \mbh ascribed to
the possible BLR inclination effect, we argue that this result may not
be affected qualitatively.
This is because the Komossa et al.\ sample, though
composed of mainly steep spectrum sources
(thus relatively larger inclination),
led to similar results.
Furthermore, in a few objects where the host galaxies are resolved,
their small \mbh are supported by a lack of massive galactic bulges,
as predicted by the \mbh--bulge relation \citep[e.g.][]{mag98}.
Our result also implies that the inverse relation between $R$ and \redd
as suggested previously
\citep[see, e.g.][]{gre06,sik07}
may not work at the high \redd end close to the Eddington rate.
Furthermore, \citet{sik07} argued that
RL AGN hosted by elliptical galaxies are much
radio-louder than AGN hosted by disk galaxies.
Though this statement may be true in general,
it is perhaps  more appropriate in terms of RL fraction,
i.e. the fraction of RL objects
 is lower in AGN hosted by disk galaxies compared to that of AGN
hosted by ellipticals;
in addition, the difference in radio-loudness is now less than
what was thought previously \citep{sik07}.

\subsection{On the nature of radio-loud NLS1 galaxies}

\subsubsection{A previously unknown population of radio-loud AGN?}

On the black hole mass--radio-loudness plane
our RL NLS1 galaxies form a natural extension to  low \mbh of classical RLQ/BLRGs
(Broad Line Radio Galaxies) with roughly constant radio-loudness
\citep[Figure\,\ref{fig:mhb-rdln}; see also Figure\,3 in ][]{sik07}.
This is not surprising, since the  transition between
BLAGN and NLS1 AGN  is {\em smooth}
in the line-width of the broad lines, and, as a consequence,
in black hole mass and luminosity.
If the \mbh are underestimated  systematically (see \S\,\ref{sect:mbh}),
the loci of the radio-loud NLS1 AGN will move even closer to those of normal RL AGN.
In this regard, it is natural to treat RL NLS1 galaxies as the low--\mbh end of
normal RL AGN.

On the other hand, however, RL NLS1 galaxies show some peculiarities
in observed properties that are unusual in comparison with,
or at an extreme of,  normal RL AGN.
The most distinctive features are in the radio properties.
Even for steep-spectrum sources, the radio sources
of RL NLS1 galaxies appear to be compact,
with the size less than several tens of kpc
or even smaller wherever higher resolution observations are available.
In contrast,  the radio sources of  RLQ/BLRGs are about 100--400\,kpc
in a similar redshift and luminosity range as our radio-loud NLS1 AGN \citep{sin93}.
Moreover, most of the RL NLS1 galaxies have radio power in the range of
FR\,I radio galaxies, whereas  RLQ/BLRGs tend to be associated
with FRII type sources\footnote{There are exceptions, though;
see, e.g. \citet[][]{blu01}.} \citep[e.g.][]{lai83}.

In the following aspects
the two types are not mutually exclusive
but overlap to some extent, yet they seem to differ in a systematic way.
Firstly, There seems to be a higher fraction of HBL/HFSRQ-like blazars
in optically selected RL NLS1 galaxies than in normal
RLQ/BLRGs of radio/optically
selected samples \citep{pado03},
though further confirmation using complete samples is needed.
Secondly, as commonly suggested,
NLS1 AGN are accreting at close to, or larger than
the Eddington limit \citep[e.g.][]{col04}.
At such high rates the accretion flow proceeds via a slim disk
where energy is partially advected into the black hole before
being radiated due to relatively long cooling time
\citep[e.g.][]{abr88,min00,wjm03}.
Thus the accretion processes in radio-loud NLS1 AGN may differ drastically
from those in normal RLQ/BLRGs,
where a standard thin disk is believed to be operating,
though they do overlap at  \redd$\simeq1$
 to some extent \citep[see e.g.][]{sik07}.
The final point---on the host galaxies---is only suggestive
because of a lack of sufficient evidence.
RL NLS1 nuclei, at least some, seem to be hosted
in disk or interacting galaxies,
and/or galaxies with relatively young stellar populations.
In contrast, classical RLQ/BLRGs \footnote{
\citet{ho01} showed that some Seyfert galaxies hosted in disk-dominated
spirals are in fact RL when the optical nuclear luminosities
are considered properly.
However, the vast majority of their sample objects are of substantially
low power and  sub-Eddington \citep{ho02}.
}
 are preferentially hosted
in giant elliptical/massive bulge dominated galaxies
or at least strongly passively evolving galaxies
\citep[e.g.][]{bah97,dun03,flo04,odo05},
although there was also evidence reported for star-formation
in some hosts beyond $z \ga 0.4$ \citep[e.g.][]{orn05,odo05}.

In consideration of their unusual properties,
we are left with two options regarding the nature of RL NLS1 galaxies:
either they are a downsized extension of normal  RL AGN to the lower-\mbh
(and perhaps higher-\redde) range,
or they form a previously unknown population
distinct from normal RL AGN.
The former would invoke some unknown processes that
can explain the
observed differences between RL NLS1 and normal RL AGN with
the smooth change of \mbhe.
For the latter, the operating physical processes
in likely black hole accretion and jet formation
may be radically different for the two classes.
A related, interesting question is whether the differences in the
observed properties between RL NLS1 and normal RL AGN can be naturally
explained by the same underlying physical processes
that make NLS1 different from normal AGN,
or whether some additional mechanisms are needed.
To further understand these questions,
a large sample of RL NLS1 galaxies with the full range of radio-loudness is needed,
as well as extensive observations, especially in the radio band.
As complementary discussion,
 we also  compare them with two special types of RL AGN below, namely,
BL Lac objects and compact steep-spectrum sources.
As will be seen,  RL NLS1 galaxies may be a heterogeneous population themselves.

\subsubsection{Relationship with BL Lac objects}

One object in our sample, J0849+5108 (0846+51W1),
was formerly classified as a BL Lac object
at its high light state based on previous observations
(see Appendix\,\ref{sect:indv} for details).
As such,  it presents an interesting case
linking possibly RL NLS1 galaxies with BL Lac objects,
as pointed out by \citet{zhou05},
though its black hole mass ($2.6\times 10^7$\,\msune) is lower compared to
those typical of BL Lac objects
\citep[\mbh$\sim 10^{8-9}$\,\msune, e.g.][]{cao03}.
In fact, it has been speculated in the past decade that
concealed Seyfert nuclei exist in a few BL Lac objects, including
BL Lac itself \citep{ver95,cor00}.
It is therefore tempting to postulate that some RL NLS1 galaxies may actually be
BL Lac objects (with low \mbhe) at their low light states.

The bulk of our RL NLS1 galaxies have a BLR luminosity\footnote{
The luminosity of the broad line region is estimated from the
\hb luminosity using the method proposed by \citet{cel97}
and the quasar emission line ratios given by \citet{fra91},
yielding  $L_{\rm BLR}=25.26$\lhb \citep[see also ][]{liu06}.
}
distribution lying within $\log\,(L_{\rm BLR}$/\ulum)=43.5--44.5,
the low end of which overlaps partially with those of  BL Lac  objects
\citep[a median of $10^{43.7}$\,\ulum was given for the upper limits of
$L_{\rm BLR}$ by][]{cel97}.
Specifically,  J0849+5108 has $L_{\rm BLR}\sim 10^{43.7}$\,\ulume.
However, this is likely not the case for most of the RL NLS1 galaxies given
their relatively higher $L_{\rm BLR}$ compared to the upper limits
for BL Lac objects.
On the other hand, this possibility means that BL Lac objects
might not be a homogeneous population,
and some may just be Seyfert nuclei/NLS1 at a high state of the
non-thermal radiation from the jet.
Similarly, this concerns only BL Lac objects at the low-\mbh end,
as is the case of J0849+5108,
rather than the bulk of classical
BL Lac objects that habour massive black holes and accrete
likely via a low efficiency accretion flow \citep[e.g.][]{cao03}.

\subsubsection{Relationship with CSS sources}
Compact steep-spectrum sources are compact,
powerful radio sources with spectral peaks near 100\,MHz.
They are inherently compact with linear sizes $\leq 20$\,kpc,
having (often symmetric) double/triple
morphology with complete jets and lobes on small scales.
The most popular scenarios to explain their nature advocate that
they are either young and evolving radio sources or confined by
dense gas of their environment \citep[see][for a review]{odea98}.
Very recently, a few lines of observational evidence
point to an interesting link between RL NLS1 galaxies and CSS, which
share some degree of similarity as the following.
(1) As mentioned in \S\,1,
the radio sources of all RL NLS1 galaxies are compact,
including steep-spectrum sources.
In particular,
a few NLS1 AGN having higher resolution observations show small-size
structure: e.g.\
most of the radio emission of 2MASX\,J0324+3410 is from a region
smaller than 6.8\,mas (8.2\,pc projected size),
as seen from VLBA images at 8.4\,GHz (in preparation).
(2) The low frequency spectrum of J1047+4725,
one of the steep-spectrum objects in our sample,
appears to flatten at around a few hundred MHz as indicated by its
flat slope of $-0.28$ between 327 and 151\,MHz;
such a spectral break is characteristic of CSS.
In fact, the RL NLS1--CSS connection was advocated by
\citet{osh01} and \citet{gal06} for the case of PKS\,2004-447,
although it is unclear as to whether  PKS\,2004-447
is a genuine NSL1 galaxy \citep[see][for a short comment]{zhou03}.
(3) There are observational indications that the host galaxies
of a large fraction of CSS are interacting/merging
and may contain significant amount of dense gas
\citep{odea98}, similar to what was found in some RL NLS1 galaxies.

The radio power distribution of our RL NLS1 galaxies overlaps with
 the lower part of that of CSS \citep{odea98}.
As such, if there is an overlap between CSS and RL NLS1 galaxies,
they should be mostly the CSS having low radio power, and,
relatively small \mbhe.
Since CSS sources also possess (small scale) jets,
in cases where the jets are directed close to our line-of-sight,
we expect to see blazar-like objects.
Some of those could be the flat-spectrum RL NLS1 galaxies studied here,
in which relativistic jets are inferred to be present.
Conclusive remarks have to await future
proper and detailed comparisons of the two classes,
as well as more radio observations with higher resolution, however.

\subsubsection{Implication for jet formation in AGN}

Given the high Eddington ratios of NLS1 galaxies in general, as
commonly suggested and also found here,
our RL NLS1 AGN present a strong case where jets,
possibly at least mildly relativistic,
can be produced in high \redd systems.
This strengthens the similar situation in powerful RLQs
with much larger   \mbhe.
This contrasts clearly with black hole X-ray binaries where
jets are found at the low/hard state, despite the suggestion that
transient jets  (or mass ejection) may occur
at the so called very-high state where \redd is higher \citep{mac03,fen04}.
The formation of jets is thought to be related to the
black hole and accretion process \citep[e.g.][]{bla00,cel01,mei03,ball07}.
The finding of relativistic jets in
black hole accretion systems with high  \redd
(where the accretion proceeds via probably
a new form other than the standard disk)
and, moreover, with relatively low \mbh is important for
addressing basic questions as to how jets are formed.
Jet formation in high  \redd systems (slim disks) is far from clear,
though there are some suggestions that outflows as well as
mildly beamed and collimated emission are likely present in such systems
\citep{ohs05}.

As has been shown above \citep[see also ][]{kom06b,wha06},
apart from the continuum radiation,
the RL NLS1 galaxies resemble closely their RQ counterparts
in many ways,
e.g.\ optical emission properties (except for stronger \feii emission),
the distributions of black hole masses and  the Eddington ratios.
These facts suggest that RL NLS1 galaxies are simply
otherwise normal NLS1 galaxies but having relativistic jets,
reminiscent of the RL/RQ dichotomy of more powerful, classical AGN.
It would be interesting to find out whether there is a bimodal
distribution in the radio-loudness, as reported for the broad-line
quasar population (though this is still a matter of debate),
in future work considering objects with the whole range of radio-loudness.
The nearly indistinguishable properties other than the presence of
radio jets  possessed by the two
subsets of NLS1 galaxies, RL and RQ,
seemingly hint at a possibly important role of black hole spin
in the making of relativistic jets in RL NLS1 galaxies.

\section{Summary of conclusions and implications}

A homogeneous sample of genuine RL NLS1 galaxies
(radio-loudness \rle$\ga 100$) is presented for the first time,
comprising 23 objects drawn from SDSS spectroscopic
objects with FIRST detections.
Their NLS1 nature is discovered by analysing
systematically the spectroscopic data from the QSO
and galaxy database in the SDSS DR5.
The sample can be treated as an optically--radio selected one.
We carry out  systematic investigations into their multiwaveband
properties from radio to X-rays using data available from the archives.
The main results are the following.

All radio sources of the sample are compact,
less than a few tens of kilo-parsec at least.
Several with VLBI observations are
unresolved or marginally resolved at scales of about tens of parsec.
Among those with available radio spectral indices,
the majority are flat-spectrum sources
($\alpha>-0.5$ around 1.4--5\,GHz),
and a few even have inverted spectra ($\alpha>0)$.
Radio variability is found to be common in most of
the flat-spectrum sources.
We were able to estimate the brightness temperatures
of the radio sources, that are
as high as $T_{\rm B}>10^{11}$\,K
and even exceeding the inverse Compton limit ($10^{12}$\,K)
in four objects.
The high brightness temperatures indicate the presence of
relativistic beaming.
The majority of RL NLS1 galaxies has a low-frequency (isotropic) radio power
falling in the range of FR\,I type radio galaxies
(\llf$\la 10^{25}$\,\whzsre),
and the remaining minority in the range of the FR\,II type.
When isotropic, rather than Doppler boosted, emission is considered,
a large fraction of the objects can be considered as
intrinsically radio-intermediate.

The observed optical luminosities are systematically in excess of
the  thermal continua predicted from the \hb line luminosities by
a factor of up to several.
The optical slopes of our RL objects,
while spanning a wide range from moderately red to blue,
are  systematically bluer than that of radio-quiet NLS1 AGN.
We interpret this effect as the contribution from a (non-thermal)
component, presumably from relativistic jets, to the optical band
superimposing the central ionising continuum seen in radio-quiet NLS1 AGN.
The optical \feii emission in the
radio-loud NLS1 in our sample is on average
stronger than `normal' NLS1 galaxies.
Except this, the two subsets of NLS1 galaxies seem to be indistinguishable
in other emission line properties.

On the  radio--optical--X-ray effective spectral indices plane
(\alpox vs.\ \alproe),
our objects have broad-band SEDs well consistent with
those of blazars, i.e.\ both the HFSRQ and LFSRQ types
as well as SSRQs.
Of particular interest, the fraction of the candidate HFSRQ-type objects
seems to be high,
though their synchrotron component peaks are only moderately high
(at around the UV).
This finding extends the recently discovered strongly-lined (FSRQ)
HBL-type blazars \citep{pado03} to NLS1 galaxies.

The continuum shape in the soft X-ray band
exhibits a large variety among our RL NLS1 galaxies.
There is evidence in one, or possibly two objects for a flat
X-ray spectral component ($\Gamma\la 1.8$), which we interpret as
an inverse-Compton origin from jets as in FSRQs.
Most of the objects with estimation of the X-ray spectral slopes
show steep spectra.
The steep X-ray spectrum is also consistent with what is expected
for HFSRQs, that originates from
the high energy end of the synchrotron component.
Based on this circumstantial evidence,
we suggest that such a component may be a significant contributor
to the observed X-ray emission,
though the X-ray component as seen in radio-quiet NLS1 galaxies
may also play a role.

In view of the independent lines of evidence presented here,
as well as those found in previous studies for several individuals
(either inclusive or exclusive of our sample),
we  suggest that there likely exists
a population of RL NLS1 galaxies showing
the broad-band continuum emission characteristic of blazars,
though alternative explanations cannot be ruled out completely.
In this scenario,  beamed non-thermal emission from jets,
that are at least mildly relativistic,
dominates the high-frequency radio band,
and contributes significantly to the optical and possibly the X-ray bands.
Our finding extends the radio-loud quasar population
into a narrow line-width regime of the broad emission lines.
We postulate that all genuine radio-loud NLS1 galaxies possess small scale,
at least mildly relativistic jets,
reminiscent of classical radio-loud AGN.
In this scenario, steep-spectrum NLS1 sources
(perhaps including some of those with smaller radio-loudness
within $R=$10--100) are simply objects of the same population
whose radio jets are aligned
at a relatively large angle to the line-of-sight.

The host galaxies of two nearby RL NLS1 nuclei, whose images are resolved,
have disk morphology, one even in a merger, in line with the
finding for the previously published blazar-NLS1 hybrid object
2MASX\,J0324+3410 \citep{zhou07}.
In several objects imprints of young stellar populations
are likely to be present in form of
high-order Balmer absorption lines in the SDSS spectra.

Based on the commonly used relation of black hole mass
with broad line width and luminosity,
our RL NLS1 galaxies have estimated \mbh in the range of $10^6-10^8$\,\msune,
provided that this \mbh scaling relation is applicable to NLS1 and
the BLR is not largely flattened.
The thus estimated Eddington ratios
range from \redd=0.5 to 3, and peak at \redde$\approx 1$
(assuming $L_{\rm bol}=9 \lambda L_{\lambda5100}$).
RL NLS1 galaxies deviate from the known trends of
radio-loudness versus the Eddington ratio and black hole mass,
and are populated in a region in the radio-loudness--\mbh diagram
where normal AGN are seldom found, i.e.\ moderate \mbh and large $R$
(\mbh$\simeq 10^{6-7.5}$\,\msune,  $R\simeq 10^{1-3}$),
confirming and improving upon the previous results suggested by \citet{kom06b}.
As such, RL NLS1 show strong excess radio emission with respect to their \mbhe,
when compared to the previously suggested radio power--\mbh relation  for normal AGN
\citep[e.g.][]{lacy01}.
Our results imply that large radio-loudness ($R=10^{1-3}$)
can be produced in AGN with  moderate \mbh down to $\sim10^6-10^7$\,\msun
and accreting at/close to the Eddington rate;
and some of these AGN are hosted in late-type galaxies,
some even in interacting systems.
The presence of relativistic jets in such systems may have interesting
implications for understanding  jet formation in AGN in general.

In spite of a likeness  to normal RL AGN in many aspects,
RL NLS1 galaxies show some extreme or unusual   properties,
though some of the evidence are only circumstantial.
These include compact radio morphology, relatively low radio power,
a high incidence of HBL/HFSRQ-type SEDs, host galaxies of probably late-type
 and/or with young stellar populations,
and the common properties characteristic of NLS1 galaxies,
e.g.\  smaller black hole masses, higher Eddington ratios,
and stronger \feii emission.
All of these lead us to postulate that either
they are an extension of normal RL AGN to the lower-\mbh range,
and all
those differences can be explained as a consequence of decreasing \mbhe;
or they form a previously unknown population
distinct from normal radio-loud AGN.
As one interesting possibility,
we speculate that low-power CSS  might be the parent population
of the flat-spectrum RL NLS1 galaxies with beamed jet emission.
For the future, a much larger sample and extensive multi-waveband observations
are essential for understanding the nature of radio-loud NLS1 galaxies.

Finally, if the broad-band radiation mechanisms
in flat-spectrum radio-loud NLS1 AGN are indeed the same as in blazars,
we expect some of these objects to be $\gamma$-ray emitters
in the GeV band.
This can be tested by observations of GLAST
in the near future.

\acknowledgments
We thank the anonymous referee for helpful comments and suggestions.
This work is supported by the Chinese Natural Science Foundation
through projects No.\ NSF10533050 and NSF10473013,
and the BaiRen Project of the Chinese Academy of Sciences.
Funding for the creation and the distribution of the
SDSS Archive has been provided
by the Alfred P. Sloan Foundation, the Participating Institutions,
the National Aeronautics and Space Administration, the National
Science Foundation, the U.S. Department of Energy, the Japanese
Monbukagakusho, and the Max Planck Society. The SDSS is managed by
the Astrophysical Research Consortium (ARC) for the Participating
Institutions. The Participating Institutions are The University of
Chicago, Fermilab, the Institute for Advanced Study, the Japan
Participation Group, The Johns Hopkins University, Los Alamos
National Laboratory, the Max-Planck-Institute for Astronomy (MPIA),
the Max-Planck-Institute for Astrophysics (MPA), New Mexico State
University, Princeton University, the United States Naval
Observatory, and the University of Washington.
This research has made use of the NASA/IPAC Extragalactic Database (NED) which is operated by the Jet Propulsion Laboratory, California Institute of Technology, under contract with the National Aeronautics and Space Administration.

\begin{appendix}
\section{Optical spectral modeling}
\label{sect:sdssspec}

The starlight and nuclear continua are modeled as
\begin{equation}\label{eq1}
S(\lambda)=A_{host}(E_{B-V}^{host},\lambda)~A(\lambda)+A_{nucleus}(E_{B-V}^{nucleus},\lambda)~[bB(\lambda)+ c_{\rm b} C_{\rm b}(\lambda)  +
c_{\rm n} C_{\rm n}(\lambda) + dD(\lambda)]
\end{equation}
where $S(\lambda)$ is the observed spectrum.
$A(\lambda)=\sum_{i=1}^6 a_{i}~IC_{i}(\lambda,\sigma_{*})$
represents the starlight component modeled by our 6 synthesized
galaxy templates, which had been built up from the spectral template
library of Simple Stellar Populations (SSPs) of
\citet{bc03} using our new method based on the Ensemble
Learning Independent Component Analysis (EL-ICA) algorithm.
The details of the galaxy
templates and their applications are presented in \citep{lu06}.
$A(\lambda)$ was broadened by convolving with a Gaussian of width
$\sigma_{*}$  to match the stellar velocity dispersion of the host galaxy;
in this process the velocity difference between emission and  absorption
lines is taken into account.
The un-reddened nuclear continuum is assumed to be
$B(\lambda)=\lambda^{-1.7}$ as given in \citet{fra96}.
We modeled the optical \feii emission, both broad and narrow,
using the  spectral data of the \feii
multiplets for I\,Zw\,I in the $\lambda\lambda$\,3535--7530\AA\
range provided by \citet[][Table\,A1,A2]{veron04}.
We assume that the  broad \feii lines ($C_{\rm b}$ in Eq.1)
have the same profile as the
broad \hb line, and the narrow \feii lines ($C_{\rm n}$),
 both permitted and forbidden, have the same profile as the that of the
narrow \hb component, or of \oiiie$\lambda5007$ if \hb is weak.
$D(\lambda)$ represents the templates of higher-order Balmer
emission lines ($10\leq n\leq 50$) and Balmer continuum generated in
the same way as in \citet{diet03}.
$A_{host}(E_{B-V}^{host},\lambda)$ and
$A_{nucleus}(E_{B-V}^{nucleus},\lambda)$ are the color excesses due
to possible extinction of the host galaxy and the nuclear region,
respectively, assuming the extinction curve for the Small Magellanic
Cloud of \citet{pei92}. The fitting is performed by minimizing
the $\chi^{2}$ with $E_{B-V}^{host}$, $E_{B-V}^{nucleus}$, $a_{i}$,
$\sigma_{*}$, $b$, $c_{\rm b}$, $c_{\rm n}$,
and $d$ being free parameters.

\section{Notes on individual objects}
\label{sect:indv}

{\bf J0849+5108} (0846+51W1): This object has drawn substantial attention
ever since its first identification with a BL Lac object
(at a bright state) by \citet{arp79},
at a claimed high redshift of $z=1.86$\footnote{
This redshift value, given by the authors
based on the spectra with low resolution and
signal-to-noise ratios, turned out to be wrong.}.
Its radio flux also showed large amplitude variations up to 50\%
at 1.4\,GHz.
This object exhibited
a remarkable optical brightening with an amplitude of $\Delta\,V\sim5$
within less than a month
\citep{arp79} and high and variable optical polarization
\citep{moore81,sitko84}, making it one of the
optically violent variables (OVV) and highly polarized QSOs (HPQ).
Night-to-night and even intra-night optical variability was also found
\citep{sitko84,stalin05}.
At maximum light outburst, the continuum is featureless,
whereas its spectra taken at normal/minimum light states
show strong emission lines \citep{arp79,stickel89}.
Thus it is a transition between a quasar and a  BL Lac object,
i.e.\ appeared to be a BL Lac object only at its flaring state.
The large and variable polarization in the optical band
suggests strongly that this emission is due to synchrotron radiation
\citep{sitko84}.
The conspicuous characteristic emission lines identified in the
SDSS spectrum of J0849+5108 give a redshift of $z=0.583$.
A detailed study of its SDSS spectrum and other properties
was presented by \citet{zhou05}.
The optical spectral slope obtained from fitting the SDSS spectrum
is red, with \alpwvle$\simeq -0.2$.

{\bf J0948+0022}: The object was first discovered for its radio
emission  in various radio surveys ($S_{\rm 5GHz}\sim$0.2--0.3\,Jy).
It was identified with a NLS1 galaxy, based on its SDSS spectrum,
by \citet{wilm02} and \citet{zhou03},
and was studied in detail as one of the few previously known
very RL NLS1 galaxies.
The radio source has an inverted radio spectrum and
was unresolved with VLBI at milli-arcsec resolution.
These observations were explained in terms of
relativistic beaming \citep{doi06}.
A very high brightness temperature ($\sim 10^{13}$\,K)
inferred from radio flux variations suggest the presence
of a jet with relativistic beaming \citep{zhou03}.
Long-term optical variability with an amplitude of about 1\,mag
in the $B$-band was also noted in  \citet{zhou03}.

{\bf J1505+0326}: It has radio fluxes peaked at $\sim 7$\,GHz \citep{tin05}.
Its VLBI images are dominated by a very compact component,
marginally resolved at 2.3\,GHz band \citep{dal98}.
The radio emission at centimeter wavelengths shows variability.

{\bf J1633+4718} (RXJ\,16333+4719):
The object was identified with a NLS1 galaxy in optical identification of
ROSAT All-Sky Survey (RASS) sources \citep{moran96,wis97}.
The AGN (see Figure\,\ref{fig:sdssspec} for its SDSS spectrum)
is actually one of the two nuclei, separated by 4\arcsec,
residing in an interacting/merger system.
The other nucleus  has a starburst spectrum \citep{bade95}.
The radio source, which is associated with the AGN,
is unresolved with VLBI at milli-arcsec resolution \citep{doi07}.
Its spectral shape above 5\,GHz is inverted, as measured by
simultaneous multiwavelength observations by \citet{neum94}.
The system was noticed to be associated
with the IRAS source IRAS\,16319+4725 (RA=16h33m23.03s, Dec=+47d19m03.5s),
whose positional uncertainty---much larger than the separation
of the two optical nuclei---encompasses both of the nuclei
\citep{bol92,moran96,cond98b}.
The \oiii doublet shows a noticeable blue wing in its SDSS spectrum.

{\bf J1644+2619}:
Its radio source has an inverted spectrum and
was unresolved with VLBI at milli-arcsec resolution  \citep{doi07}.

\section{Radio emission at low frequencies}
\label{sect:lowfemi}

In the above sections, evidence has been presented for the
blazar-like properties of the (flat-spectrum) RL NLS1 galaxies,
arguing for the presence of relativistic jets in them.
As is known in normal radio quasars/galaxies, radio jets are associated
with large scale radio lobes.
These show diffuse radio emission with steep spectrum
 \citep[\alpr$\simeq -0.77$; e.g.][]{kel66}
which dominates the radio fluxes at low frequencies
(several hundred MHz),
via synchrotron radiation from optically thin plasma.
Here we examine the radio properties  of our sample at low frequencies,
though the data are sparse in general, especially at around 100\,MHz.
The radio emission is searched for using the
WENSS\footnote{The Westerbork Northern Sky Survey}
survey \citep{ren97} at 327\,MHz and using NED for surveys at
lower frequencies.
We find 15 objects with detections at 327\,MHz,
for which the 1.4\,GHz to 327\,MHz spectral indices
\agm can be computed.
We found a median  \agm of $-0.47$, and $\sim 47\%$
being steeper than $-0.5$ with the steepest as $-0.83$,
systematically steeper than the 1.4--5\,GHz indices
(3/11 are steep-spectrum).
This means that
the spectral indices steepen progressively towards low frequencies.

This behavior can well be explained by onset of a
steep-spectrum component dominating the low frequencies,
as commonly seen in normal RL AGN.
Therefore, we postulate that in these objects there exists
diffuse, optically thin emission,
presumably from putative radio lobes that are
powered by (relativistic) jets.
If this is the case, those blazar-like (flat-spectrum) NLS1 AGN
must possess  the general core--halo morphology as seen in some
blazars \citep[e.g.][]{sta82,bro86,mur93}.
Given the compact morphology
unresolved at the FIRST resolution (smaller than several tens kpc),
there are two possibilities:
either they have
large scale extended emission---similar to the classical radio sources---but
are viewed  close to the radio axis,
or are intrinsically compact,
similar to  CSS (compact steep-spectrum sources).

At the even lower frequency of 151\,MHz,
4 objects have detected radio emission, all
from the 6C/7C survey \citep[e.g.][]{bal85,hal88}, namely
J1047+4725 (1.81\,Jy), J1305+5116 (0.32\,Jy),
J1443+4725 (0.5\,Jy), and J1548+3511 (0.26\,Jy).
The 151--327\,MHz indices are \amme=$-0.28$, $-0.54$,  $-0.62$, $+0.09$,
respectively, and the average is $-0.34$.
These values are much flatter than the `canonical' slope
for diffuse optically thin emission
from powerful sources \citep[$<-0.7$, e.g.][]{kel66},
indicating that the spectra have already turned over,
or started to turn over, in this frequency range
due possibly to synchrotron self-absorption.
In fact, the radio spectra of J1047+4725 and  J1548+3511
show a clear trend of turnover at low frequencies.
Considering the relatively large redshifts of the sample,
the break frequencies in some sources could be as high as
several hundred MHz in the source rest frame.
Spectral turnovers at such frequencies are often seen in radio sources with
small scale structure, such as in CSS \citep[e.g.][]{odea97}.
The previously reported extreme NLS1/blazar hybrid object 2MASX\,J0324+3410
\citep{zhou07} was also detected at 151\,MHz (1.02\,Jy);
its radio slope also steepens significantly towards low frequencies,
\amme=$-0.44$ \citep[compared to \alpre$\simeq 0.1$
at high frequencies, ][]{neum94},
suggesting the dominance of a steep-spectrum component at low frequencies.

\section{Effects of refractive interstellar scintillation
on radio flux variations}
\label{sect:riss}

For compact  radio sources (milli-arcsec scale),
it is known that refractive interstellar scintillation (RISS),
caused by inhomogeneities in the interstellar medium,
may give rise to variations of radio flux densities
\citep[e.g.][]{rick86,rick90, bla86}.
RISS is wavelength-dependent and concerns
predominantly the low frequency  domain (below 1\,GHz).
At centimeter wavelengths, RISS mostly causes ``flickering'' at a lever of
a few percent, as was found in observations \citep{hee84}
and is also consistent with the theoretical models \citep{rick86,bla86}.
Although variations at a larger amplitude were sometimes
ascribed to RISS, they were limited to a small number of
individual objects which ubiquitously
are at low Galactic latitude ($<10^{\circ}$)
or even have a line-of-sight passing through the Galactic plane
\citep{hje86,den87}.
Monitoring observations of radio sources at frequencies above 1\,GHz
showed that both intrinsic and extrinsic (RISS)
variability contribute to observed radio flux variations,
with the former having timescales from months to years
and the latter being faster and having rms amplitude typically $<10\%$
\citep[e.g.][]{laz01,mitc94,rick06}.

In consideration of the large variation amplitudes
in some of the objects of our sample and their high Galactic latitudes
(b=$31^{\circ}-51^{\circ}$, Table\,\ref{tbl:radvar}),
we argue that the variations are
predominantly intrinsic for the majority of the variables reported here,
instead of arising from RISS.
This is especially true for variations at 5\,GHz or even higher frequencies.
Furthermore,  the brightness temperatures deduced from variability at 1.4\,GHz
are in the same range as those from 5\,GHz (Table\,\ref{tbl:radvar});
this favours the intrinsic variability explanation,
since RISS at  1.4\,GHz  is expected to be   more pronounced
than at 5\,GHz due to its strong wavelength dependence.
We therefore suggest that,
although a non-negligible role played by RISS may not be ruled out
in individual objects, especially those with relatively low amplitudes,
our results for the sample as a whole should not be affected significantly.

\section{On the X-ray radiation mechanisms of radio-loud NLS1 AGN}
\label{app:xray}

As one of the defining characteristics,
the X-ray spectra of FSRQs are flat
and extend from the soft to the hard X-ray band
\citep[$\Gamma\sim1.6$ with a dispersion $\sigma=0.1-0.2$; see e.g.][]{ree00},
which is interpreted as
inverse Compton radiation from the relativistic jets and beamed.
In most RLQs the contribution from a hot disk corona
to the observed X-rays is negligible (in the hard X-ray band $>2$\,keV),
except the steep-spectrum soft X-ray excess below 1\,keV
\citep[e.g.][]{bri97,yuan00}.
From the  data analysis above,
we find a flat X-ray spectrum in J1633+4718 with $\Gamma=1.37\pm0.49$,
and possibly in  J0849+5108 with
$\Gamma=1.77^{+0.44}_{-0.60}$ (90\% errors).
Their concave optical to X-ray spectral shapes
(see Figure\,\ref{fig:aox-ax}) also agree with
the inverse Compton scenario.
For J0849+5108, the LBL-like SED and flat X-ray spectrum are well expected,
based on its OVV/HPQ properties (see Appendix\,\ref{sect:indv}).
The prominent soft X-ray excess in J1633+4718 is also typical of
normal NLS1 galaxies.
It is puzzling, however, that  J1633+4718 shows a broad band SED
similar to that of HFSRQs, for which a
steep X-ray spectrum and a convex \alpoxe--X-ray shape are
generally expected.
Future multi-waveband observations, especially in the
X-ray band, are needed to confirm this peculiarity.

As discussed above, in the presence of an excess soft X-ray component,
the effective spectral slopes derived from the hardness ratios
are generally steep.
The same is expected to be true for radio-loud NLS1 galaxies,
as found in J1633+4718 above.
For reference, FSRQs at redshifts ($z<1$) are found to have
 $\Gamma\sim2.1$ with a dispersion $\sigma\simeq0.2$
in the ROSAT PSPC band \citep{bri97,yuanthe}.
Therefore, we expect that a few other objects, e.g.
those having FSRQ/LBL-type SEDs yet steep $\Gamma$ values in the PSPC bandpass
(Figure\,\ref{fig:alprox})
would similarly have a flat underlying X-ray continuum.

HBLs are commonly  observed to show steep X-ray spectra
\citep[e.g.][]{wor90,bri96,pado96,wol98}, which are
intrinsically curved,
progressively steepening with increasing photon energy
\citep[e.g.][]{per05}.
Their X-ray emission is thought to be attributed to
the high-energy tail of the synchrotron radiation,
i.e. the synchrotron bump reaches the soft X-ray band \citep{ghi89}.
There are only a few flat-spectrum radio-loud NLS1 AGN
studied previously with multi-waveband data
and, interestingly, most of these show HBL/HFSRQ-like SEDs.
For example,
RXJ16290+4007, formerly classified as a FSRQ,
  was suggested by \citet{pado02} to be
the first confirmed HFSRQ for its modeled synchrotron peak at
$2\times10^{16}$\,Hz and a steep soft X-ray spectrum $\Gamma\simeq3$.
2MASX\,J0324+3410  also shows a continuum SED
\citep[square in Figure\,\ref{fig:alprox}; ][]{zhou07}
and a luminosity closely resembling that of Mkn\,421,
a well-known HBL;
the object was claimed to be even marginally detected
in TeV $\gamma$-rays \citep{fal04},
as seen in some nearby bright HBL.
In the X-ray band, 2MASX\,J0324+3410 also shows a steep spectrum
\citep[$\Gamma\simeq2.2$, ][]{zhou07}.

Most of the objects falling into the  HBL locus in Figure\,\ref{fig:alprox}
show steep X-ray spectra wherever available ($\Gamma=2.0-3.3$).
Meanwhile, their \alpoxe--\alpx SED shapes are either convex or
close to a straight line (i.e.\ power-law; see Figure\,\ref{fig:aox-ax}).
Thus, based on the similarity with HBL in both the SED and X-ray
spectral shape,  we  suggest their X-ray emission to be
the high-energy tail of the synchrotron component, as in HBL.
There are a few objects lying close to the border line,
known as intermediate BL Lac objects (IBL)
where the synchrotron peaks are around $10^{14}-10^{15}$\,Hz.
In them, the high-energy synchrotron tail
may still reach the soft X-rays, as found in a few IBL objects
\citep[e.g.][]{tag00,fer06}.
\end{appendix}

\clearpage
\begin{landscape}
\input{tab1}
\clearpage
\end{landscape}
\clearpage
\input{tab2}
\clearpage
\input{tab3}
\input{tab4}

\end{document}

%% file: tab1.tex
\begin{deluxetable}{lcccccrcrcccccccrrrr}
\tablecaption{Sample of very radio-loud NLS1 galaxies \label{tbl:sample}}
\tablewidth{0pt} 
\tablehead{ 
\colhead{No.} &
\colhead{name} &
\colhead{SDSS name} & 
\colhead{$Z$} & 
\colhead{$u$} &
\colhead{$g$} &
\colhead{$M_{\rm B}$} &
\colhead{$f_{\rm 20cm}$ } &
\colhead{$f_{\rm 6cm}$  } &
\colhead{$f_{\rm 92cm}$ } &
\colhead{$\alpha_{\rm r}$} &
\colhead{$\log R$} &
\colhead{$\lambda f_{5100\AA}$ \tablenotemark{a}} &
\colhead{\alpwvle} &
\colhead{fwhm(\hbe) } &
\colhead{$f$(\hb) \tablenotemark{b}} &
\colhead{$f$(O{III}) \tablenotemark{b}} &
\colhead{\rfe} &
\colhead{$m_{NUV}$} &
\colhead{$m_{FUV}$} \\
\colhead{} & \colhead{}  & \colhead{} & \colhead{} &
\colhead{mag} & \colhead{mag}  & \colhead{mag} & \colhead{mJy} & 
\colhead{mJy} & \colhead{mJy}  & \colhead{}    & \colhead{}    &
\colhead{}    & \colhead{}     & \colhead{\kmps}& \colhead{}&
\colhead{}& \colhead{} & \colhead{mag}& \colhead{mag} \\
\colhead{(1)} & \colhead{(2)}  & \colhead{(3)} & \colhead{(4)} &
\colhead{(5)} & \colhead{(6)}  & \colhead{(7)} & \colhead{(8)} & 
\colhead{(9)} & \colhead{(10)} & \colhead{(11)}& \colhead{(12)}&
\colhead{(13)}& \colhead{(14)} & \colhead{(15)}& \colhead{(16)}&
\colhead{(17)}& \colhead{(18)} & \colhead{(19)}& \colhead{(20)}
}
\startdata
 1&J0814+5609 & J081432.11+560956.6 & 0.509 & 18.05 & 17.89 &-24.1 &  69/60  & 43 & 72  & -0.33 & 2.53&  458 & -2.43& 2164$\pm$ 59 &  870$\pm$16 &  54$\pm$ 7 & 1.04 &      &        \\
 2&J0849+5108 & J084957.98+510829.0 & 0.583 & 19.80 & 18.85 &-23.1 & 344/266 & 161& 202 &$\sim0$\tablenotemark{c}& 3.16&  187 & -0.20& 1811$\pm$191 &  135$\pm$12 &  42$\pm$ 3 & 0.23 &      &        \\
 3&J0850+4626 & J085001.17+462600.5 & 0.523 & 19.67 & 19.02 &-22.8 &  21/16  & ...& 51  &  ...  & 2.23&  409 & -0.86& 1251$\pm$ 61 &  255$\pm$10 & 133$\pm$ 5 & 1.48 & 20.80& 21.29  \\
 4&J0902+0443 & J090227.16+044309.6 & 0.532 & 19.38 & 18.92 &-23.0 & 153/157 & 106& ... & -0.30 & 3.02&  520 & -0.53& 2089$\pm$205 &  341$\pm$20 & 192$\pm$ 6 & 1.10 & 20.28&        \\
 5&J0948+0022 & J094857.32+002225.5 & 0.584 & 18.59 & 18.36 &-23.9 & 108/70  & 295& ... &  0.82 & 2.55&  510 & -1.87& 1432$\pm$ 87 &  341$\pm$12 &  34$\pm$ 4 & 1.22 & 18.61& 19.04  \\
 6&J0953+2836 & J095317.09+283601.5 & 0.657 & 19.19 & 18.94 &-23.6 &  45/43  & ...& ... &  ...  & 2.71&  240 & -1.73& 2162$\pm$201 &  215$\pm$14 &  52$\pm$ 5 & 1.23 &      &        \\
 7&J1031+4234 & J103123.73+423439.3 & 0.376 & 19.64 & 19.36 &-21.6 &  17/19  & ...& 44  &  ...  & 2.34&  284 & -0.55& 1642$\pm$ 75 &  379$\pm$14 & 131$\pm$ 8 & 1.07 &      &        \\
 8&J1037+0036 & J103727.45+003635.6 & 0.595 & 20.04 & 19.32 &-22.8 &  27/28  & ...& ... &  ...  & 2.66&  225 & -1.71& 1357$\pm$107 &  173$\pm$ 9 &  25$\pm$ 4 & 1.58 & 21.25& 22.99  \\
 9&J1047+4725 & J104732.68+472532.1 & 0.798 & 19.16 & 18.95 &-24.0 & 734/789 & 404& 1463& -0.51 & 3.87&  293 & -1.30& 2153$\pm$267 &  421$\pm$32 & 325$\pm$ 8 & 0.66 &      &        \\
10&J1110+3653 & J111005.03+365336.3 & 0.630 & 21.13 & 20.59 &-21.8 &  19/23  & ...& 45  &  ...  & 2.97&   74 & -1.28& 1300$\pm$339 &   93$\pm$20 &  32$\pm$ 4 & 0.88 & 21.35& 21.85  \\
11&J1138+3653 & J113824.54+365327.1 & 0.356 & 20.49 & 19.52 &-21.2 &  13/12  & ...& 26  &  ...  & 2.34&  378 &  ... & 1364$\pm$ 99 &  393$\pm$20 &  72$\pm$10 & 0.69 &      &        \\
12&J1146+3236 & J114654.28+323652.3 & 0.465 & 19.08 & 18.62 &-23.0 &  15/17  & ...& 26  &  ...  & 2.11&  427 & -1.54& 2081$\pm$183 &  632$\pm$39 & 149$\pm$ 6 & 0.78 &      &        \\
13&J1238+3942 & J123852.12+394227.8 & 0.622 & 19.89 & 19.55 &-22.8 &  10/10  & ...& ... &  ...  & 2.23&  175 & -1.19&  910$\pm$200 &  103$\pm$18 &  68$\pm$ 4 & 1.05 &      &        \\
14&J1246+0238 & J124634.65+023809.0 & 0.362 & 18.52 & 18.31 &-22.7 &  37/36  & ...& ... &  ...  & 2.38&  601 & -1.89& 1425$\pm$ 68 &  527$\pm$12 & 115$\pm$ 5 & 0.84 & 19.01& 19.58  \\
15&J1305+5116 & J130522.75+511640.3 & 0.785 & 17.45 & 17.28 &-25.6 &  84/87  & 46 & 211 & -0.50 & 2.34& 1264 & -1.68& 1925$\pm$ 53 & 1982$\pm$43 & 487$\pm$13 & 0.65 & 17.45& 18.14  \\
16&J1435+3131 & J143509.49+313147.8 & 0.501 & 19.87 & 19.43 &-22.3 &  39/43  & ...& 137 &  ...  & 2.87&  306 & -1.70& 1719$\pm$253 &  346$\pm$36 & 137$\pm$ 6 & 1.32 &      &        \\
17&J1443+4725 & J144318.56+472556.7 & 0.703 & 18.20 & 18.11 &-24.6 & 165/166 & 72 & 309 & -0.67 & 3.07&  534 & -2.06& 1848$\pm$113 &  337$\pm$18 & 105$\pm$13 & 1.45 & 18.09& 18.57  \\
18&J1505+0326 & J150506.48+032630.8 & 0.408 & 18.78 & 18.52 &-22.7 & 365/395 & 859& ... &  0.66 & 3.19&  389 & -0.83& 1082$\pm$113 &   95$\pm$ 8 & 106$\pm$ 4 & 1.53 &      &        \\
19&J1548+3511 & J154817.92+351128.0 & 0.478 & 18.18 & 17.91 &-23.9 & 141/141 & 107& 279 & -0.22 & 2.84&  545 & -2.32& 2035$\pm$ 52 &  721$\pm$12 & 309$\pm$ 7 & 1.24 & 18.41& 18.69  \\
20&J1633+4718 & J163323.58+471859.0 & 0.116 & 17.94 & 17.38 &-20.9 &  63/69  & 35 & 107 & -0.30 & 2.22& 1868 & -1.70&  909$\pm$ 43 &  902$\pm$21 & 919$\pm$ 8 & 1.02 & 18.20& 19.48  \\
21&J1634+4809 & J163401.94+480940.2 & 0.494 & 20.18 & 19.53 &-22.2 &   8/14  & ...& 16  &  ...  & 2.31&  257 & -1.60& 1609$\pm$ 79 &  287$\pm$11 &  87$\pm$ 3 & 0.98 & 21.03& 22.30  \\
22&J1644+2619 & J164442.53+261913.2 & 0.144 & 18.03 & 18.02 &-20.8 &  88/128 & 99 & ... & -0.07 & 2.65& 1788 & -1.00& 1507$\pm$ 42 & 1108$\pm$22 & 120$\pm$ 4 & 0.75 & 18.50& 18.65  \\
23&J1722+5654 & J172206.03+565451.6 & 0.425 & 18.66 & 18.31 &-23.1 &  37/43  & ...& 119 &  ...  & 2.37&  544 & -1.41& 1385$\pm$ 36 &  542$\pm$11 & 323$\pm$ 4 & 0.58 &      &       
\enddata
\tablecomments{
Col.(2): abbreviated name of object used in this paper;
Col.(3): SDSS name;
Col.(4): redshift;
Col.(5): SDSS $u$ band PSF magnitude corrected for Galactic extinction;
Col.(6): as col.(5), but for the $g$ band;
Col.(7): derived $B$ band absolute magnitude;
Col.(8): FIRST and NVSS radio flux densities at 20\,cm (FIRST/NVSS);
Col.(9): flux density at 6\,cm;
Col.(10): flux density at 92\,cm given by WNESS;
Col.(11): radio spectral slope between 6\,cm and 20\,cm;
Col.(12): logrithm of radio-loudness defined as 
           $\equiv f_{\nu}({\rm 1.4\,GHz})/f_{\nu}(4400\AA)$;
Col.(13): flux $\lambda f$ at 5100\,\AA\ in the rest frame;
Col.(14): optical power-law slope $(S(\lambda)
          \propto \lambda^{\alpha_{\lambda}})$;
Col.(15): line width of \hb in FWHM;
Col.(16): line flux of the broad component of \hbe;
Col.(17): line flux of \oiiie;
Col.(18): FeII-to-\hb flux ratio (see text);
Col.(19): near-UV magnitude given by GALEX corrected for Galactic extinction;
Col.(20): the same as col.(19) but for the far-UV band.
}
\tablenotetext{a}{Rest frame flux $\lambda f$ in units of $10^{-15}$\,\ergs}
\tablenotetext{b}{Rest frame line flux in units of $10^{-17}$\,\ergs}
\tablenotetext{c}{Given the strong variation of the 1.4\,GHz flux, 
we adopt a representing radio slope \alpre$\sim 0$
as suggested by the radio data from \citet{arp79}.}
\end{deluxetable}

%% file: tab2.tex
\begin{deluxetable}{clrrrrrccc}
\tablecaption{X-ray data and analysis results \label{tbl:xray}}
\tablewidth{0pt} 
\tablehead{ 
\colhead{name} &
\colhead{X-ray source} & 
\colhead{obs.} &
\colhead{$T_{\rm expo}$(s)} &
\colhead{$N_{\rm H,Gal}$  } &
\colhead{C.Rate} &
\colhead{flux \tablenotemark{a}} &
\colhead{$\Gamma$} &
\colhead{$kT_{\rm bb}$} &
\colhead{remark}  \\
\colhead{} & \colhead{}  &  \colhead{} & \colhead{sec} & 
\colhead{$10^{20}$\,\unhe} & \colhead{ $10^{-2}$\,c/s}  & 
\colhead{} & \colhead{} & \colhead{eV} & \colhead{} \\
\colhead{(1)} & \colhead{(2)}  & \colhead{(3)} & \colhead{(4)} &
\colhead{(5)} & \colhead{(6)}  & \colhead{(7)} & \colhead{(8)} &
\colhead{(9)} & \colhead{(10)}  
}
\startdata
J0814+5609 & J081432.9+561008     & RS &  345 & 4.49 & 5.40   &  1.71 & $2.57^{+0.40}_{-0.46}$ & & BSC \\
---        & J081432.11+560956.6  & RP & 7343 & 4.49 & 2.04   &  0.39 & $3.36\pm0.12$          & & WGA \\
J0849+5108 & J084957.98+510829.0  & RP & 4472 & 3.00 & 1.19   &  0.27 & $1.77^{+0.32}_{-0.39}$ & & fit \\
J0850+4626 & no detection         &    &  469 & 2.64 &$<$4.51 &$<$0.98&  ...                   & &     \\
J0902+0443 & no detection         &    &  362 & 3.11 &$<$5.72 &$<$1.45&  ...                   & &     \\
J0948+0022 & J094856.9+002235     & RS &  447 & 5.22 & 4.10   &  1.16 & $2.26\pm0.64$          & & FSC \\
J0953+2836 & no detection         &    &  449 & 1.28 &$<$5.52 &$<$0.66&  ...                   & &     \\
J1031+4234 & no detection         &    &  498 & 0.99 &$<$6.00 &$<$0.60&  ...                   & &     \\
J1037+0036 & J103727.45+003635.6  & RP &10475 & 4.87 & 0.84   &  0.16 & $2.43^{+0.43}_{-0.49}$ & &     \\
---        & J103727.45+003635.6  & RP & 6158 & 4.87 & 2.56   &  0.51 & $3.23\pm0.29$          & &     \\
J1047+4725 & J104731.4+472528     & RS &  509 & 1.30 & 2.13   &  0.26 & ...                    & & FSC \\
J1110+3653 & no detection         &    &  315 & 1.69 &$<$7.22 &$<$1.06& ...                    & &     \\
J1138+3653 & no detection         &    &  327 & 1.75 &$<$6.16 &$<$0.94& ...                    & &     \\
J1146+3236 & J114654.0+323653     & RS &  452 & 1.42 & 1.67   &  0.21 & $2.37^{+0.81}_{-0.60}$ & & FSC \\
J1238+3942 & J123852.1+394253     & RS &  508 & 1.46 & 5.08   &  0.65 & $2.45^{+0.32}_{-0.28}$ & & BSC \\
J1246+0238 & no detection         &    &  160 & 2.01 &$<$11.6 &$<$1.98& ...                    & &     \\
J1305+5116 & no detection         &    &  499 & 0.98 &$<$5.82 &$<$0.58& ...                    & &     \\
J1435+3131 & J143509.2+313203     & RS &  382 & 1.14 &10.50   &  1.15 & ...                    & & BSC \\
J1443+4725 & J144319.8+472541     & RS &  587 & 1.47 & 1.92   &  0.25 & ...                    & & FSC \\
J1505+0326 & no detection         &    &  417 & 3.89 &$<$6.90 &$<$2.14& ...                    & &     \\
J1548+3511 & J154817.6+351130     & RS &  241 & 2.28 & 5.49   &  1.05 & ...                    & & FSC \\
J1633+4718 & J163323.58+471859.0  & RP & 3748 & 1.79 & 26.0   &  5.44 & $1.37\pm0.30$          &$32.5^{+3.9}_{-3.7}$ & fit \\
J1633+4718 & J163323.3+471848     & RS &  909 & 1.79 & 20.0   &  4.72 & $1.47^{+0.77}_{-0.92}$ &$30^{+12}_{-10}$ & fit \\
J1634+4809 & J163400.1+480915     & RS &  943 & 1.64 & 1.40   &  0.20 & ...                    & & FSC \\
J1644+2619 & J164443.2+261909     & RS &  518 & 5.12 & 10.6   &  2.60 & $2.03^{+0.25}_{-0.28}$ & & BSC \\
---        & as target            & Ch &  2946& 5.12 & 18.0\tablenotemark{b} & 1.05\tablenotemark{b} & $2.19\pm0.17$          & & fit \\
J1722+5654 & J172205.0+565450     & RS & 1415 & 2.10 & 8.11   &  1.33 & $2.40\pm0.15$          & & BSC \\
\enddata
\tablecomments{
Col.(1): object name;
Col.(2): name of X-ray source;
Col.(3): RS -- the RASS, RP -- ROSAT pointed observation, Ch -- Chandra observation;
Col.(4): effective exposure time in units of seconds (vegneting corrected);
Col.(5): Galactic absorption column density;
Col.(6): X-ray count rate as detected with the ROSAT PSPC (except one of J1644+2619 with the Chandra ACIS-S;
Col.(7): X-ray flux density corrected for Galactic absorption (0.1--2.4\,keV for ROSAT and  0.3--5\,keV band for Chandra data);
Col.(8): power-law photon index;
Col.(9): temperature of an additional soft X-ray blackbody component;
Col.(10): remarks on how the X-ray data analysis is performed;
          fit -- by spectral fitting; for the rest, the fluxes and
          $\Gamma$ are derived using count rates and hardness ratio
          (see text for details)
	  from RASS bright source catalogue (BSC), 
               RASS faint  source catalogue (FSC), and
	       ROSAT source catalogues from pointed observations (RXP)
	       and its WGA version (WGA),
               respectively.
}
\tablenotetext{a}{in units of $10^{-12}$\,\ergs }
\tablenotetext{b}{in the 0.3--5\,keV band for the Chandra ACIS-S detector}
\end{deluxetable}

%% file: tab3.tex
\begin{deluxetable}{ccccccccl}
\tabletypesize{\small}
\tablecaption{Radio flux variations  
and inferred brightness temperatures \label{tbl:radvar}}
\tablewidth{0pt} 
\tablehead{ 
\colhead{name}          & 
\colhead{$b$}             & 
\colhead{band}          & 
\colhead{$\Delta\,S$}   &
\colhead{$\Delta\,S/\langle S\rangle$} &
\colhead{$\Delta\,t$}   &
\colhead{$\log T_{\rm B}$} &
\colhead{$\delta_{\rm min}$} &
\colhead{reference} \\ 
\colhead{}      & 
\colhead{deg.}  & 
\colhead{GHz}   & 
\colhead{mJy}   &
\colhead{\%}    &
\colhead{days}  &
\colhead{$K$}   &
\colhead{}      &
\colhead{}    
}
\startdata
J0814+5609 & 33.7 & 1.4 &  9  & 14 &1273 & 12.0 & -   & 2 \\
J0849+5108 & 39.1 & 1.4 & 78  & 25 &1257 & 13.0 & 2.2 & 2,6\\
   ......  &      & 1.4 & 230 & 75 &$2191$& $13.0$ & 2.2 & 7,8\\
J0850+4626 & 39.3 & 1.4 &  5  & 27 &1223 & 11.8 & -   & 2,6\\
J0902+0443 & 31.2 & 4.85& 10  & 9  &1125 & 11.1 & -   & 1,3\\ 
J0948+0022 & 38.7 & 1.4 &  38 & 43 &1282 & 12.7 & 1.7 & 2,6\\
J1505+0326 & 50.3 & 4.85& 527 & 72 &1125 & 12.6 & 1.5 & 1,3\\
......     &      & 8.3 & 146 & 43 & 64  & 14.0 & 4.7 & 9 \\
J1548+3511 & 51.6 & 4.85& 38  & 44 &207  & 13.0 & 2.2 & 1,4 \\
J1633+4718 & 42.6 & 4.85& 24  & 55 &1370 & 9.9  & -   & 1,5\\
\enddata
\tablecomments{
band -- observing frequency;
$b$ -- Galactic latitude;
$\Delta S$-- absolute flux change between two epoches;
$\Delta S/\langle S \rangle$-- fractional amplitudes where 
$\langle S \rangle$ is the average of the two fluxes;
$\Delta t$--time span of the two epoches
(for an observation whose epoch is available accurately
only to month, the maximum length of dates is used for
conservative estimation); 
$T_{\rm B}$--brightness temperature (logrithm);
$\delta_{\rm min}$--minimum Doppler factor
of beaming assuming an intrinsic maximum $T_{\rm B}$
as the inverse Compton limit $10^{12}$\,k. 
Reference codes: (1) \citet{becker91}
(2) \citet{becker95} (3) \citet{gri95} (4) \citet{lang90}
(5) \citet{neum94} (6) \citet{cond98a} (7) \citet{wil76}
(8) \citet{arp79} 
(9) Goddard Geodetic VLBI GROUP's auxiliary web pages
http://lacerta.gsfc.nasa.gov/vlbi/images/
}
\end{deluxetable}

%% file: tab4.tex
\begin{deluxetable}{lrrrc}
\tablecaption{Broad-band effective spectral indices\tablenotemark{a} and 
          black hole masses\tablenotemark{b} \label{tbl:arox}}
\tablewidth{0pt} 
\tablehead{ 
\colhead{name} & 
\colhead{\alproe} & 
\colhead{\alprxe} & 
\colhead{\alpoxe} &
\colhead{$\log (M_{\rm BH}/M_{\odot})$}
}
\startdata
J0814+5609  & 0.44 &    0.87 &    1.75 & 8.0 \\
J0849+5108  & 0.65 &    0.84 &    1.23 & 7.4 \\
J0850+4626  & 0.44 & $>$0.68 & $>$1.17 & 7.2 \\
J0902+0443  & 0.60 & $>$0.78 & $>$1.16 & 7.7 \\
J0948+0022  & 0.57 &    0.78 &    1.22 & 7.5 \\
J0953+2836  & 0.48 & $>$0.74 & $>$1.29 & 7.8 \\
J1031+4234  & 0.44 & $>$0.71 & $>$1.27 & 7.3 \\
J1037+0036  & 0.50 &    0.79 &    1.39 & 7.3 \\
J1047+4725  & 0.70 &    0.94 &    1.45 & 8.1 \\
J1110+3653  & 0.56 & $>$0.68 & $>$0.92 & 7.1 \\
J1138+3653  & 0.47 & $>$0.66 & $>$1.07 & 7.1 \\
J1146+3236  & 0.39 &    0.74 &    1.47 & 7.8 \\
J1238+3942  & 0.41 &    0.65 &    1.16 & 6.8 \\
J1246+0238  & 0.41 & $>$0.68 & $>$1.25 & 7.3 \\
J1305+5116  & 0.40 & $>$0.78 & $>$1.56 & 8.5 \\
J1435+3131  & 0.52 &    0.71 &    1.11 & 7.5 \\
J1443+4725  & 0.51 &    0.86 &    1.58 & 7.8 \\
J1505+0326  & 0.70 & $>$0.86 & $>$1.19 & 6.6 \\
J1548+3511  & 0.52 &    0.80 &    1.38 & 7.9 \\
J1633+4718  & 0.43 &    0.68 &    1.19 & 6.3 \\
J1634+4809  & 0.44 &    0.74 &    1.35 & 7.4 \\
J1644+2619  & 0.51 &    0.74 &    1.22 & 6.9 \\
J1722+5654  & 0.43 &    0.69 &    1.24 & 7.4 \\
\enddata
\tablenotetext{a}{Calculated using the rest frame luminosities at 
5\,GHz, 2500\,\AA, and 2\,keV. The lower limits are computed
for X-ray non-detections from the X-ray flux limits set by the RASS.}
\tablenotetext{b}{Virial black hole masses estimated from the width and luminosity of the broad \hb lines (see text for details).}
\end{deluxetable}